\documentclass[11pt,fleqn]{article}
\usepackage{tikz}
\usetikzlibrary{arrows.meta,patterns}
\usepackage{amsmath,amssymb,amsthm}
\usepackage[margin=1in]{geometry}
\usepackage{xcolor}
\usepackage{thm-restate}
\usepackage{bm}
\usepackage{comment}
\usepackage{enumitem}
\usepackage[hypertexnames=false,bookmarksnumbered=true,final]{hyperref}
\usepackage[capitalize,sort]{cleveref}

\def\colorschemesepia{sepia}
\def\colorschemedark{dark}
\def\colorschemelight{light}

\ifx\colorscheme\undefined
\let\colorscheme\colorschemelight
\fi

\ifx\colorscheme\colorschemelight
\colorlet{textColor}{black}
\colorlet{bgColor}{white}
\fi

\ifx\colorscheme\colorschemesepia
\definecolor{textColor}{HTML}{433423}
\definecolor{bgColor}{HTML}{fbf0da}
\fi

\ifx\colorscheme\colorschemedark
\definecolor{textColor}{HTML}{bdc1c6}
\definecolor{bgColor}{HTML}{202124}
\colorlet{textHeavy}{white}
\definecolor{textRed}{HTML}{ff968c}  %
\definecolor{textGreen}{HTML}{70cc70}  %
\definecolor{textBlue}{HTML}{8cbcff}  %
\definecolor{textCyan}{HTML}{70cccc}  %
\definecolor{textMagenta}{HTML}{d982d9}  %
\definecolor{textYellow}{HTML}{bfbf69}  %
\else
\colorlet{textHeavy}{black}
\colorlet{textRed}{red!50!black}
\colorlet{textGreen}{green!50!black}
\colorlet{textBlue}{blue!50!black}
\colorlet{textCyan}{cyan!80!black}
\colorlet{textMagenta}{magenta!80!black}
\colorlet{textYellow}{yellow!60!black}
\definecolor{textPurple}{HTML}{681da8}
\fi

\colorlet{dimColor}{textColor!50!bgColor}

\ifx\colorscheme\colorschemelight\else
\pagecolor{bgColor}
\color{textColor}
\fi

\hypersetup{colorlinks,linkcolor=textRed,citecolor=textRed,urlcolor=textBlue}

\title{RARE: Response-Aware Rational Equilibrium\\ in Budget-Constrained Pricing Games}
\author{
Bhaskar Ray Chaudhury%
\thanks{University of Illinois at Urbana-Champaign, USA. B. Chaudhury's research was supported by the NSF CAREER award CCF-2441580. J.~Garg and E.~Sharma were supported by NSF grant CCF-2334461.}
\\ \texttt{\small braycha@illinois.edu}
\and
Jugal Garg\footnotemark[1]
\\ \texttt{\small jugal@illinois.edu}
\and
Xin Huang%
\thanks{Kyushu University, Japan. X.~Huang was supported by Wakate Grant Number 26K21169 and JST ERATO Grant Number JPMJER2301, Japan.}
\\ \texttt{\small huangxin@inf.kyushu-u.ac.jp}
\and
Eklavya Sharma\footnotemark[1]
\\ \texttt{\small eklavya2@illinois.edu}
}
\date{\empty}

\let\eps\varepsilon
\newcommand*{\defeq}{:=}
\newcommand*{\Th}{^{\textrm{th}}}
\newcommand*{\WLoG}{Without loss of generality}
\newcommand*{\wLoG}{without loss of generality}

\newcommand*{\vecOne}{\mathbf{1}}
\newcommand*{\vecZero}{\mathbf{0}}

\newcommand*{\floor}[1]{\lfloor #1 \rfloor}

\newcommand*{\norm}[1]{\lVert #1 \rVert}

\DeclareMathOperator*{\E}{\mathbb{E}}

\DeclareMathOperator*{\argmax}{argmax}

\newcommand*{\limRARE}{lim-RARE}

\DeclareMathOperator{\AffNR}{Aff^{\mathrm{NR}}}
\DeclareMathOperator{\AffR}{Aff^{\mathrm{R}}}
\DeclareMathOperator{\PO}{PO}
\DeclareMathOperator{\Dem}{Dem}
\DeclareMathOperator{\DemNR}{Dem^{\mathrm{NR}}}
\DeclareMathOperator{\DemR}{Dem^{\mathrm{R}}}
\DeclareMathOperator{\avg}{avg}
\DeclareMathOperator{\all}{all}

\DeclareMathOperator{\IR}{IR}
\DeclareMathOperator{\support}{support}

\newcommand*{\N}{\mathbb{N}}

\newcommand*{\Ical}{\mathcal{I}}
\newcommand*{\Icalhat}{\widehat{\Ical}}
\newcommand*{\Scal}{\mathcal{S}}

\newcommand*{\avec}{\bm{a}}
\newcommand*{\bvec}{\bm{b}}
\newcommand*{\evec}{\bm{e}}
\newcommand*{\pvec}{\bm{p}}
\newcommand*{\qvec}{\bm{q}}
\newcommand*{\vvec}{\bm{v}}
\newcommand*{\wvec}{\bm{w}}
\newcommand*{\xvec}{\bm{x}}
\newcommand*{\yvec}{\bm{y}}
\newcommand*{\zvec}{\bm{z}}

\newcommand*{\Ahat}{\widehat{A}}
\newcommand*{\uhat}{\widehat{u}}

\theoremstyle{plain}
\newtheorem{theorem}{Theorem}
\newtheorem{lemma}{Lemma}
\newtheorem{corollary}{Corollary}
\newtheorem{observation}{Observation}
\theoremstyle{definition}
\newtheorem{definition}{Definition}
\newtheorem{example}{Example}

\crefname{observation}{Observation}{Observations}

\allowdisplaybreaks

\makeatletter
\g@addto@macro{\UrlBreaks}{%
\do\/%
\do\a\do\b\do\c\do\d\do\e\do\f\do\g\do\h\do\i\do\j\do\k\do\l\do\m%
\do\n\do\o\do\p\do\q\do\r\do\s\do\t\do\u\do\v\do\w\do\x\do\y\do\z%
\do\A\do\B\do\C\do\D\do\E\do\F\do\G\do\H\do\I\do\J\do\K\do\L\do\M%
\do\N\do\O\do\P\do\Q\do\R\do\S\do\T\do\U\do\V\do\W\do\X\do\Y\do\Z%
\do\0\do\1\do\2\do\3\do\4\do\5\do\6\do\7\do\8\do\9%
}
\makeatother

\makeatletter
\@ifpackageloaded{enumitem}{%
\newenvironment*{tightemize}{\begin{itemize}[noitemsep]}{\end{itemize}}%
\newenvironment*{tightenum}{\begin{enumerate}[noitemsep]}{\end{enumerate}}%
}{%
\newenvironment*{tightenum}{\begin{enumerate}}{\end{enumerate}}%
}
\makeatother

\let\citep\cite
\let\citet\cite

\begin{document}

\pagenumbering{Alph}  %
\maketitle
\thispagestyle{empty}

\begin{abstract}
We study a pricing game in markets with budget-constrained buyers, where each seller offers a single good and sets its price. We consider both non-rivalrous goods, such as datasets, and rivalrous goods, such as physical goods. For non-rivalrous goods, each buyer chooses a utility-maximizing bundle within her budget. For rivalrous goods, an allocation rule selects a Pareto-optimal allocation subject to the buyers’ budgets and the goods’ supply constraints. Pure Nash equilibria often fail to exist in this game. A key reason is that Nash equilibrium assumes that the other sellers keep their prices fixed when one seller deviates. In practice, however, sellers harmed by a price change may respond, for example, by matching the new price.

We propose a new equilibrium notion called \emph{Response-Aware Rational Equilibrium} (RARE). At a RARE, every deviation by a seller can be deterred by a response from the other sellers, in which no responder earns less than she did right after the deviation. Every Nash equilibrium is a RARE. Assuming that
the allocation rule is continuous in prices,
we show that a RARE always exists for rivalrous goods, and for non-rivalrous goods when buyers have equal budgets. Moreover, the RARE that we construct extracts the buyers' entire budget. For non-rivalrous goods with unequal budgets, we show that an approximate RARE exists, where a deviating seller can increase her revenue by a factor of at most $b_{\mathrm{avg}}/b_{\min}$, the ratio of the average budget to the minimum budget. Our proof uses two nested applications of the KKM lemma.
We also study a natural dynamic in which sellers repeatedly raise their prices to earn the revenue that they could guarantee themselves by deviating. We show that, in the settings where we prove exact existence, this dynamic reaches a RARE with additive error $\delta$ within $\lfloor B/\delta \rfloor$ rounds, where $B$ is the buyers' total budget. The main results of this paper are formally verified in the Lean 4 proof assistant\texorpdfstring{ \cite{eklavya2026lean}}{}.

\end{abstract}

\newpage
\thispagestyle{empty}

{\small \tableofcontents}
\newpage
\pagenumbering{arabic}

\section{Introduction}
\label{sec:intro}

Understanding which prices emerge when sellers act strategically is a fundamental question in economics and computation~\cite{Tirole1988,Nisan2007,Babaioff2014}. We study a broad class of pricing games with budget-constrained buyers, considering both rivalrous and non-rivalrous settings. Demand is generated endogenously by buyers’ preferences and their choices of affordable bundles. Pure Nash prices need not exist in these markets, a difficulty also encountered in classical pricing models~\cite{dAspremontGabszewiczThisse1979,Edgeworth1897} and recent adaptations to data pricing~\cite{chaudhury2026equilibrium}. On closer inspection, some of this instability stems from myopic price deviations: a seller’s price change appears profitable when competitors’ prices are held fixed, yet its gains can disappear once competitors respond in their own interests.

The deployment of AI agents gives our model and stability question renewed relevance.
AI-delegated purchasing naturally motivates budget-constrained demand,
as reflected in work on shopping, procurement, negotiation, and paid tool use
\cite{FishEtAl2026EconEvals,WangEtAl2026ShoppingBench,LiuEtAl2026BudgetConstrainedAgents,ZhuEtAl2025AgentNegotiation}.
Payment protocols likewise support preauthorized price limits~\cite{Google2025AP2}.
On the supply side, strategic AI systems already model other agents' actions~\cite{FAIR2022},
and experiments show that anticipating rivals' price reactions can
influence LLM agents' pricing decisions~\cite{FishGonczarowskiShorrer2024}.
\emph{How strategic reasoning shapes market stability thus becomes a concrete question of agent design.}

In this paper, we investigate whether accounting for competitors’ rational responses can restore price stability in budget-constrained pricing games. In particular, we seek systematic and natural ways to expand the responses considered beyond competitors’ unchanged prices in Nash’s deviation test, allowing rational counterresponses to deter myopic deviations. Our approach differs from lookahead equilibria~\cite{MirrokniThainVetta2012,BiloFanelliMoscardelli2017} and opponent-aware learning methods such as LOLA~\cite{FoersterEtAl2018LOLA}: we characterize admissible competitor responses without predicting or simulating a particular continuation of play. Predicting such a continuation requires additional assumptions about competitors’ behavior; an equilibrium-based prediction must also address selection when several equilibria exist, and may be unavailable when no pure equilibrium exists. The remainder of this section introduces our market model and equilibrium notion, and gives an overview of its structural and computational properties.

\paragraph{Pricing games with budget-constrained buyers.} We consider a market with
$n$ buyers and $m$ sellers, where each seller sells a single \emph{divisible} good and sets its price $p_j \ge 0$. Each buyer $i$ has a budget $b_i > 0$ and a utility function $u_i$ over bundles of goods. Given these prices, each buyer purchases a bundle that maximizes her utility subject to her budget constraint, i.e., the total price of her bundle must be at most $b_i$. Seller $j$'s revenue is $p_j$ times the amount of good $j$ purchased by the buyers,
and each seller wants to maximize her revenue.
This puts the sellers in a \emph{pricing game} with each other, whose stable prices are the subject of this paper.

A buyer's budget is the maximum amount of money that she can spend on these goods. It need not be her entire wealth, and it can be smaller than her value for the goods. For instance, a research group that buys datasets using a grant,
or a firm that buys software licenses using a fixed annual allocation, may value more goods than it can afford.
Buyers behave the same way in Fisher markets~\cite{EisenbergGale1959,Devanur2008}. However, prices in a Fisher market are set to clear the market, whereas prices in our model are set by strategic sellers.

Budgets couple the sellers' revenues:
money that a buyer spends on one good is no longer available for other goods.
Hence, a seller's revenue depends on the prices of all goods, even when buyers have linear utilities.
This distinguishes our model from many standard formulations of Bertrand competition,
which analyze price competition through a specified aggregate demand function,
and from Bertrand--Edgeworth models, which additionally incorporate capacity constraints
and rationing~\cite{Bertrand1883,Edgeworth1897,KrepsScheinkman1983}.
In our model, demand arises from the choices of individual buyers,
each with her own budget and her own preferences over bundles of goods.

\paragraph{Rivalrous and non-rivalrous goods.}
We study two kinds of goods.
A \emph{rivalrous} good has a fixed supply that buyers share:
each seller has one unit of her good,
and the fraction of the good allocated to one buyer cannot be allocated to another.
A \emph{non-rivalrous} good can be supplied to any number of buyers:
each buyer can receive up to one unit of each good, irrespective of what other buyers receive.
Datasets are a prominent example of non-rivalrous goods,
since giving a copy of a dataset to one buyer does not prevent the seller
from giving a copy to another buyer.
This has motivated recent work on data markets~\cite{JonesTonetti2020,chaudhury2026equilibrium}.

For non-rivalrous goods, each buyer chooses her bundle independently of other buyers.
Formally, she picks a bundle $\zvec \in [0, 1]^m$
maximizing $u_i(\zvec)$ subject to $\pvec^T\zvec \le b_i$.
For rivalrous goods, prices need not clear the market,
so several buyers may want the same good, and its supply must be rationed among them.
We only assume that the rationing is \emph{Pareto optimal} for the buyers:
among all allocations that respect the goods' supply and the buyers' budgets,
no other allocation makes some buyer better off without making another buyer worse off.
In both the rivalrous and non-rivalrous settings,
there can be multiple optimal allocations at the same prices,
and they can give the sellers different revenues.
Hence, we also fix an \emph{allocation rule},
which picks one optimal allocation for each price vector $\pvec = (p_1, \ldots, p_m)$.
Let $R_j(\pvec)$ be seller $j$'s revenue at prices $\pvec$ under this allocation rule.
See \cref{sec:prelims} for the formal model.

\paragraph{Nash equilibria often fail to exist.}
The quintessential notion of stability in a game is pure Nash equilibrium.
In our setting, a price vector $\pvec$ is a pure Nash equilibrium if no seller
can increase her revenue by changing her price while the other sellers keep their prices fixed.
(We only study deterministic prices in this paper, so we drop the word `pure' from now on.)
The following instance exhibits this failure with just two sellers and linear buyer utilities.

\begin{example}
\label{ex:intro-no-nash}
Consider two non-rivalrous goods and two buyers $A$ and $B$, each with budget 1.
Their utilities are $u_A(\xvec) = \frac{3}{2}x_1 + x_2$ and $u_B(\xvec) = x_1 + \frac{3}{2}x_2$.
At prices $(1/2, 1/2)$, each buyer can afford both goods, so she buys both goods completely,
and each seller's revenue is 1.
Suppose seller 1 raises her price to $0.74$.
Buyer $A$ still gets more utility per unit of money from good 1 than from good 2,
so she buys good 1 completely and spends her remaining $0.26$ on good 2.
Buyer $B$ buys good 2 completely and spends her remaining $0.5$ on good 1.
Hence, seller 1's revenue increases to $1.24$, and seller 2's revenue drops to $0.76$
(see \cref{fig:response-example}).
So, $(1/2, 1/2)$ is not a Nash equilibrium.
Higher prices do not help either.
E.g., at prices $(1, 1)$, each buyer spends her entire budget on her favorite good,
so each seller's revenue is 1.
But then seller 1 can lower her price to just below $2/3$,
so that both buyers prefer good 1 and buy it completely,
which increases her revenue to almost $4/3$.
In fact, this market has no Nash equilibrium
(by \cref{thm:nasheq-uncommon} in \cref{sec:nasheq-uncommon}).
\end{example}

Prior work~\cite{chaudhury2026equilibrium} showed that even \emph{approximate} Nash equilibria may not exist
for non-rivalrous goods, when there are just two sellers and buyers have linear utilities.
We show that non-existence is the typical case, and not a pathological one.
Consider a market with two non-rivalrous goods and $n \ge 2$ buyers,
where each buyer $i$ has budget 1 and a linear utility function
$u_i(\xvec) = u_{i,1}x_1 + u_{i,2}x_2$, where $u_{i,1}, u_{i,2} \in [1, 2)$.
We show that such a market has a Nash equilibrium iff
all buyers have the same ratio $u_{i,1}/u_{i,2}$ (\cref{thm:nasheq-uncommon}),
irrespective of the allocation rule.
Hence, if the coefficients $u_{i,j}$ are sampled independently
from any continuous distribution over $[1, 2)$,
then with probability 1, there is no Nash equilibrium.

We obtain similar examples for rivalrous goods, where buyers have
separable piecewise-linear concave (SPLC) utilities.
The key idea is to split the unit supply of each rivalrous good into $n$ shares of size $1/n$,
and to make each buyer's marginal utility for a good drop to a tiny positive value
once she has received one share.
Then each buyer behaves as if she were buying a copy of a non-rivalrous good.
This allows us to carry over examples from the non-rivalrous setting.
We describe this connection in detail in \cref{sec:lin-nonrival-to-splc-rival}.

\paragraph{Sellers respond to deviations.}
Return to \cref{ex:intro-no-nash}.
After seller 1 raises her price from $1/2$ to $0.74$,
seller 2 can match the new price by also charging $0.74$.
By symmetry, each seller then earns 1 (see \cref{fig:response-example}).
Seller 2 thereby increases her revenue from $0.76$ to 1
and eliminates seller 1's gain.
Anticipating this response removes seller 1's incentive to make the initial change.
Price responsiveness also has familiar institutional forms;
for example, Best Buy offers a price-match guarantee for qualifying
competitor offers~\cite{BestBuyPriceMatch}.

\begin{figure}[t]
\centering
\begin{tikzpicture}[x=1cm,y=1cm,font=\fontsize{9}{10.8}\selectfont,
  panel/.style={draw=black!35,fill=white,rounded corners=2pt,minimum width=4.45cm,minimum height=4.55cm},
  move/.style={-{Stealth[length=2mm]},line width=.8pt,draw=black!65}]
\path[use as bounding box] (-.15,-.4) rectangle (15.05,5.5);
\node at (7.45,5.17) {$u_A=1.5x_1+x_2$, $u_B=x_1+1.5x_2$; both budgets are $1$};
\foreach \xc/\heading/\prices/\rev in {2.25/{Starting prices}/{(.50,.50)}/{(1,1)},7.45/{Seller 1 increases}/{(.74,.50)}/{(1.24,.76)},12.65/{Seller 2 matches}/{(.74,.74)}/{(1,1)}} {
  \node[panel] at (\xc,2.37) {};
  \node[font=\fontsize{9}{10.8}\selectfont\bfseries] at (\xc,4.22) {\heading};
  \node at (\xc,3.69) {$p=\prices$};
  \node[align=center] at (\xc,.59) {Seller revenues\\$R=\rev$};
}
\draw[move] (4.48,2.4)--(5.2,2.4);
\draw[move] (9.68,2.4)--(10.4,2.4);
\foreach \xc in {2.25,7.45,12.65} {
  \node[anchor=west] at ({\xc-1.87},3.05) {Buyer A spends};
  \node[anchor=west] at ({\xc-1.87},1.92) {Buyer B spends};
}
\foreach \xc/\aa/\bb in {2.25/.50/.50,7.45/.74/.50,12.65/.74/.26} {
  \foreach \yy/\share in {2.5/\aa,1.37/\bb} {
    \fill[blue!22] ({\xc-1.8},\yy-.22) rectangle ({\xc-1.8+3.6*\share},\yy+.22);
    \fill[teal!30] ({\xc-1.8+3.6*\share},\yy-.22) rectangle ({\xc+1.8},\yy+.22);
    \draw[black!60] ({\xc-1.8},\yy-.22) rectangle ({\xc+1.8},\yy+.22);
    \draw[black!60] ({\xc-1.8+3.6*\share},\yy-.22)--({\xc-1.8+3.6*\share},\yy+.22);
  }
}
\node at ({2.25-0.9},2.5) {$.50$}; \node at ({2.25+0.9},2.5) {$.50$};
\node at ({2.25-0.9},1.37) {$.50$}; \node at ({2.25+0.9},1.37) {$.50$};
\node at ({7.45-0.45},2.5) {$.74$}; \node at ({7.45+1.35},2.5) {$.26$};
\node at ({7.45-0.9},1.37) {$.50$}; \node at ({7.45+0.9},1.37) {$.50$};
\node at ({12.65-0.45},2.5) {$.74$}; \node at ({12.65+1.35},2.5) {$.26$};
\node at ({12.65-1.35},1.37) {$.26$}; \node at ({12.65+0.45},1.37) {$.74$};
\node[draw=blue!55!black,fill=blue!22,minimum width=.23cm,minimum height=.16cm,inner sep=0] at (4.35,-.12) {};
\node[anchor=west] at (4.58,-.12) {Seller 1};
\node[draw=teal!65!black,fill=teal!30,minimum width=.23cm,minimum height=.16cm,inner sep=0] at (7.01,-.12) {};
\node[anchor=west] at (7.24,-.12) {Seller 2};
\node[anchor=west] at (10.02,-.12) {IR: $1\ge .76$};
\end{tikzpicture}

\caption{The market of \cref{ex:intro-no-nash}.
Each bar shows how a buyer splits her budget between the two goods.
If seller 1 raises her price and seller 2 keeps her price fixed, then seller 1 gains.
If seller 2 responds by matching seller 1's new price,
then seller 2's revenue increases and seller 1's revenue drops back to its original value.}
\label{fig:response-example}
\end{figure}

\paragraph{Response-Aware Rational Equilibrium (RARE).}
Let $\pvec$ be a price vector, and suppose seller $j$ deviates by changing her price to $q_j$.
A \emph{response} by the other sellers is a vector $\qvec_{-j}$ of new prices for them,
and $\qvec \defeq (q_j, \qvec_{-j})$ is the price vector after the response.
We say that the response is \emph{individually rational} (IR) if no responder loses by responding, i.e.,
\[ R_k(\qvec) \ge R_k((q_j, \pvec_{-j})) \quad \text{for every responder } k \in [m] \setminus \{j\}. \]
Note that each responder compares her revenue after responding
to her revenue right after the deviation, and not to her revenue before the deviation.
A price vector $\pvec$ is a \emph{Response-Aware Rational Equilibrium} (RARE) if
every deviation can be deterred by an IR response, i.e.,
for every seller $j$ and every deviation $q_j \ge 0$,
there exists an IR response $\qvec_{-j}$ such that $R_j(\qvec) \le R_j(\pvec)$.
Equivalently, no seller has a deviation that is profitable against every IR response.
For $\kappa \ge 1$, we say that $\pvec$ is a \emph{$\kappa$-RARE} if
we only require $R_j(\qvec) \le \kappa R_j(\pvec)$, i.e.,
every deviation by a seller admits an IR response under which her revenue is at most $\kappa$ times her revenue at $\pvec$.

Every Nash equilibrium is a RARE,
since keeping the other sellers' prices unchanged is always an IR response.
However, a RARE can exist even when there is no Nash equilibrium.
In \cref{ex:intro-no-nash}, the prices $(1/2, 1/2)$ form a RARE.
To see this, suppose seller 1 deviates to price $q_1$ (deviations by seller 2 are symmetric).
If $q_1 < 1/2$, then seller 1's revenue is at most $2q_1 < 1$, so no response is needed.
If $q_1 \ge 1/2$ and seller 1's revenue increases beyond 1,
then seller 2's revenue drops below 1, since both buyers exhaust their budgets.
Seller 2 can then match seller 1's price, after which each seller's revenue is exactly 1.
This response is IR and eliminates seller 1's gain.
The same argument shows that $(t, t)$ is a RARE for every $t \ge 1/2$,
so a market can have many RARE.

\paragraph{Why must each responder weakly gain?}
RARE requires each responder to weakly gain by responding.
One could instead only require that the responders' total revenue does not decrease.
However, such a response may need some responders to sacrifice their revenue for the benefit of others,
which is not credible unless the responders can pay each other.
E.g., consider a market with three non-rivalrous goods and two buyers,
each with budget 1 and utility $x_1 + x_2 + x_3$,
where the allocation rule breaks ties between goods in favor of the lower-numbered good.
At prices $(0.8, 0.8, 0.8)$, the sellers' revenues are $(1.6, 0.4, 0)$.
If seller 2 deviates to price $0.5$, the revenues become $(1, 1, 0)$.
If sellers 1 and 3 respond by changing their prices to $0.31$ and $0.49$, respectively,
then the revenues become $(0.62, 0.4, 0.98)$ (see \cref{fig:ir-payments}).
This response eliminates seller 2's gain and increases the responders' total revenue from 1 to $1.6$,
but seller 1's revenue drops from 1 to $0.62$.
So seller 1 has no reason to take part in this response, and RARE does not allow it.

\begin{figure}[t]
\centering
\begin{tikzpicture}[x=1cm,y=1cm,font=\fontsize{9}{10.8}\selectfont]
\path[use as bounding box] (-.2,-.35) rectangle (15,5.25);
\node at (7.4,4.94) {Two identical buyers: budget $1$, $u(x)=x_1+x_2+x_3$; ties favor $1$, then $2$, then $3$};
\foreach \xc/\ttl/\price in {2.3/{Original}/{(.8,.8,.8)},7.4/{Seller 2 deviates}/{(.8,.5,.8)},12.5/{Joint response}/{(.31,.5,.49)}} {
  \node[font=\fontsize{9}{10.8}\selectfont\bfseries] at (\xc,4.13) {\ttl};
  \node at (\xc,3.57) {$p=\price$};
}
\foreach \xc/\a/\b/\c in {2.3/.8/.2/0,7.4/.5/.5/0,12.5/.31/.2/.49} {
  \fill[blue!24] ({\xc-2.05},2.22) rectangle ({\xc-2.05+4.1*\a},2.87);
  \fill[orange!25] ({\xc-2.05+4.1*\a},2.22) rectangle ({\xc-2.05+4.1*(\a+\b)},2.87);
  \ifdim\c pt>0pt
    \fill[teal!28] ({\xc-2.05+4.1*(\a+\b)},2.22) rectangle ({\xc+2.05},2.87);
    \fill[pattern=north east lines,pattern color=teal!40!black] ({\xc-2.05+4.1*(\a+\b)},2.22) rectangle ({\xc+2.05},2.87);
  \fi
  \draw[black!65] ({\xc-2.05},2.22) rectangle ({\xc+2.05},2.87);
  \draw[black!65] ({\xc-2.05+4.1*\a},2.22)--({\xc-2.05+4.1*\a},2.87);
  \ifdim\c pt>0pt\draw[black!65] ({\xc-2.05+4.1*(\a+\b)},2.22)--({\xc-2.05+4.1*(\a+\b)},2.87);\fi
  \node at (\xc,1.8) {$R/2=(\a,\b,\c)$};
}
\draw[-{Stealth[length=2mm]},black!65,line width=.8pt] (4.52,2.55)--(5.18,2.55);
\draw[-{Stealth[length=2mm]},black!65,line width=.8pt] (9.62,2.55)--(10.28,2.55);
\node[align=center,text width=4.35cm] at (2.3,.83) {All bars show revenue\\\emph{per buyer}.};
\node[align=center,text width=4.35cm] at (7.4,.83) {Seller 2 gains initially:\\$.2\longrightarrow .5$.};
\node[align=center,text width=4.35cm] at (12.5,.83) {Response deters seller 2, but\\seller 1 loses: $.5\to .31$.};
\node[draw=blue!55!black,fill=blue!24,minimum width=.23cm,minimum height=.16cm,inner sep=0] at (3.08,-.04) {};
\node[anchor=west] at (3.3,-.04) {Seller 1};
\node[draw=orange!70!black,fill=orange!25,minimum width=.23cm,minimum height=.16cm,inner sep=0] at (6.83,-.04) {};
\node[anchor=west] at (7.05,-.04) {Seller 2};
\node[draw=teal!65!black,fill=teal!28,pattern=north east lines,pattern color=teal!40!black,minimum width=.23cm,minimum height=.16cm,inner sep=0] at (10.58,-.04) {};
\node[anchor=west] at (10.8,-.04) {Seller 3};
\end{tikzpicture}

\caption{The market with three goods from the paragraph on why each responder must gain.
Each bar shows how a buyer splits her budget among the goods,
which is also each seller's revenue per buyer.
The response by sellers 1 and 3 eliminates seller 2's gain
and increases the responders' total revenue,
but seller 1's revenue drops compared to her revenue right after the deviation.
Hence, RARE does not allow this response.}
\label{fig:ir-payments}
\end{figure}

\subsection{Our Results}
\label{sec:intro:results}

Throughout this paper, we assume that every buyer's utility function is continuous,
strictly monotone (i.e., getting more of any good strictly increases her utility),
and is 0 for the empty bundle.
Let $B \defeq \sum_{i=1}^n b_i$ be the buyers' total budget,
and let $b_{\avg} \defeq B/n$ and $b_{\min} \defeq \min_{i=1}^n b_i$
be the average and minimum budgets, respectively.

\paragraph{Existence of RARE.}
Our main result is that approximate RARE always exist when the allocation rule is continuous
(\cref{thm:rare-exists-continuous-corr}).
Specifically, in any market with a continuous allocation rule, a $\kappa$-RARE exists,
where $\kappa = 1$ for rivalrous goods and $\kappa = b_{\avg}/b_{\min}$ for non-rivalrous goods.
Hence, an exact RARE exists for rivalrous goods,
and for non-rivalrous goods when all buyers have the same budget.
This holds for any number of buyers and sellers, and does not require the utilities to be concave.
Moreover, the RARE that we construct has high revenue:
the sellers' total revenue is $B$ for rivalrous goods and $nb_{\min}$ for non-rivalrous goods.
Our proof uses two nested applications of
the Knaster--Kuratowski--Mazurkiewicz (KKM) lemma~\cite{KKM1929} (see \cref{sec:overview:exist}).

\paragraph{When is the allocation rule continuous?}
For non-rivalrous goods and strictly concave utilities,
each buyer has a unique optimal bundle for each price vector,
and the resulting allocation rule is continuous (\cref{thm:unique-nr-demand}).
Hence, a $(b_{\avg}/b_{\min})$-RARE exists.
For rivalrous goods, even when utilities are strictly concave,
there can be many Pareto-optimal allocations at the same prices.
A natural way to pick one of them is to maximize a \emph{welfare function} $W$ of the buyers' utilities.
When utilities are strictly concave and $W$ is concave (and continuous and strictly monotone),
there is a unique maximizer, and the resulting allocation rule is continuous (\cref{thm:unique-rival-demand}).
Hence, a RARE exists.
Such welfare functions include the (weighted) sum of utilities,
and approximations of priority orders, Nash social welfare, and leximin welfare
(see \cref{sec:exist:rival}).

Linear utilities are concave but not strictly concave.
For them, a buyer can have many optimal bundles,
and every allocation rule can be discontinuous (\cref{ex:lin-util-dem-discont}).
Still, we show that when utilities are concave, a slightly different form of $\kappa$-RARE exists,
where instead of fixing an allocation rule,
we can pick an optimal allocation separately for each deviation and response
(\cref{thm:nonrival-rare-exists,thm:rival-rare-exists}).
(For rivalrous goods, the allocation must maximize a fixed concave welfare function $W$.)
We prove this by slightly perturbing the utilities to make them strictly concave,
and then taking limits.

\paragraph{Dynamics.}
We also study a natural dynamic that leads to RARE (\cref{sec:dynamics}).
For a seller $j$ and the other sellers' prices $\pvec_{-j}$,
let $V_j(\pvec_{-j})$ be the highest revenue that seller $j$ can guarantee by deviating,
assuming that the other sellers respond with the IR response that is worst for seller $j$
(see \cref{defn:vj} for the formal definition).
Then $\pvec$ is a RARE (up to an arbitrarily small additive error)
iff $R_j(\pvec) \ge V_j(\pvec_{-j})$ for every seller $j$ (\cref{thm:rare-equiv-defn}).

We describe our dynamic for non-rivalrous goods when all buyers have the same budget.
All prices are initially 0.
In each round, some of the sellers $j$ for which $p_j < V_j(\pvec_{-j})/n$
raise their price to $V_j(\pvec_{-j})/n$.
We show that throughout the dynamic, every buyer buys every good completely.
Hence, a seller $j$ who raises her price ends up with revenue exactly $V_j(\pvec_{-j})$,
and the other sellers' revenues do not change.
In other words, each seller either keeps her price,
or raises it to get the maximum revenue that she could have guaranteed herself by deviating.

For $\delta \ge 0$, we say that $\pvec$ is a \emph{$\delta$-additive RARE}
if we only require $R_j(\qvec) \le R_j(\pvec) + \delta$ in the definition of RARE.
For continuous allocation rules, we show that if the sellers who raise their prices in each round
include the \emph{most unsatisfied} seller (i.e., a seller maximizing $V_j(\pvec_{-j})/n - p_j$),
then for every $\delta > 0$, the prices become a $\delta$-additive RARE within $\floor{B/\delta}$ rounds,
and after some round, they remain a $\delta$-additive RARE forever
(\cref{thm:dynamics-convergence}).
This covers both the case where sellers update their prices one at a time,
and the case where all unsatisfied sellers update their prices simultaneously.
The same results hold for rivalrous goods (with arbitrary budgets),
where each seller sells one unit of her good, so she raises her price to $V_j(\pvec_{-j})$ instead.

\paragraph{Lean verification and AI assistance.}
Our results on the existence of RARE, the convergence of dynamics to RARE,
and the non-existence of Nash equilibrium
have been formally verified in the Lean 4 proof assistant \cite{eklavya2026lean}. We used generative AI tools to help develop RARE and its theoretical analysis; details are provided in the AI Methodology Statement at the end of the paper.

\paragraph{Open questions.}
Our work raises several questions.
Does an exact RARE always exist for non-rivalrous goods when buyers have different budgets?
For rivalrous goods, which other natural allocation rules admit a RARE?
How efficiently can one compute $V_j(\pvec_{-j})$ or a RARE?
Finally, RARE can be defined for any game, so it would be interesting to study it beyond pricing games.

\subsection{Further Related Work}
\label{sec:intro:related}

Our work connects two strands of research: markets with budget-constrained
buyers, which motivate our economic model, and approaches to strategic
anticipation, which motivate our equilibrium notion. We discuss data markets,
classical response-based equilibrium concepts, and learning and conditional
strategies, before relating our setting to the competitive-equilibrium
framework of Fisher markets.

\paragraph{Data markets.}
Markets for data and information allow sellers to monetize information used
for prediction and other decision-making tasks. Their study encompasses
strategic behavior, incentives, revenue optimization, and externalities.
Most closely related to our work, \citet{chaudhury2026equilibrium} study pricing
in markets for non-rivalrous data with budget-constrained buyers. They show
that Nash equilibria may fail to exist, and obtain approximate stability by
allowing richer pricing functions while restricting the permitted deviations.
We instead retain the original pricing strategies and broaden the
counterresponses considered when evaluating deviations; our model also
accommodates rivalrous goods.
Other work studies optimal information and data sales
\citep{admati1986monopolistic,admati1990direct,bergemann2018design,
BabaioffKP12,cai2020sell,mehta2021sell,chaudhury2026revenue,
chaudhury2026nonlinear}, the design of data marketplaces
\citep{AgarwalDS19}, the economic consequences of non-rivalry
\citep{JonesTonetti2020}, and data externalities and their implications for
equilibria and auctions
\citep{AgarwalDHR20,Hossain024,bergemann2022economics}.
Further work examines privacy and participation incentives for contributing
data
\citep{fallah2024optimal,cummings2023optimal,fallah2022bridging,
MurhekarYCLM23}, and stable outcomes in non-monetary data exchange economies
\citep{BhaskaraGIKMS24,akrami2025theoretical,song2025existence}.
See \citet{pei2020survey} for a broader survey of data pricing.

\paragraph{Anticipating responses.}
Anticipating competitors' reactions has a long history in economics,
including conjectural-variations models of oligopoly, where firms optimize
against conjectured rival responses; Bresnahan's consistent-conjectures
approach requires these conjectures to agree locally with rivals'
reaction functions \citep{Bresnahan1981}.
In markets with asymmetric information, Wilson's anticipatory equilibrium
evaluates a new offer accounting for the withdrawal of existing offers that
become unprofitable \citep{Wilson1977}.
Riley's reactive equilibrium instead allows profitable counteroffers that
undermine the profitability of a new offer \citep{Riley1979};
see \citet{Riley2001} for a survey.
Related deterrence ideas appear in sequential stability for graph models
of conflict, where profitable unilateral moves are deterred by subsequent
counter-moves that improve the responding players' own outcomes \citep{FangHipelKilgour1989}.
Nash-2 equilibrium similarly evaluates profitable deviations against
opponents' profitable responses \citep{Sandomirskaia2019};
RARE shares this basic deterrence principle, but requires individual
rationality at the resulting joint response: every competitor must
weakly improve her revenue relative to the post-deviation profile,
including competitors whose prices remain unchanged.

\paragraph{Lookahead, opponent-aware learning, and conditional strategies.}
Lookahead equilibria evaluate moves through predicted continuations of play,
typically with a specified horizon and response
process~\cite{MirrokniThainVetta2012,BiloFanelliMoscardelli2017}.
There is also substantial research on opponent-aware learning methods. LOLA differentiates through an opponent's predicted
gradient step, allowing an agent to shape the opponent's learning to improve its own return~\cite{FoersterEtAl2018LOLA}. However, when both agents use LOLA, each models the other as an ordinary gradient learner, making their predictions inconsistent with the actual updates. COLA addresses this mismatch by learning mutually consistent update functions: each agent anticipates an opponent who is also shaping its learning~\cite{WilliEtAl2022COLA}. POLA addresses a different limitation, replacing parameter-space gradient steps with proximal optimization based on distances between policies~\cite{ZhaoEtAl2022POLA}.

Repeated games and program equilibria support conditional behavior through histories of play and access to opponents' code or
simulations~\cite{FudenbergMaskin1986,Tennenholtz2004,Cooper2025}.
RARE instead characterizes admissible joint responses directly through individual rationality at the resulting joint outcome, without predicting a particular continuation or prescribing competitors' learning rules. Its stability criterion is defined in the original one-shot game, independently of the dynamics used to reach it.

\paragraph{Fisher markets and competitive equilibrium.}
Fisher markets provide a classical model of allocating rivalrous goods
to budget-constrained buyers. Competitive equilibrium combines buyers'
utility maximization at given prices with market clearing.
Our model shares the budget-constrained demand perspective, but studies
prices chosen strategically by revenue-maximizing sellers.
\citet{LiTang2024} study seller competition in linear
Fisher markets using stable allocations and a maximin equilibrium
criterion, and consider price-matching responses in a duopoly extension.
Their maximin benchmark addresses uncertainty about allocation selection,
whereas RARE evaluates deviations against competitors' individually
rational price responses.
On the computational side, linear Fisher equilibria admit polynomial-time
algorithms \citep{Devanur2008,Orlin10,Vegh12}, whereas computing an
approximate Fisher equilibrium is PPAD-hard even for separable
piecewise-linear concave utilities \citep{ChenT09}.
Related hardness results for more general exchange markets appear in
\citet{ChenDDT09,GargMVY17}.

\section{Technical Overview}
\label{sec:overview}

In this section, we sketch the key ideas behind our main results:
the existence of RARE (\cref{sec:overview:exist}) and our dynamic (\cref{sec:overview:dynamics}).
To keep things simple, we only consider non-rivalrous goods and buyers with equal budgets,
and we normalize the common budget to 1.
Our proofs for rivalrous goods and for unequal budgets use the same ideas (see \cref{sec:exist}).

\subsection{Existence of RARE}
\label{sec:overview:exist}

Fix a continuous allocation rule, and let $R_j(\pvec)$ be seller $j$'s revenue at prices $\pvec$.
Let $R(\pvec) \defeq (R_1(\pvec), \ldots, R_m(\pvec))$ be the vector of all sellers' revenues.
The main difficulty in proving that a RARE exists is that a single price vector
must handle every seller and every deviation simultaneously.
A separate fixed-point argument for each seller would not give us this.
Our proof uses two nested applications of the KKM lemma (\cref{thm:kkm}):
an \emph{inner KKM}, which, for fixed prices, constructs a response that restores every seller's revenue,
and an \emph{outer KKM}, which finds prices at which every bounded deviation has such a response.

\paragraph{Restricting the search to the simplex.}
Let $\Delta_m \defeq \{\pvec \in \mathbb{R}_{\ge 0}^m: \sum_{j=1}^m p_j = 1\}$.
At any prices $\pvec \in \Delta_m$, every buyer can afford all the goods,
so by strict monotonicity of utilities, she buys all of them completely.
Hence, each seller $j$'s revenue is $np_j$.
If seller $j$ deviates to a lower price $q_j < p_j$,
then her revenue is at most $nq_j < np_j$, irrespective of the other sellers' prices.
So keeping the other sellers' prices unchanged deters the deviation.
Hence, by looking for a RARE in $\Delta_m$, we only need to handle deviations to higher prices.

Now suppose seller $j$ deviates to a higher price $q_j > p_j$.
Each responder $k$ still has price $p_k$ right after the deviation, so she earns at most $np_k$.
Hence, any response $\qvec_{-j}$ that gives each responder at least her original revenue is IR.
If it also leaves the deviator earning no more than her original revenue, it deters the deviation.
Therefore, it suffices to find prices $\pvec^* \in \Delta_m$ such that
\begin{equation}
\label{eq:overview-restoration}
\begin{aligned}
& \forall j \in [m], \; \forall q_j > p^*_j, \; \exists \qvec_{-j} \ge \pvec^*_{-j} \; \text{ such that }
\\ & R_j(\qvec) \le n p^*_j \quad\text{and}\quad R_k(\qvec) \ge n p^*_k \quad \forall k \neq j.
\end{aligned}
\end{equation}
This is stronger than RARE, since each responder recovers at least her revenue from before the deviation,
instead of merely her revenue immediately after the deviation.
We also use this stronger property to analyze our dynamic (\cref{sec:overview:dynamics}).

\paragraph{Which sellers can gain by raising their price?}
At prices $\pvec \in \Delta_m$, all goods are sold completely,
but buyers may like some goods much more than others,
so some sellers may be able to raise their price and earn more.
E.g., suppose there are two goods with prices $(1/2, 1/2)$,
and every buyer has utility $2x_1 + x_2$, so she likes good 1 much more than good 2.
If seller 1 raises her price to $s \in (1/2, 1)$,
then every buyer still prefers good 1, so she buys it completely
and spends only her remaining budget $1 - s$ on good 2.
Hence, seller 1's revenue increases to $ns$,
and seller 2 cannot undo this by raising her price.
On the other hand, if seller 2 raises her price,
then every buyer still buys good 1 completely and spends her remaining budget $1/2$ on good 2,
so seller 2's revenue does not increase.
Thus, at prices $(1/2, 1/2)$, seller 1 can gain by raising her price, but seller 2 cannot.

\emph{In general, at any prices $\pvec \in \Delta_m$, some sellers may be able to gain by raising their price in a way that the other sellers cannot undo; however, at least one seller with a positive price has no such deviation!}
We say that a seller \emph{labels} $\pvec$ if she cannot gain this way.
Our goal is to find prices that are labeled by every seller.
Intuitively, a seller can gain by raising her price
when her good is more sought after than the other goods at the current prices.
The key insight is that being more sought after is a relative property:
not every good can be more sought after than the others.
Hence, every price vector should be labeled by at least one seller (with a positive price).
We make this precise in the outer KKM below.
We do not need to understand which goods are more sought after,
which can depend in complicated ways on the buyers' utilities.
Instead, the KKM lemma uses this insight to give us prices that are labeled by every seller.

\paragraph{KKM as a continuous version of Sperner's lemma.}
Sperner's lemma labels the vertices of a triangulation of a simplex,
restricts which labels can appear on the simplex's faces,
and guarantees a small simplex that has all the labels.
The KKM lemma~\cite{KKM1929} is a continuous analogue of Sperner's lemma.
Let $C_1, \ldots, C_m$ be closed subsets of $\Delta_m$,
and think of $C_j$ as the set of points that have label $j$; a point can have multiple labels.
Suppose that for every non-empty $S \subseteq [m]$,
every point on the $S$-face of $\Delta_m$ (i.e., every point whose support is contained in $S$)
has some label from $S$.
Then some point has all the labels, i.e., $\bigcap_{j=1}^m C_j \ne \emptyset$.
E.g., each vertex must have its own label, each edge must be covered by the labels of its endpoints,
and points in the interior can have any label (see \cref{fig:kkm-simplex}).

\begin{figure}[t]
\centering
\begin{tikzpicture}[x=1cm,y=1cm,font=\fontsize{9}{10.8}\selectfont]
\path[use as bounding box] (-.2,-.35) rectangle (15,5.45);
\coordinate (v1) at (1.0,.7);\coordinate (v2) at (7.0,.7);\coordinate (v3) at (4.0,4.9);
\coordinate (ps) at (4.0,2.10);
\fill[blue!16] (v1)--(4,.7)--(ps)--(2.5,2.8)--cycle;
\fill[orange!20] (v2)--(5.5,2.8)--(ps)--(4,.7)--cycle;
\fill[teal!17] (v3)--(2.5,2.8)--(ps)--(5.5,2.8)--cycle;
\draw[black!65,line width=.8pt] (v1)--(v2)--(v3)--cycle;
\draw[black!45,dashed] (4,.7)--(ps)--(2.5,2.8);
\draw[black!45,dashed] (ps)--(5.5,2.8);
\node at (2.65,1.4) {$C_1$};\node at (5.35,1.4) {$C_2$};\node at (4,3.44) {$C_3$};
\fill[black] (ps) circle (2.8pt);
\node[anchor=north west,fill=white,inner sep=2pt] at (4.11,2.03) {$\pvec^*$};
\node[anchor=north east] at (.95,.61) {$(1,0,0)$};
\node[anchor=north west] at (7.05,.61) {$(0,1,0)$};
\node[anchor=south] at (4,4.97) {$(0,0,1)$};
\node[anchor=west,text width=6.4cm,align=left] at (8.55,4.15)
 {$\displaystyle\Delta_3 = \{\pvec \in \mathbb{R}_{\ge 0}^3: \textstyle\sum_{j=1}^3 p_j = 1\}$\\[3pt]Every buyer can afford all the goods.};
\node[anchor=west,text width=6.4cm,align=left] at (8.55,2.56)
 {$C_j$: every deviation by seller $j$ to a higher price can be met by a response that restores every seller's revenue.};
\node[anchor=west,text width=6.4cm,align=left] at (8.55,.99)
 {Inner KKM proves the covering condition.\\Outer KKM gives $\pvec^* \in C_1 \cap C_2 \cap C_3$.};
\end{tikzpicture}

\caption{A schematic KKM cover of $\Delta_3$ for three sellers.
Each face is covered by the labels of its vertices, so some point $\pvec^*$ lies in all three sets.
In the outer KKM, $\pvec \in C_j$ means that every (bounded) deviation by seller $j$ to a higher price
can be met by a response that restores every seller's revenue.}
\label{fig:kkm-simplex}
\end{figure}

\paragraph{Inner KKM: raising prices to restore revenues.}
Fix prices $\pvec \in \Delta_m$, and for simplicity, assume that all prices are positive.
For each seller $j$, pick an arbitrary cap $z_j > p_j$ on her price.
We show that there exist prices $\qvec$ in the box $\prod_{j=1}^m [p_j, z_j]$
such that some seller's price reaches her cap, i.e., $q_i = z_i$ for some seller $i$,
and every seller gets her original revenue back, i.e., $R(\qvec) = n\pvec$
(\cref{thm:prop-box}; see \cref{fig:kkm-box}).
However, we cannot control which seller's price reaches her cap.

For any direction $\theta \in \Delta_m$, we raise the prices from $\pvec$ in direction $\theta$
until some seller's price reaches her cap, i.e., we go to the point
\[ y(\theta) \defeq \pvec + \tau(\theta)\theta, \quad \text{where} \quad
    \tau(\theta) \defeq \bigg(\max_{j \in [m]} \frac{\theta_j}{z_j - p_j}\bigg)^{-1}. \]
The map $y$ is continuous.
Label $\theta$ with $j$ if $R_j(y(\theta)) \ge np_j$, i.e., if seller $j$ gets her original revenue back.
(Note that here we label directions, which is different from labeling prices as above.)
Since the allocation rule is continuous, the label sets are closed.
To check the covering condition, let $\theta$ lie on the $S$-face of $\Delta_m$.
Then every seller outside $S$ keeps her old price, so she earns at most her old revenue.
If every seller in $S$ also earned less than her old revenue,
then the total revenue at $y(\theta)$ would be less than $n$.
But since $y(\theta) \ge \pvec$, the sum of prices at $y(\theta)$ is at least 1,
so by strict monotonicity of utilities, every buyer spends her entire budget,
and the total revenue is exactly $n$.
Hence, $\theta$ has some label from $S$.
By KKM, some direction $\theta$ has all labels, i.e., $R_j(y(\theta)) \ge np_j$ for every seller $j$.
Since the entries of both $R(y(\theta))$ and $n\pvec$ sum to $n$,
we get $R(y(\theta)) = n\pvec$.

\begin{figure}[t]
\centering
\begin{tikzpicture}[x=1cm,y=1cm,font=\fontsize{9}{10.8}\selectfont,
  ray/.style={draw=blue!55!black,line width=.65pt},
  axis/.style={-{Stealth[length=2mm]},draw=black!65}]
\path[use as bounding box] (-.2,-.3) rectangle (15,5.4);
\coordinate (p) at (1.05,.72);
\fill[blue!4] (p) rectangle (7.1,4.7);
\draw[black!60] (p) rectangle (7.1,4.7);
\draw[axis] (.7,.42)--(7.55,.42) node[right] {$q_1$};
\draw[axis] (.7,.42)--(.7,5.03) node[above] {$q_2$};
\foreach \x/\y in {7.1/1.7,7.1/2.8,7.1/3.8,6.2/4.7,4.6/4.7,2.8/4.7} {\draw[ray] (p)--(\x,\y);}
\draw[blue!65!black,line width=1.1pt] (p)--(7.1,3.8);
\fill[orange!75!black] (7.1,3.8) circle (2.5pt);
\node[anchor=west,fill=white,inner sep=2pt] at (7.26,3.8) {$y(\theta)$};
\fill[black] (p) circle (2.4pt);
\node[anchor=north west] at (1.03,.64) {$\pvec = (0.5, 0.5)$};
\node[anchor=south east] at (7.1,5.09) {$\zvec = (0.95, 0.9)$};
\draw[blue!60!black,-{Stealth[length=2mm]}] (2.2,1.68) arc[start angle=42,end angle=70,radius=1.55];
\node[fill=white,inner sep=1pt] at (2.6,2.09) {$\theta$};
\fill[teal!70!black] (4.6,4.7) circle (2.5pt);
\node[anchor=south,fill=white,inner sep=1pt] at (3.75,4.79) {$\qvec$: revenues restored};
\node[anchor=west,text width=6.1cm,align=left] at (8.82,4.15)
 {\textbf{Choose a direction $\theta$.}\\Raise prices from $\pvec$ in direction $\theta$ until some seller's price reaches her cap.};
\node[anchor=west,text width=6.1cm,align=left] at (8.82,2.53)
 {\textbf{Label by revenue.}\\Label $\theta$ with $j$ if $R_j(y(\theta)) \ge np_j$.};
\node[anchor=west,text width=6.1cm,align=left] at (8.82,.87)
 {\textbf{Intersect the labels.}\\KKM gives $\qvec$ with $R(\qvec) = n\pvec$, where some seller's price is at her cap.};
\end{tikzpicture}

\caption{The inner KKM for two sellers.
We fix prices $\pvec$ and caps $\zvec$, and raise the prices from $\pvec$ in different directions $\theta$
until they reach the upper boundary of the box $\prod_{j=1}^m [p_j, z_j]$.
KKM gives a direction in which every seller's revenue is restored,
and some seller's price is at her cap.
The marked point $\qvec$ is schematic.}
\label{fig:kkm-box}
\end{figure}

\paragraph{Outer KKM: one price vector for all sellers.}
We now apply KKM to the price simplex $\Delta_m$ itself.
We only need to consider deviations where each seller $j$'s price is at most some large bound $L_j$
(otherwise good $j$ becomes so expensive that no buyer will spend a non-negligible amount on it
if the other goods can exhaust her budget, making the deviation pointless;
we formalize this in \cref{sec:exist:overpricing}),
and we will find responses where each responder $k$'s price is at most $L_k$.
Formally, let $C_j$ be the set of prices $\pvec \in \Delta_m$ that are labeled by seller $j$, i.e.,
for every deviation $q_j \in [p_j, L_j]$ by seller $j$,
there is a response $\qvec_{-j} \in \prod_{k \ne j} [p_k, L_k]$ such that $R(\qvec) = n\pvec$.
Then any point in $\bigcap_{j=1}^m C_j$ satisfies \cref{eq:overview-restoration} for bounded deviations.
To show that $\bigcap_{j=1}^m C_j$ is non-empty using the KKM lemma,
we must prove that the sets $C_1, \ldots, C_m$ are closed and satisfy the covering condition.

Let $\pvec \in \Delta_m$, let $S$ be the set of sellers with positive prices,
and suppose $\pvec \notin C_j$ for every $j \in S$.
Then each seller $j \in S$ has a \emph{bad deviation} $z_j \in (p_j, L_j]$
for which no response restores every seller's revenue.
Use these bad deviations as caps in the inner KKM (sellers outside $S$ keep their price 0).
It gives prices $\qvec$ with $R(\qvec) = n\pvec$, where $q_i = z_i$ for some seller $i \in S$.
But then $\qvec_{-i}$ is a response to seller $i$'s bad deviation that restores every seller's revenue,
which is a contradiction.
Hence, $\pvec$ has some label from $S$, so the covering condition holds.
The label sets are closed, since the bounded responses lie in a compact set
and the allocation rule is continuous.
Thus, KKM gives us prices $\pvec^* \in \bigcap_{j=1}^m C_j$ (\cref{thm:rare-exists-continuous-2}).
These ideas form the core of \cref{thm:rare-exists-continuous}.

\subsection{Dynamics}
\label{sec:overview:dynamics}

Recall from \cref{sec:intro:results} that prices $\pvec$ form a RARE,
up to an arbitrarily small additive error,
iff $R_j(\pvec) \ge V_j(\pvec_{-j})$ for every seller $j$.
Also recall that in our dynamic, all prices are initially 0,
and in each round, some of the sellers $j$ for which $p_j < V_j(\pvec_{-j})/n$
raise their price to $V_j(\pvec_{-j})/n$.

The key to the analysis is the price vector $\pvec^* \in \Delta_m$
that satisfies \cref{eq:overview-restoration}, which acts as a ceiling for the dynamic
(see \cref{fig:dynamics-path}).
Specifically, for any prices $\pvec \le \pvec^*$ (coordinatewise) and any seller $j$,
we have $V_j(\pvec_{-j}) \le np^*_j$ (\cref{thm:vbound-pstar}).
To see this, suppose seller $j$ deviates to price $q_j$ at prices $\pvec$.
If $q_j \le p^*_j$, then her revenue is at most $nq_j \le np^*_j$, irrespective of any response.
Otherwise, \cref{eq:overview-restoration} gives a response $\qvec_{-j}$ satisfying
$R_j(\qvec) \le np^*_j$ and $R_k(\qvec) \ge np^*_k$ for every responder $k \neq j$.
Each responder $k$'s revenue at $\qvec$ is then
\[ R_k(\qvec) \ge np^*_k \ge np_k \ge R_k((q_j, \pvec_{-j})). \]
So, $\qvec_{-j}$ is also an IR response at prices $\pvec$,
and it limits the deviator's revenue to $np^*_j$.
(Here we use the stronger property in \cref{eq:overview-restoration}:
responders recover at least their pre-deviation revenues at $\pvec^*$.)

\begin{figure}[t]
\centering
\begin{tikzpicture}[x=1cm,y=.9cm,font=\fontsize{9}{10.8}\selectfont,
  dynstep/.style={draw=blue!65!black,line width=1.2pt,-{Stealth[length=2mm]}}]
\path[use as bounding box] (-.2,-.4) rectangle (15,5.5);
\coordinate (o) at (1,.6);
\coordinate (pstar) at (4.6,2.6);
\fill[black!3] (o)--(7,.6)--(1,5.6)--cycle;
\fill[blue!8] (o) rectangle (pstar);
\draw[black!55] (1,5.6)--(7,.6);
\draw[blue!55!black,densely dotted,line width=.8pt] (1,2.6)--(pstar)--(4.6,.6);
\draw[-{Stealth[length=2mm]},black!65] (.75,.6)--(7.35,.6) node[right] {$p_1$};
\draw[-{Stealth[length=2mm]},black!65] (1,.35)--(1,5.72) node[above] {$p_2$};
\node[anchor=north] at (1,.5) {$0$};
\draw[dynstep] (o)--(2.2,.6);
\draw[dynstep] (2.2,.6)--(2.2,1.4);
\draw[dynstep] (2.2,1.4)--(3.25,1.4);
\draw[dynstep] (3.25,1.4)--(3.25,1.82);
\fill[blue!65!black] (3.25,1.82) circle (2.5pt);
\node[anchor=south east,fill=white,inner sep=2pt] at (3.14,1.98) {May stop here};
\fill[black] (pstar) circle (2.8pt);
\node[anchor=south west,fill=white,inner sep=2pt] at (4.68,2.63) {$\pvec^* = (0.6, 0.4)$};
\node[rotate=-40,fill=white,inner sep=1pt] at (2.7,4.23) {$p_1 + p_2 = 1$};
\node[anchor=west,text width=6.3cm,align=left] at (8.7,4.4)
 {\textbf{Invariant}\quad $\vecZero \le \pvec^{(t)} \le \pvec^*$\\The prices never exceed $\pvec^*$.};
\node[anchor=west,text width=6.3cm,align=left] at (8.7,2.69)
 {\textbf{Potential}\quad $\sum_{j=1}^m p_j$\\The sum of prices never decreases and is at most 1.};
\end{tikzpicture}

\caption{A possible path of our dynamic for two sellers,
where $\pvec^{(t)}$ denotes the prices after round $t$.
The prices start at $\vecZero$ and never exceed $\pvec^*$,
so they stay in the shaded box below the line $p_1 + p_2 = 1$,
and the sum of prices never decreases.
In this path, the sellers raise their prices one at a time.
The dynamic may stop at prices other than $\pvec^*$.}
\label{fig:dynamics-path}
\end{figure}

By induction, the prices never exceed $\pvec^*$.
In particular, the sum of prices is always at most 1,
so every buyer buys every good completely, and each seller $j$'s revenue is always $np_j$.
Hence, the sum of prices is a potential function that never decreases and is at most 1.
After round $t$, let $g_t$ be the largest amount by which some seller $j$'s $V_j(\pvec_{-j})$
exceeds her revenue $R_j(\pvec)$ (or 0 if there is no such seller).
Then the prices after round $t$ are a $\delta$-additive RARE for every $\delta > g_t$.
If the most unsatisfied seller raises her price in round $t+1$,
then the sum of prices increases by at least $g_t/n$.
Hence, $\sum_{t \ge 0} g_t \le n = B$.
So $g_t < \delta$ for some $t \le \floor{B/\delta}$,
and $g_t \ge \delta$ for at most $\floor{B/\delta}$ values of $t$.
This gives us \cref{thm:dynamics-convergence}.

\subsection{Price discipline and broader implications}

The equilibrium profiles constructed on the simplex $\beta\Delta_m$,
as well as every price profile generated by our dynamics,
admit no profitable unilateral price reduction,
\emph{even when competitors' prices are held fixed}.
Indeed, each seller already sells the maximum possible quantity,
so lowering her price cannot be compensated by additional demand.
This provides a desirable \emph{price-discipline} property:
any remaining Nash instability arises from price increases, rather than profitable undercutting.
In this respect, our selected outcomes differ from
collusive pricing patterns documented for algorithmic pricing agents,
in which sellers forgo immediately profitable unilateral price cuts
to avoid subsequent price wars.
Such collusive pricing has been observed for reinforcement-learning agents
\cite{CalvanoEtAl2020,Klein2021},
language-model agents~\cite{FishGonczarowskiShorrer2024},
and in field data~\cite{assad2024algorithmic,musolff2026algorithmic}. Supracompetitive pricing can also arise without explicitly encoded threats~\cite{arunachaleswaran2025algorithmic}.

Our proof may also be of independent interest as a guide to defining
equilibrium notions in other pricing games where pure Nash equilibria fail to exist.
It suggests beginning with a price domain that combines buyer-spending exhaustion
with the absence of profitable undercutting, leaving only upward deviations to address.
The economic intuition is that, once buyers exhaust their budgets,
a seller can increase her revenue only by taking spending away from competitors.
This suggests seeking an economically justified notion of counterresponse
under which, at every price profile in this domain,
at least one seller cannot secure a gain by raising her price.
Our inner KKM argument establishes such a local stability property.
Thus, the broader lesson is a route to discovering new notions of price equilibrium---identify
a suitable price domain, characterize rational responses that ensure some
seller is stable at each profile, and exploit this structure to establish simultaneous stability.
Exploring this approach beyond our market model is an interesting direction for future work.

\paragraph{Organization.}
\Cref{sec:prelims} formally defines our model and RARE.
\Cref{sec:revenue-props} states some properties of the revenue function, and a way to unify the analysis of the rivalrous and the non-rivalrous settings.
\Cref{sec:exist} proves the existence of RARE, and \cref{sec:dynamics} analyzes our dynamic.
\Cref{sec:exist-extra} contains the proofs omitted from \cref{sec:exist}.
\Cref{sec:lim-rare-equiv} proves \cref{thm:rare-equiv-defn}, i.e.,
the relationship between $V_j$ and RARE.
\Cref{sec:nasheq-uncommon} shows that Nash equilibria are uncommon.

\section{Preliminaries}
\label{sec:prelims}

\paragraph{Notation.}
For any non-negative integer $t$, let $[t] \defeq \{1, 2, \ldots, t\}$.
For any vector $\zvec \in \mathbb{R}^m$, $z_j$ denotes the $j\Th$ coordinate
and $\zvec_{-j} \in \mathbb{R}^{[m] \setminus \{j\}}$ denotes the remaining coordinates.
For any $s \in \mathbb{R}$, we write $(s, \zvec_{-j})$ to denote the vector in $\mathbb{R}^m$
whose $j\Th$ coordinate is $s$ and the remaining coordinates are given by $\zvec_{-j}$.
Define $\support(\zvec) = \{j \in [m]: z_j > 0\}$.

\subsection{Utilities, Budgets, and the Market}

We now formally define the market model we study.
It is very similar to a Fisher market \cite{vazirani2007combinatorial}.

A market instance is a tuple
$\Ical \defeq ([n], [m], (u_i)_{i=1}^n, (b_i)_{i=1}^n)$.
Here $[n]$ is the set of buyers and $[m]$ is the set of (divisible) goods. We assume $n \ge 1$ and $m \ge 2$ throughout the paper. Each buyer $i \in [n]$ has a \emph{budget} $b_i > 0$
and a utility function $u_i: [0, 1]^m \to \mathbb{R}_{\ge 0}$,
where $u_i(\zvec)$ is buyer $i$'s value for a bundle containing
a $z_j$ fraction of each good $j \in [m]$.

Let $B \defeq \sum_{i=1}^n b_i$ denote the total budget,
$b_{\avg} \defeq B/n$ be the average budget across buyers,
and $b_{\min} \defeq \min_{i=1}^n b_i$ be the minimum budget across buyers.

For each buyer $i$, we assume the following:
\begin{tightenum}
\item The value of an empty bundle is zero, i.e., $u_i(\vecZero) = 0$.
\item $u_i$ is continuous, so a small change to the bundle doesn't change the utility much.
\item Getting more of any good increases utility. We model this as $u_i$ being strictly monotone,
    i.e., for any two distinct vectors $\xvec, \yvec \in [0, 1]^m$
    such that $x_j \le y_j$ for all $j$, we have $u_i(\xvec) < u_i(\yvec)$.
\end{tightenum}

A useful special case is \emph{linear utilities}:
each buyer $i$ gets utility $u_{i,j} > 0$ from each good $j$,
and $u_i(\zvec) = \sum_{j=1}^m u_{i,j} \cdot z_j$.

\subsection{Buyers' Demands}

Having defined the buyers' utilities and budgets, we now describe their demand model, i.e., what they spend their money on given the goods' prices. The demand model depends on whether the goods are rivalrous or not. We begin with the simpler case where the goods are non-rivalrous, i.e., where each good can be copied and given to multiple buyers.

\paragraph{Demand for non-rivalrous goods.}
Suppose good $j$ is priced at $p_j \ge 0$.
Given the vector $\pvec \defeq (p_1, \ldots, p_m)$ of prices,
each buyer $i$ purchases a utility-maximizing bundle subject to her budget constraint:
formally, buyer $i$'s \emph{demand set} is
\[ \Dem_i(\pvec; \Ical) \defeq \argmax_{\zvec \in [0, 1]^m:\,\pvec^T\zvec \le b_i} u_i(\zvec). \]
Let $\DemNR(\pvec; \Ical) \defeq \prod_{i=1}^n \Dem_i(\pvec; \Ical)$ be the
\emph{combined non-rivalrous demand set}, i.e.,
$\xvec \in \DemNR(\pvec; \Ical)$ denotes that each buyer $i$ purchased an $x_{i,j}$ fraction of each good $j$.
When $\Ical$ is clear from context, we simply write $\Dem_i(\pvec)$ and $\DemNR(\pvec)$
instead of $\Dem_i(\pvec; \Ical)$ and $\DemNR(\pvec; \Ical)$.

\paragraph{Demand for rivalrous goods.}
When the goods are non-rivalrous, the buyers can make purchase decisions independently.
But when the goods are rivalrous and the prices are low enough, multiple buyers may want the same good.
For example, let there be two goods having prices \$1 each, and two buyers having budgets \$2 each.
Then both buyers would like to purchase both goods entirely, but that is not possible, due to the rivalrous nature of the goods.

We need a method to decide how each good is distributed among the buyers. We study a broad class of such methods, based on a property we believe any reasonable method should satisfy: \emph{Pareto-optimality among buyers}.
We now define this formally.

For any allocation $\xvec \in [0, 1]^{n \times m}$,
let $\xvec_i$ denote buyer $i$'s bundle $(x_{i,1}, \ldots, x_{i,m})$. We say that an allocation $\xvec \in [0, 1]^{n \times m}$ \emph{Pareto dominates}
another allocation $\yvec \in [0, 1]^{n \times m}$, denoted as $\xvec \succ_{\Ical} \yvec$, if
for each buyer $i \in [n]$, we have $u_i(\xvec_i) \ge u_i(\yvec_i)$
and for some buyer $i \in [n]$, we have $u_i(\xvec_i) > u_i(\yvec_i)$.
For any set $\Scal \subseteq [0, 1]^{n \times m}$ of allocations,
define its \emph{Pareto optimal subset} as allocations in $\Scal$ that are not Pareto dominated
by any other allocation in $\Scal$, i.e.,
\[ \PO(\Scal; \Ical) \defeq \{\xvec \in \Scal: \yvec \not\succ_{\Ical} \xvec \text{ for all } \yvec \in \Scal\}. \]
We are now ready to define our demand model.
Let the set of \emph{affordable rivalrous allocations} be
\[ \AffR(\pvec; \Ical) \defeq \bigg\{\xvec \in [0, 1]^{n \times m}:
    \bigg(\sum_{i=1}^n x_{i,j} \le 1 \;\forall j \in [m]\bigg) \text{ and }
    \bigg(\sum_{j=1}^m p_jx_{i,j} \le b_i \;\forall i \in [n]\bigg)\bigg\}. \]
Then the \emph{combined rivalrous demand} is defined as
the Pareto-optimal subset among affordable allocations, i.e.,
$\DemR(\pvec; \Ical) \defeq \PO(\AffR(\pvec; \Ical); \Ical)$.
When $\Ical$ is clear from context, we omit it from $\DemR$, $\AffR$, and $\PO$.

In fact, one can show that such a description is also true for non-rivalrous goods.

\begin{observation}
If we define the set of \emph{affordable non-rivalrous allocations} as
\[ \AffNR(\pvec; \Ical) \defeq \prod_{i=1}^n \{\zvec \in [0, 1]^m: \pvec^T\zvec \le b_i\}, \]
then $\DemNR(\pvec) = \PO(\AffNR(\pvec))$.
\end{observation}

Note that when there is only one buyer, the rivalrous and non-rivalrous settings coincide.

\paragraph{Allocation rule.}
The demand sets ($\DemR(\pvec)$ and $\DemNR(\pvec)$) capture the buyers' purchasing preferences,
but they still don't specify their behavior completely.
For example, with two identical goods having prices \$1 each,
and a buyer with budget \$1 and linear utilities,
it is unclear how the buyer would split her budget among the goods,
since every split is utility-maximizing for her.
Thus, we specify their behavior using an \emph{allocation rule}.

An \emph{allocation rule} $A: \mathbb{R}_{\ge 0}^m \to [0, 1]^{n \times m}$ is a function that maps a price vector $\pvec$ to an allocation in the demand set (either $\DemR(\pvec)$ or $\DemNR(\pvec)$, depending on whether the goods are rivalrous or not).

For non-rivalrous goods, the allocation rule specifies how each buyer breaks ties among
utility-maximizing bundles. For rivalrous goods, the allocation rule additionally
specifies how we break ties among multiple buyers who want the same good.

\subsection{Sellers' Revenues and Equilibria}

Now that we can model how buyers react to prices,
we can talk about the revenue from each good.

Each good has a corresponding seller.
For prices $\pvec \in \mathbb{R}_{\ge 0}^m$ and allocation $\xvec \in [0, 1]^{n \times m}$,
each seller $j$ earns revenue $R_j(\pvec, \xvec) \defeq p_j\sum_{i=1}^n x_{i,j}$.
Under an allocation rule $A$, define seller $j$'s revenue as $R^A_j(\pvec) \defeq R_j(\pvec, A(\pvec))$.
When the rule $A$ is clear from context, we write $R_j(\pvec)$ instead of $R^A_j(\pvec)$.

The sellers are in a pricing game: each seller $j$ picks a price $p_j \ge 0$
for her good, and receives a payoff $R^A_j(\pvec)$. Which prices will the sellers pick?
A well-studied concept for this is Nash equilibrium.

\begin{definition}[Nash equilibrium]
\label{defn:nasheq}
For market $\Ical = ([n], [m], (u_i)_{i=1}^n, (b_i)_{i=1}^n)$
and allocation rule $A: \mathbb{R}_{ \ge 0}^m \to [0, 1]^{n \times m}$,
a price vector $\pvec \in \mathbb{R}_{ \ge 0}^m$ is a \emph{Nash equilibrium}
if for any seller $j \in [m]$ and any deviation $s \in \mathbb{R}_{\ge 0}$,
we have $R^A_j((s, \pvec_{-j})) \le R^A_j(\pvec)$, i.e.,
the deviator $j$ does not gain by deviating.
\end{definition}

In \cref{sec:nasheq-uncommon}, we show that Nash equilibria do not always exist.
Thus, we propose a new equilibrium notion, called RARE,
based on the idea that when a seller deviates to a new price,
the other sellers may respond by changing their own prices.

\begin{definition}[RARE]
Let $\Ical = ([n], [m], (u_i)_{i=1}^n, (b_i)_{i=1}^n)$ be a market
and $A: \mathbb{R}_{\ge 0}^m \to [0, 1]^{n \times m}$ be an allocation rule.
For $\delta \ge 0$ and $\kappa \ge 1$, a price vector $\pvec \in \mathbb{R}_{\ge 0}^m$
is a $\delta$-additive $\kappa$-multiplicative Response-Aware Rational Equilibrium
($\delta$-additive $\kappa$-RARE)
if for any seller $j \in [m]$, and any deviation $q_j \in \mathbb{R}_{\ge 0}$,
there exists a response $\qvec_{-j} \in \mathbb{R}_{\ge 0}^{[m] \setminus \{j\}}$ such that
\begin{enumerate}
\item $R_j^A(\qvec) \le \kappa \cdot R_j^A(\pvec) + \delta$,
    i.e., seller $j$ cannot gain much by deviating if others respond.
\item No responder loses by responding, i.e., $R_k^A(\qvec) \ge R_k^A((q_j, \pvec_{-j}))$
    for each responder $k \in [m] \setminus \{j\}$.
\end{enumerate}
We drop the specifier `$\delta$-additive' when $\delta = 0$,
and we write RARE instead of 1-RARE.
\end{definition}

\subsection{The Simplex and the KKM Lemma}

A key ingredient in our results is the KKM lemma, which describes conditions under which
subsets of a simplex have a common intersection.
To state this lemma, we first introduce some notation.

For any $\beta \in \mathbb{R}_{> 0}$ and $m \in \N$,
let $\beta\Delta_m \defeq \{\xvec \in \mathbb{R}_{\ge 0}^m: \sum_{j=1}^m x_j = \beta\}$
be the standard $(m-1)$-dimensional simplex scaled by a factor $\beta$,
and let $\beta\Delta_m^{\circ} \defeq \{\xvec \in \mathbb{R}_{> 0}^m: \sum_{j=1}^m x_j = \beta\}$
be its relative interior.

\begin{lemma}[Knaster--Kuratowski--Mazurkiewicz \cite{KKM1929}]
\label{thm:kkm}
Let $m \ge 1$, and let $C_1, \ldots, C_m$ be closed subsets of
the scaled simplex $\beta\Delta_m$ for some $\beta > 0$.
Suppose every vector $\xvec \in \beta\Delta_m$ is covered by the sets
$\{C_j: j \in \support(\xvec)\}$, i.e., $\xvec \in C_j$ for some $j \in \support(\xvec)$.
Then some point of $\beta\Delta_m$ lies in all the sets, i.e., $\bigcap_{j=1}^m C_j \ne \emptyset$.
\end{lemma}

\section{Properties of the Revenue Function}
\label{sec:revenue-props}

In our market setting, the sellers are in a pricing game,
where each seller's payoff is given by her revenue function $R_j$.
Analyzing this pricing game, therefore, requires understanding the revenue function, which is the topic of this section.

\subsection{A Closed Form for the Total Revenue}
\label{sec:tot-rev-closed-form}

Analyzing equilibria is often simpler for \emph{fixed-sum games},
i.e., where the sum of the players' payoffs does not depend on the players' actions.
For example, for finite two-player \emph{fixed-sum games} with explicitly given payoff matrices, a mixed (non-pure) Nash equilibrium
can be efficiently computed using linear programming \cite{dantzig1951proof}.

For our market model, the pricing game among the sellers is similar in spirit to a fixed-sum game,
since the sellers compete for the same buyers' budgets.
Indeed, if prices are high enough to exhaust all buyers' budgets,
the sellers' revenues would sum to $B$, the total buyer budget.
For lower prices, however, buyers may not exhaust their budgets,
making the game non-fixed-sum. We study the total revenue
$R_{\all}(\pvec, \xvec) \defeq \sum_{j=1}^m R_j(\pvec, \xvec)$
and give a closed-form expression for it,
which precisely quantifies this difference between the low-price and high-price regimes.

\begin{observation}
\label[observation]{thm:rev-vs-F}
For any prices $\pvec \in \mathbb{R}_{\ge 0}^m$,
\begin{tightenum}
\item (Non-rivalrous goods) If $\xvec \in \DemNR(\pvec)$,
    we have $R_{\all}(\pvec, \xvec) = F(\sum_{j=1}^m p_j)$,
    where $F(y) \defeq \sum_{i=1}^n \min(b_i, y) \le \min(B, ny)$.
\item (Rivalrous goods) If $\xvec \in \DemR(\pvec)$,
    we have $R_{\all}(\pvec, \xvec) = \min(B, \sum_{j=1}^m p_j)$.
\end{tightenum}
In either case, we get $R_{\all}(\pvec, \xvec) \le B$.
\end{observation}
\begin{proof}
In the non-rivalrous setting, each buyer $i \in [n]$ either exhausts her budget
or purchases every good completely, contributing $\min(b_i, \sum_{j=1}^m p_j)$
to the total revenue. Thus, $R_{\all}(\pvec, \xvec) = F(\sum_{j=1}^m p_j)$.

In the rivalrous setting, the buyers collectively either exhaust their budgets
or purchase every good completely
(since utilities are strictly monotone and $\xvec$ is Pareto optimal).
Hence, $R_{\all}(\pvec) = \min(B, \sum_{j=1}^m p_j)$.
\end{proof}

Since $R_{\all}(\pvec, \xvec)$ does not depend on $\xvec$ when $\xvec$ lies in the demand set,
we simply write $R_{\all}(\pvec)$ instead when it is clear from context
whether the goods are rivalrous or not.
We also conclude that $R_{\all}$ is monotonic in prices.

\begin{observation}
\label{thm:rall-mono}
Let $\pvec, \qvec \in \mathbb{R}_{\ge 0}^m$ be price vectors
such that $\sum_{j=1}^m p_j \le \sum_{j=1}^m q_j$.
Then $R_{\all}(\pvec) \le R_{\all}(\qvec)$
(for both the rivalrous and non-rivalrous settings).
\end{observation}

\subsection{Unifying the Rivalrous and Non-Rivalrous Settings}
\label{sec:unify}

We now slightly refine the analysis of \cref{sec:tot-rev-closed-form}.

When prices are low enough that each good is fully sold,
each seller $j$'s revenue is $p_j$ in the rivalrous setting
and $n \cdot p_j$ in the non-rivalrous setting.

Now consider the regime where prices are large.
In the rivalrous setting, when the prices sum to at least $B$,
we get $R_{\all}(\pvec) = \min(B, \sum_{j=1}^m p_j) = B$.
In the non-rivalrous setting, when prices sum to at least $b_{\min}$,
then every buyer contributes at least $b_{\min}$ to the total revenue,
giving us $R_{\all}(\pvec) \ge n \cdot b_{\min}$.

The above properties capture almost everything about the revenue functions that we need
in our existence results (\cref{sec:exist}) and the dynamics (\cref{sec:dynamics}).
Note how both the rivalrous and non-rivalrous settings behave similarly,
with the differences showing up only in some constants
(the factor next to $p_j$: $1$ vs $n$, and the price sum threshold: $B$ vs $b_{\min}$).
Indeed, our analyses for the existence results and the dynamics
are the same for both settings up to these constants.

To save ourselves the trouble of writing these results separately for each setting,
we introduce parameters $\nu$, $\beta$, and $\kappa$, and use them hereafter
to perform a unified analysis of both settings.

\begin{center}
\begin{tabular}{c|c|c|c}
\text{Market} & $\nu$ & $\beta$ & $\kappa \defeq B/(\nu\beta)$
\\ \hline Rivalrous & 1 & $B$ & 1
\\ Non-rivalrous & $n$ & $b_{\min}$ & $b_{\avg}/b_{\min}$
\end{tabular}
\end{center}

We now restate the above properties of the revenue functions using these parameters.
We write $\Dem(\pvec)$ instead of $\DemR(\pvec)$ or $\DemNR(\pvec)$ to refer to the demand set
in general without specifying whether the goods are rivalrous or non-rivalrous.

\begin{observation}[Properties of the revenue function]
\label{thm:accounting}
In both the rivalrous and non-rivalrous settings,
for any prices $\pvec$ and any $\xvec \in \Dem(\pvec)$, we have
\begin{enumerate}
\item \label{item:rev-props:ub}For each seller $j \in [m]$,
    we have $R_j(\pvec, \xvec) \le \nu p_j$, i.e., her revenue is
    at most what she would earn by selling her good fully.
\item \label{item:rev-props:low-prices}If $\sum_{j=1}^m p_j \le \beta$, then every good
    is fully sold, so each seller $j$'s revenue $R_j(\pvec, \xvec)$ is $\nu p_j$.
\item \label{item:rev-props:high-prices}$\sum_{j=1}^m p_j \ge \beta \implies R_{\all}(\pvec) \ge \nu \beta$.
\end{enumerate}
\end{observation}

\section{Existence of RARE}
\label{sec:exist}

In this section, we prove the existence of (approximate) RARE.
We first show that when we fix an allocation rule that is \emph{continuous},
then a $\kappa$-RARE always exists, where $\kappa$ is 1 for the rivalrous setting
and $b_{\avg}/b_{\min}$ for the non-rivalrous setting.

Then, we ask \emph{which} allocation rules are continuous,
and that leads us to different answers
for the rivalrous and non-rivalrous settings:
\begin{enumerate}
\item \textbf{Non-rivalrous setting}:
    When buyers' utility functions are strictly concave, the demand set is a singleton at all prices,
    so there is a unique allocation rule. Moreover, this allocation rule is continuous.
    Thus, $(b_{\avg}/b_{\min})$-RARE always exists for strictly concave utilities.
\item \textbf{Rivalrous setting}:
    Even with strictly concave utilities, the demand set may not be a singleton.
    But we identify a large class of natural allocation rules that are continuous,
    proving that RARE exists for all of them.
\end{enumerate}

For linear utilities, we present instances where no allocation rule is continuous.
Nevertheless, we prove that for concave (but not necessarily strictly concave) utilities,
a variant of RARE holds.

\subsection{Refining the Goal}

A price vector $\pvec \in \mathbb{R}_{\ge 0}^m$ is $\kappa$-RARE iff
\begin{quote}
for every seller $j \in [m]$ and deviation $q_j \ge 0$,
\\ there exists an individually rational response
$\qvec_{-j} \in \mathbb{R}_{\ge 0}^{[m] \setminus \{j\}}$, such that
\\ $R_j(\qvec) \le \kappa R_j(\pvec)$.
\end{quote}

We now sketch how to make the problem simpler by restricting our search space.
First, we restrict our search for a $\kappa$-RARE price vector $\pvec$ to the simplex $\beta\Delta_m$.
This puts it at the intersection of the low-price and high-price regimes from \cref{thm:accounting}:
prices are low enough that buyers can afford all goods,
but high enough to extract a $1/\kappa$ fraction of the total budget $B$.

Next, we show how to restrict the deviation $q_j$ and the responses $\qvec_{-j}$.
Fix an allocation rule and pick any price vector $\pvec \in \beta\Delta_m$.
By \cref{thm:accounting}.\ref{item:rev-props:low-prices},
each seller $k \in [m]$ earns revenue $R_k(\pvec) = \nu p_k$.
If $q_j < p_j$, then the deviator $j$'s revenue will drop from $\nu p_j$ to at most $\nu q_j$,
regardless of what the response $\qvec_{-j}$ is.
Thus, we need $q_j \ge p_j$ for the deviation to make sense.

For every respondent $k \neq j$, her revenue immediately after $j$'s deviation is
$R_k((q_j, \pvec_{-j})) \le \nu p_k$.
If every respondent $k \neq j$ raises her price to $q_k \ge p_k$,
the total revenue rises (or remains the same), i.e., $R_{\all}(\qvec) \ge R_{\all}(\pvec)$.
Then a response that preserves every seller $k$'s market share $R_k(\cdot)/R_{\all}(\cdot)$
would be individually rational, since
\[ R_k(\qvec) = \frac{R_k(\pvec)}{R_{\all}(\pvec)} R_{\all}(\qvec)
    \ge R_k(\pvec) = \nu p_k \ge R_k((q_j, \pvec_{-j})). \]
Moreover, we have $R_{\all}(\pvec) = \nu\beta$ and $R_{\all}(\qvec) \le B$, giving us
\[ R_j(\qvec) = R_j(\pvec) \frac{R_{\all}(\qvec)}{R_{\all}(\pvec)} \le \kappa R_j(\pvec). \]
The original market share of each seller $k \in [m]$ is $R_k(\pvec)/R_{\all}(\pvec) = p_k/\beta$.
Hence, for $\pvec \in \beta\Delta_m$ to be a $\kappa$-RARE, it suffices to show that
\begin{quote}
for every seller $j \in [m]$ and deviation $q_j \in [p_j, \infty)$,
\\ there exists a response $\qvec_{-j} \in \prod_{k \neq j} [p_k, \infty)$, such that
\\ for every seller $k \in [m]$, we have $R_k(\qvec)/R_{\all}(\qvec) = p_k/\beta$.
\end{quote}

As hinted in \cref{sec:overview}, we will prove that such a $\pvec$ exists using KKM.
To do so, for each seller $j$, we construct the set $C_j$ of price vectors
where every deviation to a higher price has a market-share-preserving response:
\[ C_j \defeq \bigg\{\pvec \in \beta\Delta_m:\; \forall q_j \in [p_j, \infty),\;
    \exists \qvec_{-j} \in \!\!\!\!\!\prod_{k \in [m] \setminus \{j\}} \!\!\! [p_k, \infty) \text{ such that }
    \forall k \in [m], \frac{R_k(\qvec)}{R_{\all}(\qvec)} = \frac{p_k}{\beta} \bigg\}. \]
A price vector belonging to $\bigcap_{j=1}^m C_j$ would be $\kappa$-RARE.
To show that this intersection is non-empty, we use KKM.
We can prove that the sets $(C_j)_{j \in [m]}$ satisfy the covering condition
by using an inner KKM, as outlined in \cref{sec:overview}.
It remains to show that each $C_j$ is closed. But $C_j$ need not be closed!
We give a counterexample in \cref{sec:unbounded-unclosed}.

We work around this obstacle by restricting the deviation and response sets to be compact.
In particular, we pick a large constant $L_k \gg B$ for each seller $k$,
and we require $q_k \le L_k$ for all $k \in [m]$.
With this change, the set
\begin{equation}
\label{eq:cover-sets}
C_j \defeq \bigg\{\pvec \in \beta\Delta_m:\; \forall q_j \in [p_j, L_j],\;
    \exists \qvec_{-j} \in \!\!\!\!\!\prod_{k \in [m] \setminus \{j\}} \!\!\! [p_k, L_k] \text{ such that }
    \forall k \in [m], \frac{R_k(\qvec)}{R_{\all}(\qvec)} = \frac{p_k}{\beta} \bigg\}
\end{equation}
becomes closed (we prove this in \cref{thm:rare-exists-continuous-2}).

It remains to show how to construct a response $\qvec_{-j}$ for deviations $q_j > L_j$.
If a seller $j$ deviates to a sufficiently large price, then the other sellers should be able to
easily out-compete her and drive down her revenue below $\nu p_j$.
But it's not obvious how to do this while maintaining individual rationality.
We do this by using another KKM!

\subsection{Market-Share-Restoring Responses Using KKM}
\label{sec:exist:kkm}

To apply KKM on the sets $(C_j)_{j \in [m]}$ as defined by \cref{eq:cover-sets},
we must prove the covering condition, i.e., for any $\pvec \in \beta\Delta_m$,
we have $\pvec \in C_j$ for some $j \in \support(\pvec)$.
Intuitively, this means that some seller $j$ with a positive price
should be unable to increase her market share.

We begin by proving a more general result: that some seller should be unable to \emph{gain}
by increasing prices, for an arbitrary definition of gain, not just `increase in market share'.
We prove this using KKM.
(This general result will be useful again later when analyzing responses to very high deviations,
i.e., the $q_j > L_j$ case.)

\begin{lemma}
\label{thm:general-inner-kkm}
Let $J \neq \emptyset$ be a finite set of sellers. Let $\ell_j < h_j$ for all $j \in J$.
Let $U$ be the union of the $|J|$ upper faces of the box $\prod_{j \in J} [\ell_j, h_j]$, i.e.,
\[ U \defeq \bigg\{\xvec \in \prod_{j \in J} [\ell_j, h_j]: x_i = h_i \text{ for some } i \in J\bigg\}. \]
For every $j \in J$, let $g_j: U \to \mathbb{R}$ be a continuous function
(the gain function) where for all $\xvec \in U$, the following hold:
\begin{itemize}
\item (Total gain is non-negative) $\sum_{j \in J} g_j(\xvec) \ge 0$.
\item (Non-positive gain at $\ell_j$) For all $j \in J$, $x_j = \ell_j \implies g_j(\xvec) \le 0$.
\end{itemize}
Then some $\xvec^* \in U$ satisfies $g_j(\xvec^*) \ge 0$ for all $j \in J$,
i.e., we get a non-negative gain for all $j \in J$.
\end{lemma}
\begin{proof}
For any $\theta \in \Delta_J \defeq \{\xvec \in \mathbb{R}_{\ge 0}^J: \sum_{j \in J} x_j = 1\}$,
shoot a ray from $\bm{\ell} = (\ell_j)_{j \in J}$ in the direction $\theta$,
and let $y(\theta)$ be the point on $U$ that the ray hits. Formally,
\[ y(\theta) \defeq \bm{\ell} + \tau(\theta)\theta, \quad \text{where} \quad
    \tau(\theta) \defeq \bigg(\max_{j \in J} \frac{\theta_j}{h_j - \ell_j}\bigg)^{-1}. \]
For all $j \in J$, define
\[ S_j \defeq \{\theta \in \Delta_J: g_j(y(\theta)) \ge 0 \}. \]
We now use KKM (\cref{thm:kkm}) on the simplex $\Delta_J$
to show that $\bigcap_{j \in J} S_j \neq \emptyset$.

\textbf{Closedness.}
For all $j \in J$, $y$ and $g_j$ are continuous, so $g_j \circ y$ is continuous.
Since $S_j$ is the superlevel set of $g_j \circ y$, it is closed
(see \cref{thm:superlevel-closed} in \cref{sec:exist-extra:continuous} for details).

\textbf{Covering.}
Suppose the covering condition doesn't hold,
i.e., $\exists \theta \in \Delta_J$ such that $\theta \not\in S_j$
for all $j \in \support(\theta) \defeq \{j \in J: \theta_j > 0\}$.
Hence, $\theta_j > 0 \implies g_j(y(\theta)) < 0$ for all $j \in J$.
For $j \in J \setminus \support(\theta)$, we have
$\theta_j = 0 \implies y(\theta)_j = \ell_j \implies g_j(y(\theta)) \le 0$.
Add all of these inequalities to get $\sum_{j \in J} g_j(y(\theta)) < 0$,
which is a contradiction. Thus, the covering condition holds.

By KKM, $\bigcap_{j \in J} S_j \neq \emptyset$. Let $\theta^* \in \bigcap_{j \in J} S_j$.
Let $\xvec^* \defeq y(\theta^*)$. Then $\xvec^* \in U$ and $g_j(\xvec^*) \ge 0$ for all $j \in J$.
\end{proof}

We now prove that on the upper boundary of the box $\prod_{j=1}^m [p_j, z_j]$,
one can find prices $\qvec$ that restore the market shares.
We do this by applying \cref{thm:general-inner-kkm} with
the gain function set to the increase in market share.

\begin{restatable}[market shares on box boundary]{lemma}{thmPropBox}
\label{thm:prop-box}
Let $A: \mathbb{R}_{\ge 0}^m \to [0, 1]^{n \times m}$ be a continuous allocation rule.
Let $\pvec \in \beta\Delta_m$ be a price vector
and $M^+ \defeq \{j \in [m]: p_j > 0\}$ be the positive-priced goods.

Let $\zvec \in \prod_{j \in M^+} (p_j, \infty) \times \{0\}^{[m] \setminus M^+}$
be a vector of upper bounds on increased prices.
Then there is another vector $\qvec$ on the boundary of the box $\prod_{j=1}^m [p_j, z_j]$
that preserves market shares, i.e., all of the following hold:
\begin{tightenum}
\item $p_j \le q_j \le z_j$ for all $j \in [m]$.
\item $q_i = z_i$ for some $i \in M^+$.
\item $R_{\all}(\qvec) \ge R_{\all}(\pvec) = \nu\beta$.
\item $\displaystyle \frac{R^A_j(\qvec)}{R_{\all}(\qvec)} = \frac{p_j}{\beta} = \frac{R^A_j(\pvec)}{R_{\all}(\pvec)}$
    for all $j \in [m]$.
\end{tightenum}
\end{restatable}
\begin{proof}
We get $R^A_j(\pvec) = \nu p_j$ by \cref{thm:accounting},
so $R_{\all}(\pvec) = \nu \sum_{j=1}^m p_j = \nu \beta$.

Let
\[ U \defeq \bigg\{\qvec \in \prod_{j \in M^+} [p_j, z_j] \times \{0\}^{[m] \setminus M^+}:
    q_i = z_i \text{ for some } i \in M^+\bigg\} \]
be the upper faces of the price-increase box,
and for each $j \in M^+$, let the gain function $g_j$ be the increase in market share, i.e.,
\[ g_j(\qvec) \defeq \frac{R_j(\qvec)}{R_{\all}(\qvec)} - \frac{p_j}{\beta}. \]
$R_{\all}$ is monotonic in prices (\cref{thm:rall-mono}),
so we get $R_{\all}(\qvec) \ge R_{\all}(\pvec) = \nu\beta$ for all $\qvec \in U$.
Then $g_j$ is continuous and $\sum_{j \in M^+} g_j(\qvec) = 0$.
Moreover, if $q_j = p_j$, then
\[ R_j(\qvec) \le \nu p_j = R_j(\pvec)
    \implies \frac{R_j(\qvec)}{R_{\all}(\qvec)} \le \frac{R_j(\pvec)}{R_{\all}(\pvec)}, \]
so $g_j(\qvec) \le 0$.

Hence, we can apply \cref{thm:general-inner-kkm}.
This gives us $\qvec \in U$ such that $g_j(\qvec) \ge 0$ for all $j \in M^+$.
Since $\sum_{j \in M^+} g_j(\qvec) = 0$, we have $g_j(\qvec) = 0$ for all $j \in M^+$.
Hence, $R_j(\qvec)/R_{\all}(\qvec) = p_j/\beta$ for all $j \in M^+$,
and this also holds for $j \in [m] \setminus M^+$ since $R_j(\qvec) = 0 = p_j$.
\end{proof}

We now apply KKM again to show that the sets $(C_j)_{j \in [m]}$ (as defined by \cref{eq:cover-sets})
have a common point $\pvec$. The covering condition for the KKM follows from \cref{thm:prop-box}.
At such a $\pvec$, for any seller $j \in [m]$ and deviation $q_j \in [p_j, L_j]$,
there is a market-share-restoring response.

\begin{lemma}
\label{thm:rare-exists-continuous-2}
Let $A: \mathbb{R}_{\ge 0}^m \to [0, 1]^{n \times m}$ be a continuous allocation rule
and $L_j \ge \beta$ for each seller $j \in [m]$.
There exist prices $\pvec \in \beta\Delta_m$ such that for every seller $j \in [m]$,
and every deviation $q_j \in [p_j, L_j]$ by seller $j$,
there exists a response $\qvec_{-j} \in \prod_{k \neq j} [p_k, L_k]$ such that
market shares are preserved, i.e.,
\[ \frac{R_k^A(\qvec)}{R_{\all}(\qvec)} = \frac{p_k}{\beta}
    \text{ for every seller } k \in [m]. \]
\end{lemma}
\begin{proof}
For every $j \in [m]$, let $C_j$ be the price vectors where every bounded deviation by seller $j$
can be met with a market-share-restoring response, i.e.,
\[ C_j \defeq \bigg\{\pvec \in \beta\Delta_m:\; \forall q_j \in [p_j, L_j],\;
    \exists \qvec_{-j} \in \!\!\!\!\!\prod_{k \in [m] \setminus \{j\}} \!\!\! [p_k, L_k] \text{ such that }
    \forall k \in [m], \frac{R_k(\qvec)}{R_{\all}(\qvec)} = \frac{p_k}{\beta} \bigg\}. \]
Our goal is to show that $\bigcap_{j=1}^m C_j \neq \emptyset$.
We prove this using KKM (\cref{thm:kkm}).

\textbf{Closedness.}
We first show that $C_j$ is closed. Let $\pvec^{(t)} \in C_j$ converge to $\pvec \in \beta\Delta_m$.
Fix any $q_j \in [p_j, L_j]$, and set $q_j^{(t)} \defeq \max(q_j, p_j^{(t)})$.
Since $\pvec^{(t)} \in C_j$, there is a response
$\qvec_{-j}^{(t)} \in \prod_{k \ne j} [p_k^{(t)}, L_k]$ such that
\[ \frac{R_k(\qvec^{(t)})}{R_{\all}(\qvec^{(t)})}
    = \frac{p_k^{(t)}}{\beta} \qquad \text{for every } k \in [m]. \]
The responses lie in the compact box $\prod_{k \ne j} [0, L_k]$,
so they must have a convergent subsequence.
Let $\qvec_{-j}$ be the limit of that subsequence.

By \cref{thm:rall-mono}, $R_{\all}(\qvec^{(t)}) \ge R_{\all}(\pvec^{(t)})$,
since $\qvec^{(t)} \ge \pvec^{(t)}$ coordinatewise.
Continuity of $A$ implies continuity of all revenue functions,
so taking limits gives us $R_{\all}(\qvec) \ge R_{\all}(\pvec) = \nu\beta$,
$q_k \ge p_k$ for all $k \in [m]$, and
\[ \frac{R_k(\qvec)}{R_{\all}(\qvec)} = \frac{p_k}{\beta}
    \qquad \text{for every }k \in [m]. \]
Thus, $\qvec_{-j} \in \prod_{k \neq j} [p_k, L_k]$ is a market-share-restoring response
to deviation $q_j$ and initial prices $\pvec$.
Since $q_j$ was arbitrary, we get $\pvec \in C_j$. Thus $C_j$ is closed.

\textbf{Covering.}
Fix $\pvec \in \beta\Delta_m$ and let $S \defeq \{j \in [m]: p_j > 0\}$ be its support.
Assume that the covering condition doesn't hold, i.e., $\pvec \not\in C_j$ for all $j \in S$.
For each $j \in S$, $\pvec \not\in C_j$ implies
\[ \exists q^{(j)}_j \in [p_j, L_j] \text{ such that }
    \forall \qvec^{(j)}_{-j} \in \!\!\!\!\!\prod_{k \in [m] \setminus \{j\}} \!\!\! [p_k, L_k],\;
    \exists k \in [m] \text{ such that } \frac{R_k(\qvec^{(j)})}{R_{\all}(\qvec^{(j)})} \neq \frac{p_k}{\beta}. \]
Let $\zvec \in \mathbb{R}^m$ be a vector where $z_j \defeq q^{(j)}_j$ for $j \in S$
and $z_j = 0$ for $j \in [m] \setminus S$.

Suppose $z_j = p_j$ for some $j \in S$. Then for $\qvec^{(j)}_{-j} = \pvec_{-j}$ and every $k \in [m]$, we get
\[ \frac{R_k(\qvec^{(j)})}{R_{\all}(\qvec^{(j)})} = \frac{R_k(\pvec)}{R_{\all}(\pvec)} = \frac{p_k}{\beta}, \]
which is a contradiction. Hence, $z_j > p_j$ for all $j \in S$.

By \cref{thm:prop-box}, there exist prices $\qvec$ such that
\begin{tightenum}
\item $p_j \le q_j \le z_j$ for all $j \in [m]$.
\item $q_i = z_i$ for some $i \in S$.
\item $R_{\all}(\qvec) \ge R_{\all}(\pvec) = \nu\beta$.
\item $\displaystyle \frac{R^A_j(\qvec)}{R_{\all}(\qvec)} = \frac{p_j}{\beta}$ for all $j \in [m]$.
\end{tightenum}
Picking $\qvec^{(i)}_{-i} = \qvec_{-i}$ gives us a contradiction.
Thus, the covering condition holds, so KKM applies.
\end{proof}

\subsection{The Futility of Large Deviations}
\label{sec:exist:overpricing}

\Cref{thm:rare-exists-continuous-2} gives us prices $\pvec$ such that
if any seller $j$ deviates to price $q_j \in [p_j, L_j]$,
there is a market-share-restoring response.
We now show that by defining $L_j$ appropriately,
we can construct a deterring response for $q_j \in (L_j, \infty)$.
To do this, we introduce additional parameters.

\paragraph{Security Parameters.}
The first of these are \emph{security parameters} $\sigma_j$ and $\rho_j$ for each seller $j$.
These say that if seller $j$ prices her good at $\sigma_j$, her revenue is at least $\rho_j$.
The following lemma proves that these parameters always exist.

\begin{restatable}{lemma}{thmSecurity}
\label{thm:security}
For any seller $j \in [m]$, there exist constants $\rho_j > 0$ and $\sigma_j \ge \rho_j/\nu$ such that
for any response $\qvec_{-j} \in \mathbb{R}_{\ge 0}^{[m]\setminus\{j\}}$ by the other sellers
and any allocation $\xvec \in \Dem((\sigma_j, \qvec_{-j}))$, we have
$R_j((\sigma_j, \qvec_{-j}), \xvec) \ge \rho_j$.
\end{restatable}
\begin{proof}[Proof sketch]
Intuitively, if seller $j$ sets a sufficiently small price for her good,
then every other seller falls into two categories:
\begin{tightenum}
\item \emph{cheap sellers}: these sellers also have a sufficiently low price,
    so they can extract only a small amount of revenue from the buyers,
    which leaves some budget for $j$.
\item \emph{expensive sellers}: these sellers charge a high price, so buyers prefer $j$ to these sellers.
\end{tightenum}
Thus, seller $j$ is guaranteed a minimum revenue.
We defer the full proof to \cref{sec:exist-extra:market-params}.
\end{proof}

\paragraph{Overpricing Parameter.}
Next, we study the \emph{overpricing parameter} $L_j$ for each seller $j$.
If the other sellers are trying to extract all budget $B$,
and if seller $j$ prices her good more than $L_j$, then she earns almost nothing.

\begin{restatable}{lemma}{thmOverpricing}
\label{thm:overpricing}
For any seller $j \in [m]$, define $\sigma_j$ and $\rho_j$ as in \cref{thm:security}.
Then there exists a constant $L_j \ge \max(B, \sigma_j)$ such that
for any prices $\qvec \in \mathbb{R}_{\ge 0}^m$ and allocation $\xvec \in \Dem(\qvec)$, we have
\[ q_j > L_j \text{ and } \max_{k \in [m] \setminus \{j\}} q_k = B
    \implies R_j(\qvec, \xvec) \le \frac{\rho_j}{2\kappa}. \]
\end{restatable}
\begin{proof}[Proof sketch]
Intuitively, if seller $j$'s price is too high, buyers will prefer the other sellers,
and since their prices are large enough to extract almost all buyer budget,
they will do so, leaving almost nothing for seller $j$.
We defer the full proof to \cref{sec:exist-extra:market-params}.
\end{proof}

The security parameters and the overpricing parameters together show
why it doesn't make sense to price too high:
seller $j$ can earn $\rho_j$ by pricing at $\sigma_j$,
but if she deviates to a price $q_j > L_j$,
then for some response $\qvec_{-j}$, she earns $\rho_j/(2\kappa) < \rho_j$.
It remains to show that such a response $\qvec_{-j}$ can be individually rational,
which we show in \cref{thm:large-dev-deter} using \cref{thm:general-inner-kkm},
which in turn uses KKM.

\begin{lemma}
\label{thm:large-dev-deter}
Let $A: \mathbb{R}_{\ge 0}^m \to [0, 1]^{n \times m}$ be a continuous allocation rule.
Define $(L_j)_{j \in [m]}$ as in \cref{thm:overpricing}.
Let $\pvec \in \beta\Delta_m$ be a price vector and $j \in [m]$ be a seller such that $p_j \ge \beta\rho_j/B$.
For every deviation $q_j > L_j$ by seller $j$, there exists a response
$\qvec_{-j} \in \prod_{k \neq j} [p_k, B]$ such that
\begin{tightenum}
\item ($j$ loses) $R^A_j(\qvec) < \nu p_j$.
\item (others gain) $R^A_k(\qvec) \ge \nu p_k$ for all $k \in [m] \setminus \{j\}$.
\end{tightenum}
\end{lemma}
\begin{proof}
Since $p_j > 0$, for all $k \in [m] \setminus \{j\}$, we get $p_k \le \beta - p_j < B$. Let
\[ U \defeq \bigg\{\qvec \in \{q_j\} \times \!\!\!\!\! \prod_{k \in [m] \setminus \{j\}} \!\!\! [p_k, B]:
    \max_{k \in [m] \setminus \{j\}} q_k = B\bigg\}. \]
For all $k \in [m] \setminus \{j\}$, define the gain function $g_k(\qvec) \defeq R_k(\qvec) - \nu p_k$.
Then $g_k$ is continuous.
If $q_k = p_k$ for some $k \in [m] \setminus \{j\}$, we have $R_k(\qvec) \le \nu p_k$, so $g_k(\qvec) \le 0$.

By \cref{thm:overpricing} (overpricing), for any $\qvec \in U$,
we get $R_j(\qvec) \le \rho_j/(2\kappa)$.
Also, $p_j \ge \beta\rho_j/B = \rho_j/(\kappa\nu)$.
Hence, $R_j(\qvec) \le \nu p_j / 2 < \nu p_j$.

Since $R_{\all}$ is monotonic in prices (\cref{thm:rall-mono}),
we get $R_{\all}(\qvec) \ge R_{\all}(\pvec) = \nu\beta$. Thus,
\[ \sum_{k \in [m] \setminus \{j\}} R_k(\qvec) = R_{\all}(\qvec) - R_j(\qvec)
    > \nu\beta - \nu p_j = \sum_{k \in [m] \setminus \{j\}} \nu p_k. \]
Hence, $\sum_{k \in [m] \setminus \{j\}} g_k(\qvec) > 0$.
Thus, we can apply \cref{thm:general-inner-kkm}.
This gives us $\qvec \in U$ such that $g_k(\qvec) \ge 0$ for all $k \in [m] \setminus \{j\}$,
i.e., $R_k(\qvec) \ge \nu p_k$.
\end{proof}

\subsection{RARE for Continuous Allocation Rules}
\label{sec:exist:continuous}

We are now ready to formally prove our main result:
the existence of a $\kappa$-RARE price vector
when the allocation rule is continuous.
In fact, we show something stronger:
\begin{tightenum}
\item We show that the sellers together extract a $1/\kappa$ fraction of the total buyer budget.
\item We show that the $\kappa$-RARE price vector $\pvec$ lies in the simplex $\beta\Delta_m$,
    and show a positive lower bound on each coordinate.
\item We show that the response $\qvec_{-j}$ to seller $j$'s deviation
    lies in the box $\prod_{k \neq j} [p_k, L_k]$.
\end{tightenum}
Note that we do not require the utilities to be concave.

\begin{theorem}
\label{thm:rare-exists-continuous}
Let $\Ical = ([n], [m], (u_i)_{i=1}^n, (b_i)_{i=1}^n)$ be a market instance
and $A: \mathbb{R}_{\ge 0}^m \to [0, 1]^{n \times m}$ be a continuous allocation rule.
Define parameters $(\rho_j)_{j=1}^m$ and $(L_j)_{j=1}^m$ as in \cref{sec:exist:overpricing}.
Then there exists a price vector $\pvec$ in the truncated simplex
$\{\yvec \in \beta\Delta_m: y_j \ge \beta\rho_j/B \text{ for all } j \in [m]\}$
such that for every seller $j \in [m]$ and every deviation $q_j \in \mathbb{R}_{\ge 0}$ by seller $j$,
there exists a response $\qvec_{-j} \in \prod_{k \in [m]\setminus\{j\}} [p_k, L_k]$ such that
\begin{tightenum}
\item (deviator $j$ doesn't gain much) $R_j^A(\qvec) \le \kappa \cdot \nu p_j = \kappa \cdot R_j^A(\pvec)$.
\item (responder $k$ doesn't lose) $R_k^A(\qvec) \ge \nu p_k \ge R_k^A((q_j, \pvec_{-j}))$
    for every seller $k \in [m] \setminus \{j\}$.
\end{tightenum}
\end{theorem}
\begin{proof}
Let $\pvec \in \beta\Delta_m$ be the prices obtained from
the double KKM argument (\cref{thm:rare-exists-continuous-2}).
Fix a seller $j \in [m]$. We first show that $p_j \ge \beta\rho_j/B$.

Suppose $p_j < \beta\rho_j/B$. Recall from \cref{thm:security} that $\rho_j/\nu \le \sigma_j$. Then
\[ p_j < \frac{\beta\rho_j}{B} = \frac{\rho_j}{\kappa\nu} \le \frac{\rho_j}{\nu} \le \sigma_j. \]
Let $q_j \defeq \sigma_j$. Since $q_j \in [p_j, L_j]$, by \cref{thm:rare-exists-continuous-2},
there exists a response $\qvec_{-j} \in \prod_{k \neq j} [p_k, L_k]$ such that
\[ R_j(\qvec) = \frac{p_j}{\beta} R_{\all}(\qvec) \le \frac{p_j}{\beta} B. \]
By \cref{thm:security} (security), we get $R_j(\qvec) \ge \rho_j$. Thus
\[ \rho_j \le \frac{p_j}{\beta} B \implies p_j \ge \frac{\beta\rho_j}{B}, \]
which is a contradiction. Hence, we have $p_j \ge \beta\rho_j/B$.

Now we show that there always exists a response $\qvec_{-j}$ satisfying the requirements of this theorem.
There are three cases, depending on the value of $q_j$.

\textbf{Case 1}: $q_j \in [p_j, L_j]$.
\\ Then by \cref{thm:rare-exists-continuous-2}, we get that there exists a response
$\qvec_{-j} \in \prod_{k \neq j} [p_k, L_k]$ such that for every $k \in [m]$, we have
\[ R_k(\qvec) = \frac{p_k}{\beta} R_{\all}(\qvec). \]
Thus,
\[ R_j(\qvec) = \frac{p_j}{\beta} R_{\all}(\qvec) \le \nu p_j \frac{B}{\nu\beta} = \kappa \nu p_j, \]
and for any $k \in [m] \setminus \{j\}$, using \cref{thm:rall-mono}, we have
\[ R_k(\qvec) = \frac{p_k}{\beta} R_{\all}(\qvec) \ge \frac{p_k}{\beta} R_{\all}(\pvec) = \nu p_k. \]
By \cref{thm:accounting}, we get $R_j(\pvec) = \nu p_j$ and $R_k((q_j, \pvec_{-j})) \le \nu p_k$.
Thus, $\qvec_{-j}$ satisfies the requirements of this theorem.

\textbf{Case 2}: $q_j < p_j$.
\\ Let $\qvec_{-j} \defeq \pvec_{-j}$. By \cref{thm:accounting}, we get
$R_j(\qvec) = \nu q_j < \nu p_j$ and for all $k \in [m] \setminus \{j\}$,
we get $R_k(\qvec) = \nu q_k = \nu p_k = R_k((q_j, \pvec_{-j}))$.
Thus, $\qvec_{-j}$ satisfies the requirements of this theorem.

\textbf{Case 3}: $q_j > L_j$.
\\ \Cref{thm:large-dev-deter} gives us a response $\qvec_{-j} \in \prod_{k \neq j} [p_k, B]$
such that $R_j(\qvec) \le \nu p_j$ and $R_k(\qvec) \ge \nu p_k$ for all $k \in [m] \setminus \{j\}$.
By \cref{thm:accounting}, we get $R_j(\pvec) = \nu p_j$ and $R_k((q_j, \pvec_{-j})) \le \nu p_k$.
Thus, $\qvec_{-j}$ satisfies the requirements of this theorem.
\end{proof}

\Cref{thm:rare-exists-continuous} implies the following simple corollary.

\begin{corollary}
\label{thm:rare-exists-continuous-corr}
For any market $\Ical$ and continuous allocation rule $A$,
there exists a $\kappa$-RARE price vector $\pvec$
at which the sellers' total revenue is $R_{\all}(\pvec) = B/\kappa$,
where $\kappa = 1$ in the rivalrous setting and
$\kappa = b_{\avg}/b_{\min}$ in the non-rivalrous setting.
\end{corollary}
\begin{proof}
\Cref{thm:rare-exists-continuous} gives us the $\kappa$-RARE price vector $\pvec \in \beta\Delta_m$.
By \cref{thm:accounting}, we get $R_{\all}(\pvec) = \nu\sum_{j=1}^m p_j = \nu\beta = B/\kappa$.
\end{proof}

\subsection{RARE for Non-Rivalrous Goods}
\label{sec:exist:non-rival}

In non-rivalrous markets, when the utility functions are strictly concave,
the demand set is a singleton, so there is a unique allocation rule.
Moreover, one can prove that such an allocation rule is continuous.
We defer the full proof to \cref{sec:exist-extra:non-rival}.

\begin{restatable}{theorem}{thmUniqueNrDemand}
\label{thm:unique-nr-demand}
Let $\Ical = ([n], [m], (u_i)_{i=1}^n, (b_i)_{i=1}^n)$ be a market instance.
Let $u_i$ be strictly concave for all $i \in [n]$.
Then for all prices $\pvec \in \mathbb{R}_{\ge 0}^m$,
the non-rivalrous demand $\DemNR(\pvec)$ is a singleton.
Moreover, the unique allocation rule $A$ is continuous.
\end{restatable}

Applying \cref{thm:rare-exists-continuous-corr} then shows that for strictly
concave utilities, a $(b_{\avg}/b_{\min})$-RARE always exists.
On the other hand, for linear utility functions,
there are instances where every allocation rule is discontinuous.

\begin{example}
\label{ex:lin-util-dem-discont}
Let there be two goods,
and a single buyer with budget 1 and utility function $u_1(\xvec) = x_1 + x_2$.
Let $\pvec \in (0, 1)^2$ be a price vector such that $p_1 + p_2 > 1$.
If $p_1 < p_2$, then $\Dem_1(\pvec) = \{(1, \frac{1-p_1}{p_2})\}$.
If $p_1 > p_2$, then $\Dem_1(\pvec) = \{(\frac{1-p_2}{p_1}, 1)\}$.
Thus, for any $\eps \in (0, 1/3)$, we have $\Dem_1((2/3-\eps, 2/3+\eps)) = \{(1, (1+3\eps)/(2+3\eps))\}$
and $\Dem_1((2/3+\eps, 2/3-\eps)) = \{((1+3\eps)/(2+3\eps), 1)\}$.
Thus, $\Dem_1$ is discontinuous.
\end{example}

However, we can still show that approximate RARE exists by adding
an infinitesimally small concave perturbation to the utilities.
We use $\vecOne$ to denote the allocation where every good is allocated to every buyer.

\begin{theorem}[\texorpdfstring{$(b_{\avg}/b_{\min})$}{(bavg/bmin)}-RARE for non-rivalrous markets]
\label{thm:nonrival-rare-exists}
Let $\Ical = ([n], [m], (u_i)_{i=1}^n, (b_i)_{i=1}^n)$ be a non-rivalrous market instance.
Suppose each utility $u_i: [0,1]^m \to \mathbb{R}_{\ge 0}$ is concave.
Let $\kappa \defeq b_{\avg}/b_{\min}$.
Then there exist prices $\pvec \in b_{\min}\Delta_m^{\circ}$ such that $\vecOne \in \DemNR(\pvec)$ and
for every seller $j \in [m]$ and deviation price $q_j \in \mathbb{R}_{\ge 0}$,
there exists a response $\qvec_{-j} \in \mathbb{R}_{\ge 0}^{[m]\setminus\{j\}}$ and allocation $\xvec \in \DemNR(\qvec)$
(where $\qvec \defeq (q_j, \qvec_{-j})$) such that:
\begin{tightenum}
\item Seller $j$'s revenue after the response is at most $\kappa$ times her original revenue, i.e.,
    $R_j(\qvec, \xvec) \le \kappa \cdot R_j(\pvec, \vecOne) = \kappa \cdot n p_j$.
\item No responder loses by responding, i.e., $R_k(\qvec, \xvec) \ge n p_k$
    for every responder $k \in [m] \setminus \{j\}$.
\end{tightenum}
\end{theorem}
\begin{proof}
For every $\eps \in [0, 1]$, define the perturbed instance
$\Ical^{(\eps)} \defeq ([n], [m], (u_i^{(\eps)})_{i=1}^n, (b_i)_{i=1}^n)$, where
\[ u_i^{(\eps)}(\xvec) \defeq u_i(\xvec) + \eps \sum_{k=1}^m x_k(2-x_k). \]
One can show that the parameters $\rho_j > 0$, $\sigma_j \ge \rho_j/\nu$,
and $L_j \ge \max(B, \sigma_j)$ can be picked to be independent of the perturbation.
See \cref{thm:robust-params} in \cref{sec:exist-extra:market-params}.

For $\eps_t = 1/t$, for each $t \in \N$ and each $i \in [n]$,
the utility function $u_i^{(\eps_t)}$ is strictly concave.
Hence, by \cref{thm:unique-nr-demand}, there is a unique continuous allocation rule
$A^{(\eps_t)}$ for each instance $\Ical^{(\eps_t)}$.
Apply \cref{thm:rare-exists-continuous} to obtain $\pvec^{(t)} \in b_{\min}\Delta_m$
and $p^{(t)}_j \ge b_{\min}\rho_j/B$.

By compactness, pass to a subsequence along which $\pvec^{(t)} \to \pvec \in b_{\min}\Delta_m$.
Relabel this subsequence and its corresponding perturbations using the same notation;
we still have $\eps_t \to 0$.
Fix this price subsequence before choosing a seller or deviation.
By the uniform lower bound, we get $p_j \ge b_{\min}\rho_j/B > 0$ for all $j \in [m]$.
Hence, every buyer can buy all the goods, so $\vecOne \in \DemNR(\pvec; \Ical)$.

Fix a seller $j \in [m]$ and deviation $q_j$. By \cref{thm:rare-exists-continuous},
we get a response $\qvec_{-j}^{(t)}$ in each instance $\Ical^{(\eps_t)}$ such that
$q^{(t)}_k \in [p_k^{(t)}, L_k]$ for all $k \in [m] \setminus \{j\}$.
Moreover, let $\xvec^{(t)} \defeq A^{(\eps_t)}(\qvec^{(t)})$ be the
unique element in $\DemNR(\qvec^{(t)}; \Ical^{(\eps_t)})$.
\Cref{thm:rare-exists-continuous} also guarantees
$R_j(\qvec^{(t)}, \xvec^{(t)}) \le \kappa n p_j^{(t)}$
and $R_k(\qvec^{(t)}, \xvec^{(t)}) \ge n p_k^{(t)}$ for all $k \in [m] \setminus \{j\}$.

The pairs $(\qvec_{-j}^{(t)}, \xvec^{(t)})$ lie in the compact set
\[ \bigg(\prod_{k \neq j} [0, L_k]\bigg) \times [0, 1]^{n \times m}. \]
Thus, there is a further subsequence along which
$(\qvec_{-j}^{(t)}, \xvec^{(t)}) \to (\qvec_{-j}, \xvec)$.
Along this same subsequence, $\pvec^{(t)} \to \pvec$ and $\eps_t \to 0$ still hold.
This further subsequence may depend on $j$ and $q_j$.
Let $\qvec \defeq (q_j, \qvec_{-j})$; since the deviation price is fixed, $\qvec^{(t)} \to \qvec$.
Then $\qvec_{-j} \in \prod_{k \neq j} [p_k, L_k]$, and by \cref{thm:demand-limit}, we get $\xvec \in \DemNR(\qvec; \Ical)$.
By the continuity of $R_j(\cdot, \cdot)$, we get $R_j(\qvec, \xvec) \le \kappa n p_j$
and $R_k(\qvec, \xvec) \ge n p_k$ for all $k \in [m] \setminus \{j\}$.
\end{proof}

\subsection{RARE for Rivalrous Goods}
\label{sec:exist:rival}

For non-rivalrous goods, strictly concave utilities determine a unique allocation rule.
For merely concave utilities, \cref{thm:nonrival-rare-exists} gives a $\kappa$-RARE guarantee
without a fixed allocation rule: a demand allocation can be chosen for each response.

For rivalrous goods, the demand set can contain multiple allocations even with strictly concave utilities.
E.g., if there are two buyers and two goods, priced such that no buyer can afford either good completely,
then the demand set $\DemR(\pvec)$ can have very different allocations.
We restrict the choice of allocation by maximizing a fixed \emph{welfare function}.
When its maximizer is unique at every price vector, this determines an allocation rule.

\begin{definition}[\texorpdfstring{$W$}{W}-rivalrous demand]
Let $W: \mathbb{R}_{\ge 0}^n \to \mathbb{R}$ be a continuous
and strictly monotone function, which we call the \emph{welfare function}.
Define the $W$-rivalrous demand at prices $\pvec \in \mathbb{R}_{\ge 0}^m$ as
\[ \Dem^W(\pvec) \defeq \argmax_{\xvec \in \AffR(\pvec)} W(u_1(\xvec_1), \ldots, u_n(\xvec_n)). \]
(The argmax is well-defined: since $W$ is continuous and $\AffR(\pvec)$ is a compact set, a maximum always exists.)
\end{definition}

Recall that $\AffR(\pvec)$ is the set of allocations where each buyer can afford her bundle.
Since $W$ is strictly monotone, the maximizer allocation $\xvec$ is buyer-PO, so $\Dem^W(\pvec) \subseteq \DemR(\pvec)$.

\Cref{ex:lin-util-dem-discont} shows that in a market with two sellers and one buyer
with linear utilities, every allocation rule can be discontinuous.
Thus, we study strictly concave utility functions.
For such utilities, when $W$ is concave, then $\Dem^W(\pvec)$ is a singleton
and the corresponding unique allocation rule is continuous.
We defer the full proof to \cref{sec:exist-extra:rival}.

\begin{restatable}{theorem}{thmUniqueRivalDemand}
\label{thm:unique-rival-demand}
Let $\Ical = ([n], [m], (u_i)_{i=1}^n, (b_i)_{i=1}^n)$ be a market instance.
Let $u_i$ be strictly concave for all $i \in [n]$.
Let $W: \mathbb{R}_{\ge 0}^n \to \mathbb{R}$ be a continuous, strictly monotone, concave function.
Then for all prices $\pvec \in \mathbb{R}_{\ge 0}^m$,
the $W$-rivalrous demand $\Dem^W(\pvec)$ is a singleton.
Moreover, the unique allocation rule $A$ is continuous.
\end{restatable}

We give several examples of concave welfare functions:
\begin{enumerate}
\item \textbf{Utilitarian social welfare}: Set $W(\xvec) = \sum_{i=1}^n x_i$.
\item \textbf{Weighted social welfare}: Set $W(\xvec) \defeq \wvec^T\xvec$, where $\wvec \in \mathbb{R}_{>0}^n$.
\item \textbf{Priority order}:
    Set $W(\xvec) \defeq \wvec^T\xvec$, where $w_i \defeq M^{-i}$ for each buyer $i \in [n]$
    and $M$ is a very large number. This approximates the requirement that we give the maximum
    possible utility to buyer 1, then the maximum possible utility to buyer 2, and so on.
    We can renumber the buyers \wLoG{} to accommodate any priority order.
\item \textbf{Weighted Nash Social Welfare}: For a very small $\eps > 0$ and $\wvec \in \Delta_n^{\circ}$,
    \[ W(\xvec) \defeq \prod_{i=1}^n x_i^{w_i} + \eps \sum_{i=1}^n x_i, \]
    gives an approximation to the Weighted Nash Social Welfare.
    Adding $\eps$ times the utilitarian social welfare ensures that $W$ is strictly monotone.
\item \textbf{Leximin Welfare}: For $0 < \eps \ll 1$,
    \[ W(\xvec) \defeq \sum_{i=1}^n \eps^{i-1} x_{[i]}, \]
    where $x_{[i]}$ is the $i\Th$ smallest coordinate in $\xvec$,
    gives an approximation to the Leximin Welfare.
\end{enumerate}
For all of the above welfare functions, there is a corresponding unique allocation rule
(when utilities are strictly concave) which is continuous.
For that allocation rule, a RARE always exists by \cref{thm:rare-exists-continuous-corr}.

When utilities are merely concave, $\Dem^W(\pvec)$ may contain several allocations,
so fixing $W$ need not determine a unique allocation rule.
The following theorem gives a version of RARE without a fixed allocation rule.
It fixes $W$ and the initial allocation $\xvec$, and allows a welfare-maximizing allocation $\zvec$
to be chosen separately for each seller's deviation and response.
We obtain this guarantee by perturbing the utilities to make them strictly concave and taking a limit.
The proof is similar to \cref{thm:nonrival-rare-exists},
so we defer the full proof to \cref{sec:exist-extra:rival}.

\begin{restatable}{theorem}{thmRivalRareExists}
\label{thm:rival-rare-exists}
Let $\Ical = ([n], [m], (u_i)_{i=1}^n, (b_i)_{i=1}^n)$ be a market instance.
Suppose each utility $u_i: [0,1]^m \to \mathbb{R}_{\ge 0}$ is concave.
Fix a continuous, strictly monotone, concave function $W: \mathbb{R}_{\ge 0}^n \to \mathbb{R}$.
Then there exist prices $\pvec \in B\Delta_m^{\circ}$ and an allocation $\xvec \in \Dem^W(\pvec)$
such that for every seller $j \in [m]$ and deviation price $q_j \in \mathbb{R}_{\ge 0}$,
there exists a response $\qvec_{-j} \in \mathbb{R}_{\ge 0}^{[m]\setminus\{j\}}$ and allocation $\zvec \in \Dem^W(\qvec)$
(where $\qvec \defeq (q_j, \qvec_{-j})$) such that:
\begin{tightenum}
\item $j$'s revenue after the response is at most her original revenue, i.e.,
    $R_j(\qvec, \zvec) \le R_j(\pvec, \xvec) = p_j$.
\item No responder loses by responding, i.e., $R_k(\qvec, \zvec) \ge p_k$
    for every responder $k \in [m] \setminus \{j\}$.
\end{tightenum}
\end{restatable}

\section{Dynamics}
\label{sec:dynamics}

In this section, we look at a simple dynamic that leads to approximate RARE prices.

\subsection{Limiting RARE}
\label{sec:lim-rare}

We begin by giving an alternative formulation of RARE,
which would be more useful for analyzing dynamics.
To do so, we define each seller $j$'s \emph{optimal deviation revenue} $V_j(\pvec_{-j})$.

\begin{definition}[Optimal deviation revenue]
\label{defn:vj}
Let $\Ical = ([n], [m], (u_i)_{i=1}^n, (b_i)_{i=1}^n)$ be a market,
$A: \mathbb{R}_{\ge 0}^m \to [0, 1]^{n \times m}$ be an allocation rule,
and $\pvec \in \mathbb{R}_{\ge 0}^m$ be a price vector.
When seller $j$ deviates to price $q_j \ge 0$,
define the other sellers' set of individually rational (IR) responses as
\[ \IR^A_j(q_j, \pvec_{-j}) \defeq \{\qvec_{-j} \in \mathbb{R}_{\ge 0}^{[m] \setminus \{j\}}:
    R^A_k(\qvec) \ge R^A_k((q_j, \pvec_{-j}))\; \forall k \in [m] \setminus \{j\}\}. \]
Note that this set is non-empty, since $\pvec_{-j} \in \IR^A_j(q_j, \pvec_{-j})$.
Define the deviator $j$'s \emph{worst-case revenue} at deviation $q_j$ as
\[ D^A_j(q_j, \pvec_{-j}) \defeq \inf_{\qvec_{-j} \in \IR^A_j(q_j, \pvec_{-j})} R^A_j(\qvec), \]
and her optimal deviation revenue as
\[ V^A_j(\pvec_{-j}) \defeq \sup_{q_j \in \mathbb{R}_{\ge 0}} D^A_j(q_j, \pvec_{-j}). \]
When $A$ is clear from context, we simply write $\IR_j$, $D_j$, and $V_j$
instead of $\IR^A_j$, $D^A_j$, and $V^A_j$.
\end{definition}

Intuitively, it might seem that a price vector $\pvec$ is RARE iff
$R_j(\pvec) \ge V_j(\pvec_{-j})$ for every seller $j \in [m]$,
giving us an alternative formulation of RARE.
But this formulation is \emph{not} definitionally equivalent to RARE, for a subtle reason.
In particular, for every deviation $q_j$ by seller $j$,
RARE requires an IR response $\qvec_{-j}$
such that the new revenue $R_j(\qvec)$ is at most $R_j(\pvec)$.
Our new definition, instead, requires a \emph{sequence of} IR responses for which $j$'s new revenue
converges to at most $R_j(\pvec)$.
We call this new definition \emph{limiting-RARE}.

\begin{definition}[limiting-RARE]
\label{defn:lim-RARE}
Consider a market $\Ical$ and allocation rule $A$.
Then for any $\delta \ge 0$ and $\kappa \ge 1$, the price vector $\pvec$
is called $\delta$-additive $\kappa$-\limRARE{} if for every seller $j \in [m]$, we have
\[ R^A_j(\pvec) \ge \frac{V^A_j(\pvec_{-j}) - \delta}{\kappa}. \]
\end{definition}

The following result shows that RARE and \limRARE{} are the same up to
an infinitesimally small additive constant.
We defer the proof to \cref{sec:lim-rare-equiv}.

\begin{restatable}{lemma}{thmRareEquivDefn}
\label{thm:rare-equiv-defn}
For any $\delta \ge 0$ and $\kappa \ge 1$,
a price vector $\pvec$ is $\delta$-additive $\kappa$-\limRARE{} iff
for every $\eps > 0$, it is $(\delta + \eps)$-additive $\kappa$-RARE.
Moreover, a $\delta$-additive $\kappa$-RARE price vector is also
$\delta$-additive $\kappa$-\limRARE.
\end{restatable}

\subsection{The Dynamic}
\label{sec:dynamics:defn}

Fix a continuous allocation rule $A$. The sellers initially have prices $\pvec^{(0)} = \vecZero$.
At each time step $t \in \N$, a seller $j$ is called \emph{unsatisfied} if
$V^A_j(\pvec^{(t-1)}_{-j}) > \kappa\nu p^{(t-1)}_j$.
Denote the set of all unsatisfied sellers by $J^{(t)}$.
If $J^{(t)} = \emptyset$, i.e., there are no unsatisfied sellers,
then let $\pvec^{(t)} \defeq \pvec^{(t-1)}$.
Otherwise, pick an arbitrary non-empty subset $S^{(t)} \subseteq J^{(t)}$, and update the prices as
\[ p^{(t)}_j \defeq \begin{cases}
\displaystyle \frac{V^A_j(\pvec^{(t-1)}_{-j})}{\kappa\nu} & \text{ if } j \in S^{(t)}
\\[0.8em] p^{(t-1)}_j & \text{ if } j \notin S^{(t)}
\end{cases}. \]

We will soon prove that $R_j(\pvec^{(t)}) = \nu p^{(t)}_j$ for all $t$ and $j$, which means that when $\kappa = 1$,
each seller either retains her old price, or increases her price to get her optimal deviation revenue,
making this a very natural dynamic.

Note that if $J^{(t)} = \emptyset$, then the prices freeze permanently.
Otherwise, prices of goods $S^{(t)}$ strictly increased, and the other goods had the same price as the previous round.
By picking $S^{(t)}$ appropriately, we can model different types of dynamics:
if $S^{(t)} = J^{(t)}$ for all $t$, then sellers update prices synchronously,
and if $|S^{(t)}| = 1$ for all $t$, then sellers update prices one-at-a-time.

\subsection{Analyzing the Dynamic}

The $\kappa$-RARE price vector whose existence we established in
\cref{thm:rare-exists-continuous} plays a key role in the analysis.
We begin by proving a useful property of that price vector:
it helps us upper bound the optimal deviation revenue of the price vectors it dominates.

\begin{lemma}
\label{thm:vbound-pstar}
There exists $\pvec^* \in \beta\Delta_m^{\circ}$ such that for every $\pvec \in \prod_{k=1}^m [0, p^*_k]$,
and seller $j \in [m]$, we have $V^A_j(\pvec_{-j}) \le \kappa\nu p^*_j$.
\end{lemma}
\begin{proof}
Let $\pvec^*$ be the $\kappa$-RARE price vector from \cref{thm:rare-exists-continuous}.
Pick any seller $j$ and deviation $q_j \ge 0$.
By \cref{thm:rare-exists-continuous}, there exists a response $\qvec_{-j}$ such that
\begin{tightenum}
\item $R_j(\qvec) \le \kappa R_j(\pvec^*) = \kappa\nu p^*_j$.
\item For each responder $k \in [m] \setminus \{j\}$, we have $R_k(\qvec) \ge \nu p_k^*$.
\end{tightenum}

For any responder $k \in [m] \setminus \{j\}$, we have
$R_k((q_j, \pvec_{-j})) \le \nu p_k \le \nu p_k^* \le R_k(\qvec)$.
Thus, $\qvec_{-j} \in \IR_j(q_j, \pvec_{-j})$, and so,
$D_j(q_j, \pvec_{-j}) \le R_j(\qvec) \le \kappa \nu p^*_j$.
Hence, $V_j(\pvec_{-j}) \le \kappa \nu p^*_j$.
\end{proof}

We now show that in each iteration of the dynamic, the prices are upper bounded by $\pvec^*$.

\begin{lemma}
Let $(\pvec^{(t)})_{t \in \N \cup \{0\}}$ be the dynamic defined in \cref{sec:dynamics:defn}.
Let $\pvec^*$ be as defined in \cref{thm:vbound-pstar}.
Then for all $t \in \N \cup \{0\}$ and $j \in [m]$, we have
$p^{(t)}_j \le p^{(t+1)}_j \le p^*_j$ and $R_j(\pvec^{(t)}) = \nu p^{(t)}_j$.
\end{lemma}
\begin{proof}
The monotonicity of $\pvec^{(t)}$ in $t$ follows easily from its definition,
i.e., $p^{(t)}_j \le p^{(t+1)}_j$ for all $j \in [m]$.

For all $t \in \N$ and $j \in [m]$, using \cref{thm:vbound-pstar}, we get
\[ p^{(t)}_j \le \max\bigg(p^{(t-1)}_j, \frac{V_j(\pvec^{(t-1)}_{-j})}{\kappa\nu}\bigg)
    \le \max(p^{(t-1)}_j, p^*_j). \]
Then we can prove $p^{(t)}_j \le p^*_j$ for all $j \in [m]$ and $t \in \N \cup \{0\}$ by induction on $t$,
since $\pvec^{(0)} = \vecZero$.

Moreover,
\[ \sum_{j=1}^m p^{(t)}_j \le \sum_{j=1}^m p^*_j = \beta, \]
so $R_j(\pvec^{(t)}) = \nu p^{(t)}_j$ for all $j$ by \cref{thm:accounting}.
\end{proof}

Observe that if the dynamic reaches equilibrium (i.e., the prices stop changing),
then the prices are $\kappa$-\limRARE{}.

\begin{observation}
Let $\pvec^{(0)} = \vecZero$, and for $t \in \N$, let $p^{(t)}$, $J^{(t)}$, and $S^{(t)}$
be as defined in \cref{sec:dynamics:defn}.
Suppose the prices do not change in some iteration $T+1$, i.e., $\pvec^{(T+1)} = \pvec^{(T)}$.
Then $R_j(\pvec^{(T)}) = \nu p^{(T)}_j \ge V_j(\pvec^{(T)}_{-j})/\kappa$ for all $j \in [m]$,
i.e., $\pvec^{(T)}$ is $\kappa$-\limRARE{}.
\end{observation}

Note that the prices $\pvec^{(T)}$ above need not be the same as $\pvec^*$.
There may be multiple $\kappa$-\limRARE{} price vectors.

Even if the dynamic does not reach equilibrium, we show that after sufficiently many rounds,
we get $\kappa$-RARE with a small additive approximation
if $S^{(t)}$ always includes the \emph{most unsatisfied} seller.

\begin{theorem}
\label{thm:dynamics-convergence}
Let $\pvec^{(0)} = \vecZero$, and for $t \in \N$, let $p^{(t)}$, $J^{(t)}$, and $S^{(t)}$
be as defined in \cref{sec:dynamics:defn}.
In every iteration $t \in \N$ where $J^{(t)} \neq \emptyset$, suppose $S^{(t)}$ always includes
some seller $j$ maximizing $V_j(\pvec^{(t-1)}_{-j})/(\kappa\nu) - p^{(t-1)}_j$.
Then for any $\delta > 0$:
\begin{enumerate}
\item (an approximate RARE occurs soon) $\exists t \le \floor{B/\delta}$ such that
    $\pvec^{(t)}$ is a $\delta$-additive $\kappa$-RARE.
\item (eventually prices remain approximately RARE) $\exists T \in \N \cup \{0\}$ such that
    for all $t > T$, $\pvec^{(t)}$ is a $\delta$-additive $\kappa$-RARE.
\end{enumerate}
\end{theorem}
\begin{proof}
For any $t \in \N \cup \{0\}$, let
\[ g_t \defeq \max_{j \in [m]} \max(0, V_j(\pvec^{(t)}_{-j}) - \kappa R_j(\pvec^{(t)})). \]
Then $\pvec^{(t)}$ is $g_t$-additive $\kappa$-\limRARE.
Since $R_j(\pvec^{(t)}) = \nu p^{(t)}_j$ for all $t$ and $j$,
we get $g_t = 0$ if $J^{(t+1)} = \emptyset$. Otherwise,
\begin{align*}
\kappa\nu \max_{j \in [m]} (p^{(t+1)}_j - p^{(t)}_j)
   &= \kappa\nu \max_{j \in S^{(t+1)}} (p^{(t+1)}_j - p^{(t)}_j)
    = \max_{j \in S^{(t+1)}} V_j(\pvec^{(t)}_{-j}) - \kappa R_j(\pvec^{(t)})
\\ &= \max_{j \in [m]} \max(0, V_j(\pvec^{(t)}_{-j}) - \kappa R_j(\pvec^{(t)}))
    = g_t.
\end{align*}
(Here the last equality follows from $S^{(t+1)}$ containing the most unsatisfied seller.)

Now for any $T \in \N \cup \{0\}$, sum this equality over all $t \in [T] \cup \{0\}$ to get
\[ \sum_{t=0}^T g_t = \kappa\nu \sum_{t=0}^T \max_{j \in [m]} (p^{(t+1)}_j - p^{(t)}_j)
    \le \kappa\nu \sum_{j=1}^m \sum_{t=0}^T (p^{(t+1)}_j - p^{(t)}_j)
    = \kappa\nu \sum_{j=1}^m p^{(T+1)}_j \le \kappa\nu\beta = B. \]
Pick any $\delta > 0$.
\begin{enumerate}
\item Let $T \defeq \floor{B/\delta}$. Then for some $t \in [T] \cup \{0\}$, we have
    $g_t \le B/(T+1) < \delta$, which implies that $\pvec^{(t)}$ is a $\delta$-additive $\kappa$-RARE.
    (Note that we get an approximate RARE, not just an approximate \limRARE{},
    since $g_t$ is \emph{strictly} less than $\delta$.)
\item There are at most $\floor{B/\delta}$ values of $t$ for which $g_t \ge \delta$.
    Let $T$ be the largest such that $g_T \ge \delta$ (and let $T$ be 0 if $g_t < \delta$ for all $t$).
    Then for all $t > T$, we have $g_t < \delta$, i.e., $\pvec^{(t)}$ is a $\delta$-additive $\kappa$-RARE.
    \qedhere
\end{enumerate}
\end{proof}

Intuitively, if we don't require $S^{(t)}$ to always include the most unsatisfied seller,
one seller might get stuck at a particularly bad price while the other sellers keep improving.

\appendix
\crefalias{section}{appendix}
\crefalias{subsection}{appendix}
\crefalias{subsubsection}{appendix}

\section{Details on Existence of RARE}
\label{sec:exist-extra}

\subsection{Security and Overpricing Parameters}
\label{sec:exist-extra:market-params}

Let $\evec_j$ denote the $j\Th$ standard basis vector in $\mathbb{R}^m$.
We first isolate the utility estimate used in the security and overpricing arguments.

\begin{lemma}[Uniform utility improvement]
\label{thm:uniform-utility-improvement}
Let $v: [0, 1] \times [0, 1]^m \to \mathbb{R}$ be continuous, and suppose that
$v(\eps, \cdot)$ is strictly monotone for every $\eps \in [0, 1]$.
Fix a good $k$ and $\eta \in (0, 1)$.
There exists $\delta > 0$ such that, for every $\eps \in [0, 1]$ and
$\zvec, \yvec \in [0, 1]^m$,
\[ z_k \le 1 - \eta \quad \text{and} \quad
    \norm{\yvec - (\zvec + \eta\evec_k)}_\infty < \delta
    \implies v(\eps, \yvec) > v(\eps, \zvec). \]
\end{lemma}
\begin{proof}
By continuity, strict monotonicity, and compactness,
\[ \gamma \defeq \min_{\substack{\eps \in [0, 1], \; \zvec \in [0, 1]^m \\ z_k \le 1 - \eta}}
    \bigl(v(\eps, \zvec + \eta\evec_k) - v(\eps, \zvec)\bigr) > 0. \]
Uniform continuity of $v$ gives $\delta > 0$ such that changing the bundle
by less than $\delta$ in the infinity norm changes its value by less than $\gamma/2$,
uniformly over $\eps$.
Thus, under the stated conditions,
$v(\eps, \yvec) > v(\eps, \zvec + \eta\evec_k) - \gamma/2 > v(\eps, \zvec)$.
\end{proof}

\thmSecurity*
\begin{proof}
Fix seller $j$ and use buyer 1 to secure positive revenue.
Apply \cref{thm:uniform-utility-improvement} with
$v(\eps, \zvec) \defeq u_1(\zvec)$, $k = j$, and $\eta = 1/4$, obtaining $\delta_j > 0$.
Set
\[ \sigma_j \defeq b_1 \min(1, \delta_j), \qquad
   \rho_j \defeq \frac{\sigma_j}{4}, \qquad
   \lambda_j \defeq \frac{\sigma_j}{4b_1}. \]
In particular, $0 < \lambda_j \le 1/4$ and $\lambda_j < \delta_j$.

Fix prices with $q_j = \sigma_j$ and an allocation $\xvec \in \Dem(\qvec)$.
Suppose that $\sum_i x_{i, j} < 1/4$.
Replace only buyer 1's bundle by
\[ \yvec_1 \defeq (1 - \lambda_j)\xvec_1 + \tfrac{1}{4}\evec_j. \]
This bundle lies in $[0, 1]^m$, and its cost is at most
\[ \qvec^T\yvec_1 \le (1 - \lambda_j)b_1 + \frac{\sigma_j}{4} = b_1. \]
In the rivalrous setting, the modification also preserves the stock constraints:
the total allocation of good $j$ is less than $1/2$, and every other total weakly decreases.
Since $x_{1, j} < 1/4$ and
\[ \norm{\yvec_1 - (\xvec_1 + \tfrac{1}{4}\evec_j)}_\infty
    = \lambda_j\norm{\xvec_1}_\infty < \delta_j, \]
\cref{thm:uniform-utility-improvement} gives $u_1(\yvec_1) > u_1(\xvec_1)$.
This contradicts buyer 1's optimality in the non-rivalrous setting
and Pareto optimality in the rivalrous setting.
Hence $\sum_i x_{i, j} \ge 1/4$, so $R_j(\qvec, \xvec) \ge \sigma_j/4 = \rho_j$.
\end{proof}

\thmOverpricing*
\begin{proof}
Fix seller $j$, and let
\[ a_j \defeq \frac{\rho_j}{2\kappa}, \qquad
   \eta_j \defeq \min\biggl(\frac{1}{2}, \frac{a_j}{2nB}\biggr). \]
For each buyer $i$ and good $k \ne j$, apply \cref{thm:uniform-utility-improvement}
to $v(\eps, \zvec) \defeq u_i(\zvec)$ with increment $\eta_j$ in good $k$.
Let $\delta_j > 0$ be the minimum of the resulting finitely many constants, and set
\[ L_j \defeq \max\biggl(B, \sigma_j, \frac{B}{\delta_j}\biggr). \]

Fix prices and demand $\xvec$ as in the statement, and suppose that
$R_j(\qvec, \xvec) > a_j$.
Choose a good $k \ne j$ with $q_k = B$, and a buyer $i$ with
$q_jx_{i, j} > a_j/n$.
We will replace buyer $i$'s purchases of good $j$ by $\eta_j$ units of good $k$.

First, there is enough room for this increase in good $k$.
In the non-rivalrous setting, buyer $i$'s budget constraint gives
\[ Bx_{i, k} \le b_i - q_jx_{i, j} < B - a_j/n,
    \qquad \text{so} \qquad x_{i, k} < 1 - 2\eta_j. \]
In the rivalrous setting, summing the budget constraints gives
\[ B \sum_{\ell = 1}^n x_{\ell, k}
    \le B - R_j(\qvec, \xvec) < B - a_j. \]
Thus the unsold quantity of good $k$ is greater than $a_j/B \ge 2n\eta_j$.
In either setting, adding $\eta_j$ units of good $k$ to buyer $i$ is feasible
with respect to the consumption and stock constraints.

Now replace only buyer $i$'s bundle by
\[ \yvec_i \defeq \xvec_i - x_{i, j}\evec_j + \eta_j\evec_k. \]
Its cost decreases, since $q_jx_{i, j} > a_j/n$ and $B\eta_j \le a_j/(2n)$.
Moreover,
\[ \norm{\yvec_i - (\xvec_i + \eta_j\evec_k)}_\infty
    = x_{i, j} \le \frac{b_i}{q_j} \le \frac{B}{q_j} < \delta_j. \]
By \cref{thm:uniform-utility-improvement}, this strictly improves buyer $i$'s utility.
As in the security proof, this contradicts demand in either setting.
Consequently, $R_j(\qvec, \xvec) \le a_j$.
\end{proof}

\begin{restatable}[Robust parameters]{lemma}{thmRobustParams}
\label{thm:robust-params}
Let $\Ical \defeq ([n], [m], (u_i)_{i=1}^n, (b_i)_{i=1}^n)$ be a market instance.
For every $\eps \in [0, 1]$, define the perturbed instance
$\Ical^{(\eps)} \defeq ([n], [m], (u_i^{(\eps)})_{i=1}^n, (b_i)_{i=1}^n)$, where
\[ u_i^{(\eps)}(\xvec) \defeq u_i(\xvec) + \eps \sum_{k=1}^m x_k(2-x_k). \]
There exist constants $\rho_j > 0$, $\sigma_j \ge \rho_j/\nu$, and $L_j \ge \max(B, \sigma_j)$
for every seller $j \in [m]$, independent of $\eps$,
such that for every $\eps \in [0, 1]$, seller $j \in [m]$, price vector $\qvec \in \mathbb{R}_{\ge 0}^m$,
and allocation $\xvec \in \Dem(\qvec; \Ical^{(\eps)})$, both of the following hold:
\begin{tightenum}
\item (security) $q_j = \sigma_j \implies R_j(\qvec, \xvec) \ge \rho_j$.
\item (overpricing)
    $\displaystyle q_j > L_j \text{ and } \max_{k \in [m] \setminus \{j\}} q_k = B
        \implies R_j(\qvec, \xvec) \le \frac{\rho_j}{2\kappa}$.
\end{tightenum}
\end{restatable}
\begin{proof}
For every buyer $i$, the function
\[ v_i(\eps, \zvec) \defeq u_i(\zvec) + \eps \sum_{k = 1}^m z_k(2 - z_k) \]
is jointly continuous in $(\eps, \zvec)$ and strictly monotone in $\zvec$.
Indeed, $z \mapsto z(2 - z)$ is increasing on $[0, 1]$, and $u_i$ is strictly monotone.
Thus \cref{thm:uniform-utility-improvement} applies uniformly over $\eps \in [0, 1]$.

For each seller $j$, choose $\delta_j$ using that lemma with $v_1$, good $j$,
and increment $1/4$, and set $\sigma_j = b_1 \min(1, \delta_j)$ and $\rho_j = \sigma_j/4$.
The feasible bundle modification in the proof of \cref{thm:security}
then gives the security bound for every $\eps \in [0, 1]$ with these same constants.

Next, set $a_j = \rho_j/(2\kappa)$ and
$\eta_j = \min(1/2, a_j/(2nB))$.
Apply \cref{thm:uniform-utility-improvement} to every $v_i$ and good $k \ne j$
with increment $\eta_j$, and let $\widehat{\delta}_j > 0$ be the minimum of its constants.
Set $L_j = \max(B, \sigma_j, B/\widehat{\delta}_j)$.
The feasible bundle modification in the proof of \cref{thm:overpricing}
now gives the overpricing bound for every $\eps \in [0, 1]$.
All of these parameters are independent of $\eps$.
\end{proof}

\subsection{RARE for Continuous Allocation Rule}
\label{sec:exist-extra:continuous}

\begin{lemma}
\label{thm:superlevel-closed}
Let $C \subseteq \mathbb{R}^m$ be a closed set and let $f: C \to \mathbb{R}$ be continuous.
Then the set $S = \{\xvec \in C: f(\xvec) \ge 0\}$ is closed.
\end{lemma}
\begin{proof}
Since $\mathbb{R}^m$ is a metric space, it suffices to show that
$S$ contains the limits of all its convergent sequences.

Let $(\xvec^{(k)})_{k \in \N}$ be a sequence in $S$ with $\xvec^{(k)} \to \xvec$
for some $\xvec \in \mathbb{R}^m$. We must show $\xvec \in S$.

First, each $\xvec^{(k)} \in S \subseteq C$, and $C$ is closed, so $\xvec \in C$.
In particular, $f(\xvec)$ is defined.

Second, since $f$ is continuous on $C$ and $(\xvec^{(k)})$ is a sequence in $C$
converging to $\xvec \in C$, we have $f(\xvec^{(k)}) \to f(\xvec)$.
Each $f(\xvec^{(k)}) \ge 0$, and $[0, \infty)$ is closed in $\mathbb{R}$,
so $f(\xvec) = \lim_{k \to \infty} f(\xvec^{(k)}) \ge 0$.

Hence $\xvec \in C$ and $f(\xvec) \ge 0$, i.e., $\xvec \in S$. Therefore $S$ is closed.
\end{proof}

\subsection{Limits of Affordable Allocations and Maximizers}
\label{sec:exist-extra:limits}

The following two lemmas will be used for both non-rivalrous and rivalrous demand.

\begin{lemma}[Limits of affordable allocations]
\label{thm:affordable-limits}
For each price vector $\pvec \in \mathbb{R}_{\ge 0}^m$ and buyer $i$, let
$K_i(\pvec) \defeq \{\zvec \in [0, 1]^m: \pvec^T\zvec \le b_i\}$.
Let $\mathcal{F}(\pvec)$ be either $\prod_{i=1}^n K_i(\pvec)$ or $\AffR(\pvec)$.
These sets are nonempty, compact, and convex.
If $\pvec^{(t)} \to \pvec$, then:
\begin{tightenum}
\item If $\xvec^{(t)} \in \mathcal{F}(\pvec^{(t)})$ and $\xvec^{(t)} \to \xvec$,
    then $\xvec \in \mathcal{F}(\pvec)$.
\item For every $\xvec \in \mathcal{F}(\pvec)$, there are
    $\xvec^{(t)} \in \mathcal{F}(\pvec^{(t)})$ with $\xvec^{(t)} \to \xvec$.
\end{tightenum}
\end{lemma}
\begin{proof}
Both sets contain the zero allocation and are bounded sets defined by closed linear inequalities,
so they are compact and convex.
The first claim follows by taking limits in these inequalities.

For the second claim, fix $\xvec \in \mathcal{F}(\pvec)$ and set
\[ \alpha_i^{(t)} \defeq
    \frac{b_i}{\max\{b_i, (\pvec^{(t)})^T\xvec_i\}},
    \qquad \xvec_i^{(t)} \defeq \alpha_i^{(t)}\xvec_i. \]
Each $\alpha_i^{(t)}$ lies in $(0, 1]$, and $\xvec_i^{(t)}$ is affordable at $\pvec^{(t)}$.
Scaling bundles down preserves the consumption bounds and, in the rivalrous setting,
the stock constraints. Thus $\xvec^{(t)} \in \mathcal{F}(\pvec^{(t)})$.
Since $\pvec^T\xvec_i \le b_i$ and $b_i > 0$, we have $\alpha_i^{(t)} \to 1$,
and hence $\xvec^{(t)} \to \xvec$.
\end{proof}

\begin{lemma}[Limits of maximizers]
\label{thm:maximizer-limits}
Let $\mathcal{F}$ be either of the feasible sets in \cref{thm:affordable-limits}.
Let $V_t, V: [0, 1]^{n \times m} \to \mathbb{R}$ be continuous functions such that
$V_t \to V$ uniformly.
If $\pvec^{(t)} \to \pvec$, $\xvec^{(t)} \to \xvec$, and
\[ \xvec^{(t)} \in \argmax_{\yvec \in \mathcal{F}(\pvec^{(t)})} V_t(\yvec), \]
then $\xvec \in \argmax_{\yvec \in \mathcal{F}(\pvec)} V(\yvec)$.
\end{lemma}
\begin{proof}
By \cref{thm:affordable-limits}, $\xvec \in \mathcal{F}(\pvec)$.
Fix any $\yvec \in \mathcal{F}(\pvec)$, and use the same lemma to choose
$\yvec^{(t)} \in \mathcal{F}(\pvec^{(t)})$ with $\yvec^{(t)} \to \yvec$.
Optimality gives $V_t(\xvec^{(t)}) \ge V_t(\yvec^{(t)})$.
Uniform convergence and continuity of $V$ imply that the two sides converge to
$V(\xvec)$ and $V(\yvec)$, respectively.
Thus $V(\xvec) \ge V(\yvec)$, as required.
\end{proof}

\subsection{RARE for Non-Rivalrous Goods}
\label{sec:exist-extra:non-rival}

\thmUniqueNrDemand*
\begin{proof}
Let $V(\xvec) \defeq \sum_{i=1}^n u_i(\xvec_i)$.
Maximizing $V$ over $\prod_i K_i(\pvec)$ is equivalent to maximizing each buyer's utility
over her affordable bundles, so its set of maximizers is exactly $\DemNR(\pvec)$.
The objective is continuous and strictly concave, and the feasible set is nonempty,
compact, and convex by \cref{thm:affordable-limits}.
Consequently, there is a unique maximizer $A(\pvec)$.

To prove continuity, let $\pvec^{(t)} \to \pvec$.
The allocations $A(\pvec^{(t)})$ lie in the compact set $[0, 1]^{n \times m}$.
By \cref{thm:maximizer-limits} with $V_t = V$, every convergent subsequence of these
allocations has limit in $\DemNR(\pvec) = \{A(\pvec)\}$.
It follows that the whole sequence converges to $A(\pvec)$, proving continuity.
\end{proof}

\begin{lemma}[Demand survives limits]
\label{thm:demand-limit}
Let $\Ical \defeq ([n], [m], (u_i)_{i=1}^n, (b_i)_{i=1}^n)$ be a market instance.
For every $\eps \in [0, 1]$, define the perturbed instance
$\Ical^{(\eps)} \defeq ([n], [m], (u_i^{(\eps)})_{i=1}^n, (b_i)_{i=1}^n)$, where
\[ u_i^{(\eps)}(\xvec) \defeq u_i(\xvec) + \eps \sum_{j=1}^m x_j(2 - x_j). \]
If $(\eps_t)_{t \in \N} \to 0$, $(\qvec^{(t)})_{t \in \N} \to \qvec$, $(\xvec^{(t)})_{t \in \N} \to \xvec$,
and $\xvec^{(t)} \in \DemNR(\qvec^{(t)}; \Ical^{(\eps_t)})$, then $\xvec \in \DemNR(\qvec; \Ical)$.
\end{lemma}
\begin{proof}
Define
\[ V_t(\yvec) \defeq \sum_{i=1}^n u_i^{(\eps_t)}(\yvec_i),
    \qquad V(\yvec) \defeq \sum_{i=1}^n u_i(\yvec_i). \]
Since $0 \le z(2 - z) \le 1$ on $[0, 1]$, we have
$\sup_{\yvec \in [0, 1]^{n \times m}} |V_t(\yvec) - V(\yvec)| \le nm\eps_t \to 0$.
The budgets are unchanged by perturbation, and $\xvec^{(t)}$ maximizes $V_t$
over $\prod_i K_i(\qvec^{(t)})$.
By \cref{thm:maximizer-limits}, $\xvec$ maximizes $V$ over $\prod_i K_i(\qvec)$,
which is equivalent to $\xvec \in \DemNR(\qvec; \Ical)$.
\end{proof}

\subsection{RARE for Rivalrous Goods}
\label{sec:exist-extra:rival}

\begin{lemma}[Demand survives limits]
\label{thm:demand-limit-rival}
Let $\Ical \defeq ([n], [m], (u_i)_{i=1}^n, (b_i)_{i=1}^n)$ be a market instance.
Let $W: \mathbb{R}_{\ge 0}^n \to \mathbb{R}$ be a continuous, strictly monotone function.
For every $\eps \in [0, 1]$, define the perturbed instance
$\Ical^{(\eps)} \defeq ([n], [m], (u_i^{(\eps)})_{i=1}^n, (b_i)_{i=1}^n)$, where
\[ u_i^{(\eps)}(\xvec) \defeq u_i(\xvec) + \eps \sum_{j=1}^m x_j(2 - x_j). \]
If $(\eps_t)_{t \in \N} \to 0$, $(\qvec^{(t)})_{t \in \N} \to \qvec$, $(\xvec^{(t)})_{t \in \N} \to \xvec$,
and $\xvec^{(t)} \in \Dem^W(\qvec^{(t)}; \Ical^{(\eps_t)})$, then $\xvec \in \Dem^W(\qvec; \Ical)$.
\end{lemma}
\begin{proof}
Define
\begin{align*}
V_t(\yvec) & \defeq W\bigl(u_1^{(\eps_t)}(\yvec_1), \ldots, u_n^{(\eps_t)}(\yvec_n)\bigr), \\
V(\yvec) & \defeq W\bigl(u_1(\yvec_1), \ldots, u_n(\yvec_n)\bigr).
\end{align*}
For each $i$, $u_i^{(\eps_t)}$ converges uniformly to $u_i$, with error at most $m\eps_t$.
All utility vectors lie in the compact box $\prod_i [0, u_i(\vecOne) + m]$.
Uniform continuity of $W$ on this box therefore gives uniform convergence $V_t \to V$.
The feasible set $\AffR$ depends only on prices and budgets, so it is unchanged by perturbation.
Apply \cref{thm:maximizer-limits} to conclude that $\xvec$ maximizes $V$ over $\AffR(\qvec)$,
i.e., $\xvec \in \Dem^W(\qvec; \Ical)$.
\end{proof}

\thmUniqueRivalDemand*
\begin{proof}
Let $V(\xvec) \defeq W(u_1(\xvec_1), \ldots, u_n(\xvec_n))$.
We first show that $V$ is strictly concave.
Fix distinct allocations $\xvec, \yvec$ and $\lambda \in (0, 1)$, and write
\[ \avec \defeq (u_i(\xvec_i))_{i=1}^n, \qquad
    \bvec \defeq (u_i(\yvec_i))_{i=1}^n, \qquad
    \mathbf{c} \defeq (u_i(\lambda\xvec_i + (1 - \lambda)\yvec_i))_{i=1}^n. \]
Concavity of the utilities gives $\mathbf{c} \ge \lambda\avec + (1 - \lambda)\bvec$ coordinatewise.
Since $\xvec_i \ne \yvec_i$ for some buyer $i$, strict concavity of $u_i$
makes at least one of these inequalities strict.
Strict monotonicity and concavity of $W$ now give
\begin{align*}
V(\lambda\xvec + (1 - \lambda)\yvec)
    & = W(\mathbf{c}) > W(\lambda\avec + (1 - \lambda)\bvec) \\
    & \ge \lambda V(\xvec) + (1 - \lambda)V(\yvec).
\end{align*}
Thus $V$ is continuous and strictly concave.
Since $\AffR(\pvec)$ is nonempty, compact, and convex,
its maximizer $A(\pvec)$ is unique.
Strict monotonicity of $W$ ensures that this maximizer is Pareto optimal,
so $A$ is an allocation rule.

Finally, let $\pvec^{(t)} \to \pvec$.
By compactness of $[0, 1]^{n \times m}$ and \cref{thm:maximizer-limits} with $V_t = V$,
every convergent subsequence of $A(\pvec^{(t)})$ has limit equal to the unique maximizer $A(\pvec)$.
Hence $A(\pvec^{(t)}) \to A(\pvec)$, proving continuity.
\end{proof}

\thmRivalRareExists*
\begin{proof}
For $\eps \in [0, 1]$, let $\Ical^{(\eps)}$ have utilities
\[ u_i^{(\eps)}(\zvec) \defeq u_i(\zvec) + \eps \sum_{k = 1}^m z_k(2 - z_k). \]
Choose common parameters $\sigma_j, \rho_j, L_j$ using \cref{thm:robust-params}.
For $\eps_t = 1/t$, the perturbed utilities are strictly concave.
By \cref{thm:unique-rival-demand}, their $W$-rivalrous demand defines a unique continuous
allocation rule $A^{(\eps_t)}$.
Apply \cref{thm:rare-exists-continuous} with $\nu = \kappa = 1$ and $\beta = B$ to obtain
prices $\pvec^{(t)} \in B\Delta_m$ such that $p_j^{(t)} \ge \rho_j$ for every $j$.
Let $\xvec^{(t)} \defeq A^{(\eps_t)}(\pvec^{(t)})$.
Then $R_j(\pvec^{(t)}, \xvec^{(t)}) = p_j^{(t)}$ by \cref{thm:accounting}.

By compactness, pass to a joint convergent subsequence
\[ (\pvec^{(t)}, \xvec^{(t)}) \to (\pvec, \xvec)
    \in B\Delta_m \times [0, 1]^{n \times m}. \]
Relabel this subsequence and its perturbations using the same notation;
we still have $\eps_t \to 0$.
Fix this subsequence before choosing a seller or deviation.
The lower bound gives $p_j \ge \rho_j > 0$, so $\pvec \in B\Delta_m^\circ$.
By \cref{thm:demand-limit-rival}, $\xvec \in \Dem^W(\pvec; \Ical)$.
Continuity of revenue in prices and allocations gives $R_j(\pvec, \xvec) = p_j$ for every $j$.

Fix a seller $j$ and a deviation price $q_j \ge 0$.
For each $t$, \cref{thm:rare-exists-continuous} gives a response
$\qvec_{-j}^{(t)} \in \prod_{k \ne j} [p_k^{(t)}, L_k]$.
With $\qvec^{(t)} = (q_j, \qvec_{-j}^{(t)})$ and
$\zvec^{(t)} \defeq A^{(\eps_t)}(\qvec^{(t)})$, its revenue guarantees are
\[ R_j(\qvec^{(t)}, \zvec^{(t)}) \le p_j^{(t)}, \qquad
    R_k(\qvec^{(t)}, \zvec^{(t)}) \ge p_k^{(t)} \quad \text{for every }k \ne j. \]
The pairs $(\qvec_{-j}^{(t)}, \zvec^{(t)})$ lie in the compact set
$\bigl(\prod_{k \ne j} [0, L_k]\bigr) \times [0, 1]^{n \times m}$.
Take a further joint convergent subsequence with limit $(\qvec_{-j}, \zvec)$.
This further subsequence may depend on $j$ and $q_j$, but the limits
$\pvec, \xvec$ remain fixed and $\eps_t \to 0$ still holds.
We have $\qvec_{-j} \in \prod_{k \ne j} [p_k, L_k]$ and
$\zvec \in \Dem^W(\qvec; \Ical)$ by \cref{thm:demand-limit-rival}.
Taking limits in the revenue inequalities gives
\[ R_j(\qvec, \zvec) \le p_j = R_j(\pvec, \xvec), \qquad
    R_k(\qvec, \zvec) \ge p_k \quad \text{for every }k \ne j, \]
as required.
\end{proof}

    \subsection{Unbounded Responses and Closedness}
\label{sec:unbounded-unclosed}

The proof of closedness in \cref{thm:rare-exists-continuous-2} uses the upper bounds on response prices
to extract a convergent subsequence of responses.
The following example shows that removing these bounds can make the outer KKM sets
non-closed, even when the utility function is strictly concave.

\begin{lemma}
\label[lemma]{thm:unbounded-unclosed}
There exists a market with one buyer of budget 1 and three goods,
whose utility function is continuous, strictly monotone, and strictly concave,
with the following property.
Let $A$ be any allocation rule for this market, and for each seller $j \in [3]$, define
\[ C_j^\infty \defeq \bigg\{\pvec \in \Delta_3:\; \forall q_j \in [p_j, \infty),\;
    \exists \qvec_{-j} \in \prod_{k \ne j} [p_k, \infty) \text{ such that }
    \frac{R_k^A(\qvec)}{R_{\all}(\qvec)} = p_k \;\forall k \in [3]\bigg\}. \]
Then $C_1^\infty$ is not closed.
This holds in both the rivalrous and non-rivalrous settings.
\end{lemma}
\begin{proof}
With one buyer, the rivalrous and non-rivalrous demand sets coincide.
Let the buyer's utility be
\[ u(\xvec) \defeq 2x_1 + x_3 + 2\sqrt{x_2} + \sqrt{x_3} - \frac12(x_1 - x_3)^2. \]
This function is continuous and satisfies $u(\vecZero) = 0$.
Its partial derivatives, wherever the corresponding coordinate is positive, are
\[ \frac{\partial u}{\partial x_1} = 2 - x_1 + x_3, \qquad
    \frac{\partial u}{\partial x_2} = \frac{1}{\sqrt{x_2}}, \qquad
    \frac{\partial u}{\partial x_3} = 1 + x_1 - x_3 + \frac{1}{2\sqrt{x_3}}. \]
These are positive on their domains in $[0, 1]^3$, so $u$ is strictly monotone.
All terms in $u$ are concave.
If two bundles differ in coordinate 2 or 3, the corresponding square-root term
gives a strict concavity inequality.
If they differ only in coordinate 1, the quadratic term gives a strict inequality.
Thus $u$ is strictly concave.
By \cref{thm:unique-nr-demand}, the demand set is a singleton for every price vector,
making the allocation rule $A$ unique.
Thus, we suppress the superscript $A$ in the revenue functions from here onwards.

For any $\pvec \in \Delta_3$ and $\qvec \ge \pvec$, we have
$R_{\all}(\qvec) = 1$ by \cref{thm:rev-vs-F}.
Hence, the revenue-proportion condition in $C_j^\infty$ is equivalent to
$R_k(\qvec) = p_k$ for every $k$.
We will show that
\[ \pvec^{(\eps)} \defeq \left(\frac12, \frac12 - \eps, \eps\right) \in C_1^\infty
    \quad \text{for every } 0 < \eps \le \frac1{64}, \]
but $\pvec^{(0)} = (1/2, 1/2, 0) \notin C_1^\infty$.

\paragraph{The limiting price vector is not in $C_1^\infty$.}
Let seller 1 deviate to $q_1 = 1$.
Suppose that there is a response with $q_2 \ge 1/2$ and $q_3 \ge 0$
that gives revenues $(1/2, 1/2, 0)$.
Let $\xvec$ be the buyer's optimal bundle at these prices.
The revenue requirements imply
\[ x_1 = \frac12, \qquad q_2x_2 = \frac12, \qquad q_3x_3 = 0. \]
If $q_3 > 0$, then $x_3 = 0$.
For sufficiently small $t > 0$, replacing $\xvec$ by
$\xvec - q_3t\evec_1 + t\evec_3$ preserves its cost and stays in $[0, 1]^3$.
The change in utility is $\sqrt{t} + O(t) > 0$ as $t$ tends to zero,
contradicting optimality.
Thus $q_3 = 0$, and strict monotonicity gives $x_3 = 1$.

At this bundle, the marginal utilities per unit of money satisfy
\[ \frac{1}{q_1}\frac{\partial u}{\partial x_1}(\xvec) = \frac52 > 2
    \ge 2\sqrt{x_2} = \frac{1}{q_2}\frac{\partial u}{\partial x_2}(\xvec), \]
where the last equality uses $q_2x_2 = 1/2$.
Since $x_1 < 1$ and $x_2 > 0$, shifting a sufficiently small amount of spending
from good 2 to good 1 strictly increases utility, again contradicting optimality.
Therefore $\pvec^{(0)} \notin C_1^\infty$.

\paragraph{Every $\pvec^{(\eps)}$ is in $C_1^\infty$.}
Fix $0 < \eps \le 1/64$ and any deviation $q_1 = s \ge 1/2$.
If $s = 1/2$, keeping the other prices unchanged restores all revenues.
Suppose $s > 1/2$, and let $a \defeq 1/(2s) \in (0, 1)$.
We construct an optimal bundle $(a, y, z)$ whose expenditures are $(1/2, 1/2 - \eps, \eps)$.

For $z \in [0, \eps]$, define
\[ h(z) \defeq a(2 - a + z). \]
We first choose $z \in (0, \eps)$ satisfying
\begin{equation}
\label{eq:unbounded-unclosed-z}
z(1 + a - z) + \frac12\sqrt{z} = 2\eps h(z).
\end{equation}
At $z = 0$, the left side is zero and the right side is positive.
At $z = \eps$, we have $h(\eps) \le 1 + \eps$ and
\[ \eps(1 + a - \eps) + \frac12\sqrt{\eps} \ge \frac12\sqrt{\eps}
    \ge 4\eps > 2\eps(1 + \eps) \ge 2\eps h(\eps). \]
Thus such a $z$ exists by the intermediate value theorem.

Set
\[ y \defeq \bigl((1 - 2\eps)h(z)\bigr)^2, \qquad
    \qvec \defeq \left(s, \frac{1/2 - \eps}{y}, \frac{\eps}{z}\right). \]
Since $0 < z < \eps$ and $0 < a < 1$, we have
\[ 0 < (1 - 2\eps)h(z) < (1 - 2\eps)(1 + \eps) = 1 - \eps - 2\eps^2 < 1. \]
Hence $0 < y < 1$, so $\qvec \ge \pvec^{(\eps)}$.
The bundle $\xvec \defeq (a, y, z)$ lies in $(0, 1)^3$ and exhausts the budget:
\[ (q_1a, q_2y, q_3z) = \left(\frac12, \frac12 - \eps, \eps\right). \]
Moreover, the definitions and \cref{eq:unbounded-unclosed-z} give
\begin{align*}
\frac{1}{q_1}\frac{\partial u}{\partial x_1}(\xvec) &= 2a(2 - a + z) = 2h(z),
\\ \frac{1}{q_2}\frac{\partial u}{\partial x_2}(\xvec) &= \frac{\sqrt{y}}{1/2 - \eps} = 2h(z),
\\ \frac{1}{q_3}\frac{\partial u}{\partial x_3}(\xvec) &= \frac{z(1 + a - z) + \sqrt{z}/2}{\eps} = 2h(z).
\end{align*}
Thus $\nabla u(\xvec) = 2h(z)\qvec$.
For any affordable bundle $\wvec \in [0, 1]^3$, concavity gives
\[ u(\wvec) \le u(\xvec) + \nabla u(\xvec)^T(\wvec - \xvec)
    = u(\xvec) + 2h(z)\qvec^T(\wvec - \xvec) \le u(\xvec). \]
Consequently, $\xvec$ is optimal.
By strict concavity, it is the unique optimal bundle and must be selected by $A$.
Thus $R_k(\qvec) = p_k^{(\eps)}$ for every seller $k$.
This proves $\pvec^{(\eps)} \in C_1^\infty$.
Since $\pvec^{(\eps)} \to \pvec^{(0)}$ as $\eps \to 0$,
the set $C_1^\infty$ is not closed.
\end{proof}

The constructed responses illustrate why the limiting argument fails.
For any fixed deviation $s > 1/2$, \cref{eq:unbounded-unclosed-z} implies
\[ \frac12\sqrt{z} \le 2\eps h(z) \le 2\eps(1 + \eps), \qquad
    q_3 = \frac{\eps}{z} \ge \frac{1}{16\eps(1 + \eps)^2}. \]
Thus seller 3's response price tends to infinity while her revenue tends to zero.
At $\pvec^{(0)}$, however, restoring her zero revenue forces her price to be zero.
The buyer then receives all of good 3, changing her relative preference
for goods 1 and 2 and preventing revenue restoration at the deviation $q_1 = 1$.

\section{RARE vs Limiting RARE}
\label{sec:lim-rare-equiv}

Recall the definition of \limRARE{} from \cref{sec:lim-rare}.

\thmRareEquivDefn*
\begin{proof}
Fix the allocation rule $A$.
(We omit its superscript in this proof.)
First, suppose that $\pvec$ is $\delta$-additive $\kappa$-RARE.
For every seller $j \in [m]$ and deviation $q_j \ge 0$, there exists a response
$\qvec_{-j} \in \IR_j(q_j, \pvec_{-j})$ such that
\[ D_j(q_j, \pvec_{-j}) \le R_j(\qvec) \le \kappa R_j(\pvec) + \delta. \]
Taking the supremum over $q_j \ge 0$ gives
$V_j(\pvec_{-j}) \le \kappa R_j(\pvec) + \delta$ for every seller $j$.
Thus, $\pvec$ is $\delta$-additive $\kappa$-\limRARE{}, proving the last claim.

Now suppose that $\pvec$ is $\delta$-additive $\kappa$-\limRARE{}.
Fix any $\eps > 0$, seller $j \in [m]$, and deviation $q_j \ge 0$.
The set $\IR_j(q_j, \pvec_{-j})$ is nonempty since it contains $\pvec_{-j}$,
and revenues are nonnegative, so its defining infimum $D_j(q_j, \pvec_{-j})$ is finite.
By the definition of infimum, there exists a response
$\qvec_{-j} \in \IR_j(q_j, \pvec_{-j})$ such that
\[ R_j(\qvec) < D_j(q_j, \pvec_{-j}) + \eps
    \le V_j(\pvec_{-j}) + \eps
    \le \kappa R_j(\pvec) + \delta + \eps. \]
Hence, $\pvec$ is $(\delta + \eps)$-additive $\kappa$-RARE.

Conversely, suppose that $\pvec$ is $(\delta + \eps)$-additive $\kappa$-RARE
for every $\eps > 0$.
By the first part of the proof, for every seller $j \in [m]$ and every $\eps > 0$,
\[ V_j(\pvec_{-j}) \le \kappa R_j(\pvec) + \delta + \eps. \]
Letting $\eps$ tend to zero gives $V_j(\pvec_{-j}) \le \kappa R_j(\pvec) + \delta$,
so $\pvec$ is $\delta$-additive $\kappa$-\limRARE{}.
\end{proof}

\section{Nash Equilibria are Uncommon}
\label{sec:nasheq-uncommon}

\subsection{Non-Rivalrous Goods}
\label{sec:nasheq-uncommon:non-rival}

\begin{theorem}[Nash equilibria are uncommon]
\label{thm:nasheq-uncommon}
Let $\Ical$ be a market with 2 non-rivalrous goods and $n$ buyers, where each buyer $i \in [n]$
has budget 1 and linear utility, with $u_{i,j} \in [1, 2)$ for each good $j \in [2]$.
Fix an allocation rule $A$. Then a Nash equilibrium exists iff all utility ratios are the same, i.e.,
\[ \frac{u_{1,1}}{u_{1,2}} = \frac{u_{2,1}}{u_{2,2}} = \cdots = \frac{u_{n,1}}{u_{n,2}}. \]
\end{theorem}
\begin{proof}
For $i \in [n]$, let $\lambda_i \defeq u_{i,1} / u_{i,2}$ be buyer $i$'s utility ratio.
We suppress the superscript $A$ in the revenue functions throughout the proof.
\WLoG, let $\lambda_1 \le \ldots \le \lambda_n$.
Since $u_{i,j} \in [1, 2)$ for all $i$ and $j$, we get $1 / 2 < \lambda_1$ and $\lambda_n < 2$.

We now show how the ratio of prices affects which buyers prefer which good.
At any prices $(q_1, q_2)$, if $q_1 < q_2 \lambda_k$, then for any $i$,
\[ \frac{q_1}{u_{i,1}} \cdot \frac{u_{i,2}}{q_2} = \frac{1}{\lambda_i} \cdot \frac{q_1}{q_2} < \frac{\lambda_k}{\lambda_i}. \]
Thus, buyers $\{k, \ldots, n\}$ strictly prefer good 1 over 2.
Conversely, if $q_1 > q_2 \lambda_k$, then for any $i$,
\[ \frac{u_{i,1}}{q_1} \cdot \frac{q_2}{u_{i,2}} = \lambda_i \cdot \frac{q_2}{q_1} < \frac{\lambda_i}{\lambda_k}, \]
so buyers $[k]$ strictly prefer good 2 over 1.

\paragraph{Nash equilibrium exists for equal utility ratios.}
Suppose $\lambda_i = \alpha$ for all $i \in [n]$.
Let $p_1 = \alpha / (1 + \alpha)$ and $p_2 = 1 / (\alpha + 1)$.
Then all buyers buy both goods fully, and we get $R_j(p_1, p_2) = n p_j$.

Suppose seller 1 has an improving deviation $q_1$, i.e., $R_1(q_1, p_2) > n p_1$.
Since $R_1(q_1, p_2) \le n q_1$, we get $q_1 > p_1 = \lambda_n p_2$.
Thus, all buyers strictly prefer good 2 over 1 at prices $(q_1, p_2)$,
so $R_1(q_1, p_2) = n (1 - p_2) = n p_1$, which is a contradiction.
Hence, seller 1 does not have an improving deviation.

Suppose seller 2 has an improving deviation $q_2$, i.e., $R_2(p_1, q_2) > n p_2$.
Since $R_2(p_1, q_2) \le n q_2$, we get $q_2 > p_2 = p_1 / \lambda_1$.
Thus, all buyers strictly prefer good 1 over 2 at prices $(p_1, q_2)$,
so $R_2(p_1, q_2) = n (1 - p_1) = n p_2$, which is a contradiction.
Hence, seller 2 does not have an improving deviation,
which implies that $(p_1, p_2)$ is a Nash equilibrium.

\paragraph{Nash equilibrium does not exist for unequal utility ratios.}
Assume $\lambda_1 < \lambda_n$. Let $(p_1, p_2)$ be a Nash equilibrium.
We will derive a contradiction.

Suppose $p_1 + p_2 < 1$. Then $R_1(1 - p_2, p_2) = n (1 - p_2) > n p_1 = R_1(p_1, p_2)$, which is a contradiction.
Hence, we have $p_1 + p_2 \ge 1$.

\paragraph{Both prices and both revenues are positive.}
If $p_j = 0$ for some seller $j$, the other seller $\ell$ has price $p_\ell \ge 1$.
Seller $j$ can choose
\[ 0 < q_j < \min\bigg\{1, p_\ell \min_{i \in [n]} \frac{u_{i,j}}{u_{i,\ell}}\bigg\}. \]
Every buyer then strictly prefers good $j$ and buys it fully,
giving seller $j$ positive revenue instead of zero, a contradiction.
Thus, both prices are positive. The same deviation now shows that both sellers' equilibrium revenues are positive.

\paragraph{The sum of prices exceeds the budget.}
Suppose $p_1 + p_2 = 1$. Every buyer buys both goods fully, so $R_j(p_1, p_2) = n p_j$.
If $p_1 / p_2 < \lambda_n$, choose
\[ p_1 < q_1 < \min\{1, \lambda_n p_2\}. \]
Buyer $n$ still strictly prefers good 1 and pays $q_1$ to seller 1.
Every other buyer pays seller 1 at least $1 - p_2 = p_1$:
her budget is exhausted, and her payment to seller 2 is at most $p_2$.
Hence, $R_1(q_1, p_2) \ge q_1 + (n - 1) p_1 > n p_1$, a contradiction.
Otherwise, $p_1 / p_2 \ge \lambda_n > \lambda_1$.
Choose $p_2 < q_2 < \min\{1, p_1 / \lambda_1\}$.
Buyer 1 still strictly prefers good 2, and the symmetric argument gives
$R_2(p_1, q_2) \ge q_2 + (n - 1) p_2 > n p_2$, again a contradiction.
Thus, $p_1 + p_2 > 1$.

\paragraph{No buyer has equal bang per buck for the two goods.}
Suppose $p_1 / p_2 = \lambda_k$ for some buyer $k$.
Let $c_j$ be buyer $k$'s payment to seller $j$ under $A$ at $(p_1, p_2)$.
Since her budget is exhausted,
\[ c_1 + c_2 = 1 < \min\{1, p_1\} + \min\{1, p_2\}. \]
Thus, some seller $j$ satisfies $g \defeq \min\{1, p_j\} - c_j > 0$.
Let $\ell$ be the other seller, and decrease $p_j$ to $p_j - \delta$,
where
\[ 0 < \delta < \min\{p_j, p_1 + p_2 - 1, g / n\}. \]
Choose $\delta$ small enough that every buyer with a strict bang-per-buck ordering
at the original prices retains that ordering; there are only finitely many buyers.
Buyer $k$ now strictly prefers good $j$, so her payment to seller $j$ becomes
$\min\{1, p_j - \delta\}$, an increase of at least $g - \delta$.
For any other buyer who originally strictly preferred good $j$, her payment changes
from $\min\{1, p_j\}$ to $\min\{1, p_j - \delta\}$, a decrease of at most $\delta$.
For a buyer who originally strictly preferred good $\ell$, her payment to seller $j$
remains $\max\{1 - p_\ell, 0\}$.
Finally, any other originally indifferent buyer now strictly prefers good $j$;
her old payment was at most $\min\{1, p_j\}$, so her payment decreases by at most $\delta$.
These statements hold for any allocation rule $A$.
Seller $j$'s revenue therefore increases by at least $g - n \delta > 0$, a contradiction.
Hence, $p_1 / p_2 \neq \lambda_i$ for every buyer $i$.

\paragraph{Both prices are at least the budget.}
A seller $j$ with $p_j < 1$ cannot be strictly preferred by any buyer:
otherwise, she could increase her price slightly while keeping it below 1
and preserving every buyer's strict bang-per-buck ordering.
Each buyer who prefers good $j$ would pay the increased price,
and every other buyer's payment to seller $j$ would remain $\max\{1 - p_\ell, 0\}$,
where $\ell$ is the other seller. This would be an improving deviation.
Now suppose $p_j < 1$. Since there are no ties, every buyer strictly prefers good $\ell$.
If $p_\ell < 1$, seller $\ell$ has an improving deviation by the same argument.
If $p_\ell\ge1$, every buyer spends her entire budget on good $\ell$,
leaving seller $j$ zero revenue, which we have already ruled out.
Thus, $p_1, p_2\ge1$.

\paragraph{The lower-revenue seller can profitably attract every buyer.}
The total revenue is $n$, so some seller $j$ has $R_j(p_1, p_2) \le n / 2$.
Let $\ell$ be the other seller. Since $u_{i,j} / u_{i,\ell} > 1 / 2$ for every buyer,
we can choose
\[ \frac12 < q_j < \min\bigg\{1, \min_{i \in [n]} \frac{u_{i,j}}{u_{i,\ell}}\bigg\}. \]
Since $p_\ell\ge1$, we have
\[ \frac{q_j}{u_{i,j}} < \frac{1}{u_{i,\ell}} \le \frac{p_\ell}{u_{i,\ell}}
\qquad \text{for every }i \in [n]. \]
Every buyer therefore strictly prefers good $j$ and buys it fully, giving
\[ R_j((q_j, \pvec_{-j})) = n q_j > n / 2 \ge R_j(p_1, p_2), \]
where $\pvec = (p_1, p_2)$.
This contradicts Nash equilibrium and completes the proof.
\end{proof}

\subsection{Non-Rivalrous Linear Utility to Rivalrous SPLC Utility}
\label{sec:lin-nonrival-to-splc-rival}

In \cref{sec:nasheq-uncommon:non-rival}, we proved that Nash equilibria are uncommon
in a large family of non-rivalrous markets. Now we extend this result to rivalrous markets,
by reducing the non-rivalrous markets to rivalrous markets.

Let $\Ical$ be a market where buyers have equal budgets and linear utilities over non-rivalrous goods.
We show how to construct a market $\Icalhat$ where buyers have
SPLC (separable piecewise-linear concave) utilities over rivalrous goods
such that $\pvec$ is a Nash equilibrium for $\Icalhat$ iff $\pvec/n$ is a Nash equilibrium for $\Ical$.

\begin{lemma}
\label{thm:rivalize-1}
Let $\Ical \defeq ([n], [m], (u_i)_{i=1}^n, (b)_{i=1}^n)$ be a market where each buyer $i \in [n]$
has budget $b > 0$ and a linear utility function $u_i(\zvec) = \sum_{j=1}^m u_{i,j}z_j$.

Let $u_{\min} \defeq \min_{i,j} u_{i,j}$, $u_{\max} \defeq \max_{i,j} u_{i,j}$, and $0 < \eps < u_{\min}^2/u_{\max}$.
Define each buyer $i$'s new utility function $\uhat_i$ as
\[ \uhat_i(\zvec)
    = \sum_{j=1}^m \begin{cases}u_{i,j} \cdot nz_j & \text{ if } z_j \le 1/n
        \\ u_{i,j} + \eps(nz_j - 1) & \text{ if } z_j > 1/n\end{cases}. \]
Define the new market $\Icalhat \defeq ([n], [m], (\uhat_i)_{i=1}^n, (b)_{i=1}^n)$. Then
\begin{enumerate}
\item $\uhat_i$ is strictly increasing and SPLC for each buyer $i \in [n]$.
\item For any prices $\pvec$, we have $\xvec \in \DemNR(\pvec; \Ical) \iff
    (\xvec \in [0, 1]^{n \times m} \text{ and } \xvec/n \in \DemR(n\pvec; \Icalhat))$.
\end{enumerate}
\end{lemma}
\begin{proof}
Since $0 < \eps < u_{\min}^2/u_{\max} \le u_{\min}$, each component of $\uhat_i$
is continuous and has slope $nu_{i,j}$ before $1/n$ and slope $n\eps$ afterward.
These slopes are positive and non-increasing, so $\uhat_i$ is strictly increasing and SPLC.

We first prove the forward implication. Fix arbitrary prices $\pvec \in \mathbb{R}_{\ge 0}^m$
and an allocation $\xvec \in \DemNR(\pvec; \Ical)$.
Then $\xvec \in [0,1]^{n \times m}$, and $\xvec/n$ satisfies
\[ \sum_{i=1}^n \frac{x_{i,j}}{n} \le \sum_{i=1}^n \frac{1}{n} = 1 \quad\text{for every }j,
    \qquad (n\pvec)^T\frac{\xvec_i}{n} = \pvec^T\xvec_i \le b \quad\text{for every }i. \]
Thus, $\xvec/n \in \AffR(n\pvec; \Icalhat)$.
We will prove its Pareto optimality by showing that it maximizes a positively weighted sum
of the utilities $\uhat_i$ over $\AffR(n\pvec; \Icalhat)$.

For $t \in [0,n]$, define
\[ f_{i,j}(t) \defeq u_{i,j}\min(t,1) + \eps\max(t-1,0). \]
Then $\uhat_i(\yvec_i)=\sum_j f_{i,j}(ny_{i,j})$ for every bundle $\yvec_i \in [0,1]^m$.

\paragraph{Case 1: The full bundle is affordable.}
Suppose $\sum_j p_j \le b$. Then $x_{i,j} = 1$ for all $i,j$,
since all utility coefficients are positive and $\xvec \in \DemNR(\pvec; \Ical)$.
For every $t \in [0,n]$,
\begin{align*}
f_{i,j}(t) - f_{i,j}(1)
&= u_{i,j}\min(t-1, 0) + \eps\max(t-1, 0)
\\ &= \begin{cases}u_{i,j}(t-1) & \text{ if } t \le 1 \\ \eps(t-1) & \text{ if } t > 1\end{cases}
\\ &\le u_{\min}(t-1),
\end{align*}
since $\eps \le u_{\min} \le u_{i,j}$ for all $i$ and $j$.
Consequently, every $\yvec \in \AffR(n\pvec; \Icalhat)$ obeys
\begin{align*}
& \sum_{i=1}^n \uhat_i(\yvec_i) - \sum_{i=1}^n \uhat_i(\xvec_i/n)
= \sum_{j=1}^m \sum_{i=1}^n (f_{i,j}(ny_{i,j}) - f_{i,j}(x_{i,j}))
\\ &\le \sum_{j=1}^m \sum_{i=1}^n u_{\min}(ny_{i,j} - 1)
= nu_{\min} \sum_{j=1}^m \bigg(\sum_{i=1}^n y_{i,j} - 1\bigg)
\le 0.
\end{align*}
Thus $\xvec/n$ maximizes the unweighted sum of utilities over $\AffR(n\pvec;\Icalhat)$ in this case,
so $\xvec/n \in \PO(\AffR(n\pvec; \Icalhat); \Icalhat) = \DemR(n\pvec; \Icalhat)$.

\paragraph{Case 2: The full bundle is unaffordable.}
Now suppose $\sum_j p_j > b$. Every buyer exhausts her budget at $\xvec$,
and every zero-price good $j$ satisfies $x_{i,j} = 1$ for all buyers $i$.
We will choose welfare weights using each buyer's bang-per-buck at the point where her budget is exhausted.

Let $M^+ \defeq \{j \in [m]: p_j > 0\}$ be the set of positively-priced goods.
Consider a buyer with utility coefficients $\vvec \in \mathbb{R}_{>0}^m$.
She purchases all zero-price goods $[m] \setminus M^+$,
and then positive-price goods from $M^+$ in decreasing order of $v_j/p_j$ till she spends $b$.
Let $\tau(\vvec)$ be the bang-per-buck of the last good purchased.
Formally, let $T(\vvec) \defeq \{v_j/p_j: j \in M^+\}$ be the set of bang-per-bucks, and
\[ \tau(\vvec) \defeq \max\Bigg\{t \in T(\vvec): \!\!\!\! \sum_{j \in M^+:\, v_j/p_j \ge t} \!\!\!\! p_j \ge b\Bigg\}. \]
The maximum selects the highest cutoff at which these goods still suffice to exhaust the budget.
The set $T(\vvec)$ is finite and nonempty, and its smallest element satisfies the budget inequality because $\sum_j p_j>b$.
Thus the maximum exists and is positive.
This formula groups tied goods together, so the cutoff does not depend on their order of purchase.
Zero-price goods are already purchased fully and do not enter this calculation.

For each actual buyer $i \in [n]$, let $\lambda_i \defeq \tau((u_{i,j})_{j=1}^m)$.
The goods strictly above her cutoff cost less than $b$,
but including those at the cutoff cost at least $b$:
\[ \sum_{j \in M^+:\, u_{i,j}/p_j > \lambda_i} \!\!\!\!\!\! p_j \; < \; b \;
    \le \!\!\!\! \sum_{j \in M^+:\, u_{i,j}/p_j \ge \lambda_i} \!\!\!\!\!\! p_j. \]
Thus every optimal bundle purchases all goods above the cutoff, none below it,
and spends the remaining budget on goods at the cutoff.
In particular, for all $j \in M^+$,
\begin{equation}
\label{eq:rivalize-cutoff}
u_{i,j}/p_j > \lambda_i \implies x_{i,j} = 1,
\qquad
u_{i,j}/p_j < \lambda_i \implies x_{i,j} = 0.
\end{equation}
We also have $x_{i,j} = 1$ for all $j \in [m] \setminus M^+$.

We use $1/\lambda_i$ as buyer $i$'s welfare weight.
For her original linear utility, this normalization makes each good $j \in M^+$ at her cutoff,
i.e., with $u_{i,j}/p_j=\lambda_i$, have marginal utility $u_{i,j}/\lambda_i=p_j$.
This aligns the buyers' cutoff marginal utilities with the same price vector.
Our goal is to prove that $\xvec/n$ maximizes
$\sum_i \uhat_i(\yvec_i)/\lambda_i$ over $\yvec \in \AffR(n\pvec; \Icalhat)$.

To control the positive utility slopes beyond the allotments, we first compare these weights across buyers.
Let $\lambda_0\defeq\tau(\vecOne)$ be the cutoff for a hypothetical buyer
who values one unit of every good equally and has the same budget $b$.
Increasing utility coefficients cannot decrease $\tau$: the goods in $M^+$ counted at the old cutoff
still cost at least $b$, and the smallest of their new bang-per-buck ratios
is an admissible cutoff at least as large as the old one.
Thus $\tau$ is coordinatewise non-decreasing.
The coefficients of goods in $[m] \setminus M^+$ do not affect $\tau$.
Also, multiplying all coefficients by $c>0$ multiplies all bang-per-buck ratios on $M^+$ by $c$,
giving $\tau(c\vvec) = c\tau(\vvec)$.
Since $u_{\min}\vecOne \le (u_{i,j})_{j=1}^m \le u_{\max}\vecOne$ coordinatewise, we obtain
\[ u_{\min}\lambda_0 \le \lambda_i \le u_{\max}\lambda_0. \]
Here the common budget is essential: all buyers' cutoffs are computed using the same function $\tau$.
Writing $\lambda_{\min} \defeq \min_i\lambda_i$ and $\lambda_{\max} \defeq \max_i\lambda_i$, we have
\[ \frac{\lambda_{\max}}{\lambda_{\min}} \le \frac{u_{\max}}{u_{\min}}. \]

Recall that $f_{i,j}(t) \defeq u_{i,j}\min(t,1) + \eps\max(t-1,0)$.
The normalized component $f_{i,j}/\lambda_i$ has slope $u_{i,j}/\lambda_i$ before $t=1$
and slope $\eps/\lambda_i$ afterward.
The bound on $\eps$ gives $\eps\lambda_{\max}/\lambda_{\min} < u_{\min}$.
Thus the largest normalized slope after $t=1$ is smaller than every normalized slope before $t=1$:
\begin{equation}
\label{eq:rivalize-tail-bound}
\theta\defeq\frac{\eps}{\lambda_{\min}}
    <\frac{u_{\min}}{\lambda_{\max}}
    \le\frac{u_{i,j}}{\lambda_i}\quad\forall i,j.
\end{equation}

We next find, for each good $j \in [m]$, a common slope that bounds changes in every buyer's normalized utility.
For $j \in M^+$, the cutoff conditions make $p_j$ a suitable slope below the allotment.
At $x_{i,j}=1$, however, such a bound must also account for the positive slope $\eps/\lambda_i$
beyond the allotment. We therefore define auxiliary prices $q_j\defeq\max(p_j,\theta)$ for all $j \in [m]$.
In particular, $q_j=\theta$ for $j \in [m] \setminus M^+$.
These numbers are used only to compare utilities; affordability remains defined by $\AffR(n\pvec;\Icalhat)$.
Crucially, we increase $p_j$ only for goods whose shared supply is fully allocated.
Indeed, if $x_{i,j}<1$ for some buyer $i$, then $j \in M^+$, since every zero-price good is purchased fully.
By \cref{eq:rivalize-cutoff}, $u_{i,j}/p_j \le \lambda_i$, so \cref{eq:rivalize-tail-bound} gives
\[ p_j\ge\frac{u_{i,j}}{\lambda_i}\ge\frac{u_{\min}}{\lambda_{\max}}>\theta. \]
Hence $q_j=p_j$ for every such good, or equivalently,
\begin{equation}
\label{eq:rivalize-full-supply}
q_j>p_j \implies x_{i,j}=1\text{ for every }i
    \implies \sum_i x_{i,j}=n.
\end{equation}

More precisely, the line of slope $q_j$ through the current point
$(x_{i,j},f_{i,j}(x_{i,j})/\lambda_i)$ lies above the graph of $f_{i,j}/\lambda_i$:
for every $i,j$ and $t \in [0,n]$,
\begin{equation}
\label{eq:rivalize-support}
\frac{f_{i,j}(t)-f_{i,j}(x_{i,j})}{\lambda_i}
    \le q_j(t-x_{i,j}).
\end{equation}
First consider $j \in M^+$.
If $x_{i,j}=0$, then \cref{eq:rivalize-cutoff} gives $u_{i,j}/p_j \le \lambda_i$,
so $u_{i,j}/\lambda_i\le p_j\le q_j$,
and all slopes of $f_{i,j}/\lambda_i$ are at most $q_j$.
If $0<x_{i,j}<1$, then \cref{eq:rivalize-cutoff} gives $u_{i,j}/p_j=\lambda_i$,
and $q_j=p_j=u_{i,j}/\lambda_i$,
so concavity gives the claimed inequality.
Finally, if $x_{i,j}=1$, then \cref{eq:rivalize-cutoff} gives $u_{i,j}/p_j \ge \lambda_i$,
so $p_j \le u_{i,j}/\lambda_i$. Together with \cref{eq:rivalize-tail-bound}, this yields
\[ \frac{\eps}{\lambda_i}\le\theta\le q_j
    \le\frac{u_{i,j}}{\lambda_i}, \]
so $q_j$ lies between the two slopes at the breakpoint of $f_{i,j}/\lambda_i$,
which again gives \cref{eq:rivalize-support}.

Now consider $j \in [m] \setminus M^+$. Then $x_{i,j}=1$ and $q_j=\theta$.
By \cref{eq:rivalize-tail-bound},
\[ \frac{\eps}{\lambda_i}\le q_j=\theta<\frac{u_{i,j}}{\lambda_i}. \]
Thus $q_j$ again lies between the two slopes at the breakpoint,
and the same concavity argument proves \cref{eq:rivalize-support} for these goods.

For any $\yvec \in \AffR(n\pvec;\Icalhat)$, apply \cref{eq:rivalize-support} with $t=ny_{i,j}$ and sum to obtain
\begin{align*}
\sum_i\frac{\uhat_i(\yvec_i)-\uhat_i(\xvec_i/n)}{\lambda_i}
&\le\sum_{i,j}q_j(ny_{i,j}-x_{i,j})\\
&=\sum_i\bigl((n\pvec)^T\yvec_i-\pvec^T\xvec_i\bigr)
    +n\sum_j(q_j-p_j)\left(\sum_i y_{i,j}-1\right)\\
&\le0.
\end{align*}
The equality uses \cref{eq:rivalize-full-supply}: whenever $q_j-p_j$ is positive,
the original allocation has $\sum_i x_{i,j}=n$.
The first term is non-positive because every buyer already spends $b$ at $\xvec$
and cannot spend more in an allocation from $\AffR(n\pvec;\Icalhat)$.
The second term is non-positive because $q_j-p_j\ge0$ and shared supply gives $\sum_i y_{i,j}\le1$.
Thus the budget constraints control the contribution of the original prices,
and the supply constraints control the contribution of the auxiliary price increases.
Since every weight $1/\lambda_i$ is positive, no allocation in $\AffR(n\pvec;\Icalhat)$
can Pareto dominate $\xvec/n$.
Together with the case where the full bundle is affordable, this proves the forward implication.

\paragraph{The reverse implication.}
Suppose $\xvec \in [0,1]^{n\times m}$ and
$\xvec/n \in \DemR(n\pvec;\Icalhat)\subseteq\AffR(n\pvec;\Icalhat)$.
In particular, $\pvec^T\xvec_i\le b$ for every buyer $i$.
If $\xvec\notin\DemNR(\pvec;\Ical)$, then some buyer $i$ has a bundle
$\zvec_i \in [0,1]^m$ with $\pvec^T\zvec_i\le b$ and $u_i(\zvec_i)>u_i(\xvec_i)$.
Define $\yvec_i\defeq\zvec_i/n$ and $\yvec_k\defeq\xvec_k/n$ for every $k\ne i$.
All buyers remain within budget, and for each good $j$,
\[ \sum_k y_{k,j}=\frac{z_{i,j}+\sum_{k\ne i}x_{k,j}}{n}\le1. \]
Thus $\yvec \in \AffR(n\pvec;\Icalhat)$.
Since $\uhat_i(\wvec/n)=u_i(\wvec)$ for every $\wvec \in [0,1]^m$, we have
\[ \uhat_i(\yvec_i)=u_i(\zvec_i)>u_i(\xvec_i)=\uhat_i(\xvec_i/n). \]
Every other buyer's utility is unchanged. Thus $\yvec$ Pareto dominates $\xvec/n$
within $\AffR(n\pvec;\Icalhat)$, a contradiction.
It follows that $\xvec \in \DemNR(\pvec;\Ical)$.
\end{proof}

\begin{theorem}
\label{thm:rivalize-2}
Let $\Ical \defeq ([n], [m], (u_i)_{i=1}^n, (b)_{i=1}^n)$ be a market
where every buyer has budget $b > 0$ and linear utilities.
Define $\Icalhat \defeq ([n], [m], (\uhat_i)_{i=1}^n, (b)_{i=1}^n)$ as in \cref{thm:rivalize-1}.
Let $A$ be a non-rivalrous allocation rule, i.e.,
$A(\pvec) \in \DemNR(\pvec; \Ical)$ for all $\pvec \in \mathbb{R}_{\ge 0}^m$.
Let $\Ahat(\pvec) \defeq A(\pvec/n)/n$. Then
\begin{enumerate}
\item $\Ahat$ is a valid rivalrous allocation rule for $\Icalhat$,
    i.e., $\Ahat(\pvec) \in \DemR(\pvec; \Icalhat)$ for all $\pvec \in \mathbb{R}_{\ge 0}^m$.
\item $\pvec$ is a Nash equilibrium for $(\Icalhat, \Ahat)$
    iff $\pvec/n$ is a Nash equilibrium for $(\Ical, A)$.
\end{enumerate}
\end{theorem}
\begin{proof}
Fix any $\pvec \in \mathbb{R}_{\ge0}^m$ and let $\xvec\defeq A(\pvec/n)$.
Since $\xvec \in \DemNR(\pvec/n;\Ical)$, \cref{thm:rivalize-1} gives
\[ \Ahat(\pvec)=\xvec/n \in \DemR(\pvec;\Icalhat). \]
Thus $\Ahat$ is a valid rivalrous allocation rule.

For every seller $j$,
\[ R_j^{\Ahat}(\pvec)
    =p_j\sum_i\frac{x_{i,j}}{n}
    =R_j^A(\pvec/n). \]
In particular, a unilateral deviation to price $s\ge0$ satisfies
\[ R_j^{\Ahat}((s,\pvec_{-j})) = R_j^A((s/n,\pvec_{-j}/n)). \]
As $s$ ranges over all non-negative prices, so does $s/n$.
Therefore $\pvec$ is a Nash equilibrium under $\Ahat$ iff $\pvec/n$ is a Nash equilibrium under $A$,
which is the claimed equivalence.
\end{proof}

\clearpage

\phantomsection
\addcontentsline{toc}{section}{References}

\newcommand{\etalchar}[1]{$^{#1}$}

\clearpage

\phantomsection
\addcontentsline{toc}{section}{AI Methodology Statement}
\section*{Acknowledgements}
We thank Yiling Chen, Jeffrey Linderoth, David C. Parkes, and Ariel Procaccia for helpful discussions on RARE.
These discussions took place primarily during the first author’s seminar talks on
preliminary results of this work at Harvard University and the University of Wisconsin–Madison. The research of Bhaskar Ray Chaudhury was supported by the NSF CAREER award CCF-2441580. Jugal Garg and Eklavya Sharma were supported by NSF grant CCF-2334461. The research of Xin Huang was supported by Wakate Grant Number 26K21169 and JST ERATO Grant Number JPMJER2301, Japan. 

\section*{AI Methodology Statement}
We used OpenAI's GPT-6 Astra to explore the properties of several candidate equilibrium notions,
and after several rounds of refinement, we arrived at RARE.
We used GPT-6 to assist with the development of existence proofs and the analysis of the dynamics.
The paper's main body was written by us, with some assistance from GPT-6 and Anthropic's Claude for copy-editing.
The appendices were written by GPT-6, with minor edits by us.
We used Claude Opus 5.5 to generate Lean proofs of this paper's main results.
We take full responsibility for the contents of this paper, including its proofs and references.

\end{document}